\documentclass[smallextended]{svjour3}       
\smartqed  
\usepackage{graphicx}
\usepackage{amsmath,amsfonts,amssymb,isomath}
\usepackage{float}
\usepackage{natbib}
\usepackage[usenames]{color}
\usepackage{xcolor}
\usepackage{mathptmx}      

\usepackage[colorlinks=true,citecolor=blue,urlcolor=blue,breaklinks]{hyperref}

\definecolor{light-gray}{gray}{0.95}
\definecolor{light-blue}{cmyk}{0.1,0,0,0} 
\definecolor{light-yellow}{cmyk}{0,0,0.2,0} 

\journalname{Living Reviews in Relativity}
\begin{document}

\title{Electromagnetic counterparts to massive black hole mergers
}


\author{Tamara Bogdanovi\'c  \and M. Coleman Miller \and Laura Blecha}
\authorrunning{Bogdanovi\'c, Miller, \& Blecha} 

\institute{T. Bogdanovi\'c \at
	School of Physics and\\ Center for Relativistic Astrophysics\\
              Georgia Institute of Technology\\
              837 State St. NW, Atlanta GA 30332, USA\\
              \email{tamarab@gatech.edu}          
           \and
           M.C.  Miller \at
           Department of Astronomy\\
             University of Maryland\\
             4296 Stadium Dr., College Park, MD 20742, USA\\
              \email{miller@astro.umd.edu}       
            \and
	   L. Blecha \at
	   Department of Physics \\
	   University of Florida \\
	   2001 Museum Rd., Gainesville, FL 32611, USA\\
	   \email{lblecha@ufl.edu}        
}

\date{Received: date / Accepted: date}

\maketitle

\begin{abstract}

The next two decades are expected to open the door to the first coincident detections of electromagnetic (EM) and gravitational wave (GW) signatures associated with massive black hole (MBH) binaries heading for coalescence. These detections will launch a new era of multimessenger astrophysics by expanding this growing field to the low-frequency GW regime and will provide an unprecedented understanding of the evolution of MBHs and galaxies. They will also constitute fundamentally new probes of cosmology and would enable unique tests of gravity. The aim of this Living Review is to provide an introduction to this research topic by presenting a summary of key findings, physical processes and ideas pertaining to EM counterparts to MBH mergers as they are known at the time of this writing. We review current observational evidence for close MBH binaries, discuss relevant physical processes and timescales, and summarize the possible EM counterparts to GWs in the precursor, coalescence, and afterglow stages of a MBH merger. We also describe open questions and discuss future prospects in this dynamic and quick-paced research area.

\end{abstract}

\setcounter{tocdepth}{3}
\tableofcontents

\section{Introduction}
\label{section:introduction}

Coincident detections of electromagnetic (EM) and gravitational wave (GW) signals from coalescences of massive black hole binaries\footnote{Black hole binaries with mass in the range of $10^5$--$10^{10}\,M_\odot$.} (MBHBs) have the potential to provide unparalleled understanding of the evolution of massive black holes (MBHs) in the context of large-scale structure \citep{kormendy13_araa,heckman14_araa}. Furthermore, since direct detection of GWs can yield accurate luminosity distances and EM counterparts can provide redshifts of coalescing binary systems, their combination yields fundamentally new multimessenger probes of the Universe's expansion \citep{schutz86,holz05,abbott17_hubble}. Coincident EM and GW detections can measure or place limits on differences in the arrival times of photons and gravitons from the same cosmological source, which would in turn inform us about the mass of the graviton and possible violations of the equivalence principle and Lorentz invariance in the gravitational sector \citep{kocsis08,hazboun13,yagi16,abbott16_tests}.

\smallskip
\noindent\fcolorbox{white}{white}{%
    \minipage[!t]{\dimexpr0.50\linewidth-2\fboxsep-2\fboxrule\relax}    
The outcome of these scientific endeavors directly depends on our ability to identify observed MBHB systems characterized by both messengers. At the time of this writing such detections are anticipated but are not a foregone conclusion. Their feasibility will be determined by the properties of MBHBs and environments in which they reside, as well as by the technical capabilities of EM and GW observatories.
 \endminipage}\hfill
\noindent\fcolorbox{black}{light-yellow}{%
    \minipage[!t]{\dimexpr0.47\linewidth-2\fboxsep-2\fboxrule\relax}
\begin{center} {\bf Coincident EM and GW detections can provide...} \end{center}
{ 
\noindent$\bullet$ Understanding of the evolution of MBHs in the context of large-scale structure.\\
\vspace{-2.0mm}

$\bullet$ Fundamentally new probes of the universe's expansion. \\

$\bullet$ Unique tests of the theory of gravity.\\
}
 \endminipage}\hfill\\

A basic requirement for a successful detection using any messenger is that at least some fraction of galaxy mergers lead to the formation of close MBH pairs that coalesce within the age of the Universe. Observations of dual and multiple AGNs with kiloparsec separations confirm that galactic mergers are natural sites for formation of wide MBH pairs, and in some cases multiplets.
Theoretical models and simulations of galactic mergers suggest that subsequent evolution of widely separated MBHs to smaller scales should be common. However, the timescales for this process remain uncertain, owing to the inherent challenges in modeling MBHB evolution from galactic to milliparsec scales. This, combined with a lack of concrete observational evidence for close, gravitationally bound MBH pairs with sub-parsec separations, currently precludes a definitive conclusion that galactic mergers result in MBHB coalescences. 

The link between merging galaxies and merging MBHs is nontrivial to establish in observations, because the two events are separated by hundreds of millions to billions of years and because the MBH coalescence timescale is very short relative to all other evolutionary stages of the binary. Consequently, MBH coalescences are difficult to identify in EM observations alone, without prior knowledge about their occurrence. This is where a specialized messenger, such as GWs, will play an unparalleled role in pinpointing the instance of a MBH merger. In this case, the temporal coincidence of a prompt EM counterpart with a gravitationally timed MBH merger may offer the best chance of identifying a unique host galaxy. Combined together, the EM and GWs can provide crucial new information about the link between hierarchical structure formation and MBH growth.
 
 Identification of a galaxy that is a host to a MBH coalescence also relies on uniqueness of the associated EM counterpart and the ability of astronomers to recognize it in observations. This entails a priori knowledge about the observational appearance of the EM counterpart as well as the wavelength band optimal for its detection. This is because there may be more than one, or indeed many, plausible host galaxies that are EM bright in the area on the sky associated with the GW signal. 

\smallskip
\noindent\fcolorbox{white}{white}{%
    \minipage[!t]{\dimexpr0.50\linewidth-2\fboxsep-2\fboxrule\relax}    
To be useful beyond detection, the EM and GW signatures must ultimately encode in a predictable way the information about MBHBs and environments in which they reside. If a MBHB heading for coalescence were to be detected within the next few years, the limiting factor in interpretation of its coincident GW and EM signatures would likely be the uncertainties related to its EM signatures. Indeed, the LIGO-Virgo GW community has demonstrated that mergers of 
 \endminipage}\hfill
\noindent\fcolorbox{black}{light-yellow}{%
    \minipage[!t]{\dimexpr0.47\linewidth-2\fboxsep-2\fboxrule\relax}
\begin{center} {\bf Open questions} \end{center}
{ 
$\bullet$ Do MBHBs form and coalesce and on what timescales?\\
\vspace{-2.0mm}

\noindent$\bullet$ Are there unique EM signatures associated with MBHB coalescences?\\
\vspace{-2.0mm}


$\bullet$ Do they encode the properties of MBHBs and their environments?\\
}
 \endminipage}\hfill\\
%
stellar origin black hole binaries can be successfully detected and interpreted even in the absence of any EM counterparts.  The EM counterparts associated with MBH coalescences nevertheless remain an active and quick-paced research area with realistic potential for major breakthroughs before the first multimessenger detection is made.

The aim of this Living Review is to provide an introduction to this research topic by presenting a  summary of key findings, physical processes and ideas pertaining to EM counterparts to MBH mergers as they are known at the time of this writing. We review current observational evidence for close MBHBs in Sect.~\ref{section:observations} and discuss the relevant physical processes and timescales in Sect.~\ref{section:timescales}. In Sect.~\ref{section:signatures} we summarize the possible precursor, coalescence, and afterglow EM counterparts to GWs associated with MBHBs. We describe open questions, discuss future prospects in this research area, and conclude in Sect.~\ref{section:conclusions}.
 


\section{Observational evidence for close massive black hole binaries}
\label{section:observations}

Galactic mergers are an important driver of galaxy evolution and a natural channel for formation of MBH pairs and, possibly, multiplets \citep{bbr80}. Most of the scientific attention was initially focused on the galaxies in the act of merging, and their central MBHs were considered passive participants, taken for a ride by their hosts. The realization that MBHs play an important role in the evolution of their host galaxies \citep{kormendy95_araa,1998AJ....115.2285M,ferrarese00,gebhardt00,tremaine02} spurred a number of observational and theoretical studies, undertaken over the last few decades, with an aim to determine what happens to MBHs once their host galaxies merge. 

From an observational point of view, this question is pursued through EM searches for dual and multiple MBHs with a variety of separations, ranging from $\sim$ tens of kpc to sub-parsec scales \citep[see][for a review]{derosa19}. The multi-wavelength searches for MBH systems with large separations, corresponding to early stages of galactic mergers, have so far successfully identified a few dozen dual and offset active galactic nuclei \citep[AGNs;][and others]{komossa03, koss11, koss16, liu13b, comerford15, barrows16}. MBHs with even smaller (parsec and sub-parsec) separations are representative of the later stages of galactic mergers in which the two MBHs are sufficiently close to form a gravitationally bound pair.

\smallskip
\noindent\fcolorbox{white}{white}{%
    \minipage[!t]{\dimexpr0.50\linewidth-2\fboxsep-2\fboxrule\relax}    
A MBH pair is gravitationally bound if the amount of gas and stars enclosed within its orbit is smaller than its own mass. For a wide range of MBH masses and host galaxy properties, this happens when the two holes reach separations of $\sim 1-10$\,pc (see Sect.~\ref{ssection:bound}). Throughout this paper, we refer to such systems as \emph{massive black hole binaries} and use the term ``dual MBHs'' for unbound pairs of MBHs in the same general galactic vicinity. A key characteristic of gravitationally bound MBHBs is that they are 
 \endminipage}\hfill
\noindent\fcolorbox{black}{light-yellow}{%
    \minipage[!t]{\dimexpr0.47\linewidth-2\fboxsep-2\fboxrule\relax}
\begin{center} {\bf Observational evidence for MBHBs (from more to less direct)} \end{center}
{ 
\noindent$\bullet$ Emission of gravitational waves.\\
\vspace{-2.0mm}

$\bullet$ Direct imaging of double nuclei with Very Long Baseline Interferometry. \\
\vspace{-2.0mm}

$\bullet$ Quasi-periodicity in light curves of AGNs and quasars.\\
\vspace{-2.0mm}

$\bullet$ Doppler-shifted emission lines in spectra of AGNs and quasars.\\
}
 \endminipage}\hfill\\
observationally elusive and expected to be intrinsically rare.  While the frequency of binaries is uncertain and dependent on their unknown rate of evolution on small scales (see Sect.~\ref{section:timescales}), theorists estimate that a fraction of $<10^{-2}$ AGNs at redshift $z<0.6$ may host MBHBs \citep{kelley19}, although only a small fraction can be identified as such \citep[see for example the estimates in][]{volonteri09}.

These results have two important implications: (a) any EM search for MBHBs must involve a large sample of AGNs, making the archival data from large surveys of AGNs an attractive starting point and (b) observational techniques used in the search must be able to distinguish signatures of binaries from those of AGNs powered by single MBHs. Observational techniques used to search for such systems in large archival datasets have so far largely relied on direct imaging, photometry, and spectroscopic measurements. They have recently been complemented by observations with pulsar timing arrays (PTAs), which could detect GWs from massive MBHBs. In the remainder of this section we summarize the outcomes of and future prospects for these different observational approaches. We also direct the reader to \citet{dotti12} and \citet{schnittman13} for reviews of a broad range of MBHB signatures proposed in the literature.

\subsection{Emission of gravitational waves}
\label{ssec:gw}

\begin{figure}[t]
\center
\includegraphics[width=0.65\textwidth, trim=0 0 0 0, clip]{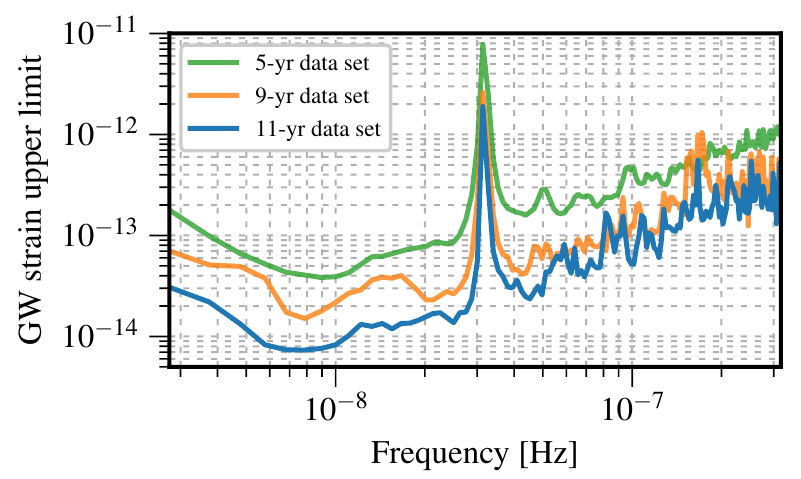} 
\includegraphics[width=0.61\textwidth,, trim=-6 0 0 0, clip]{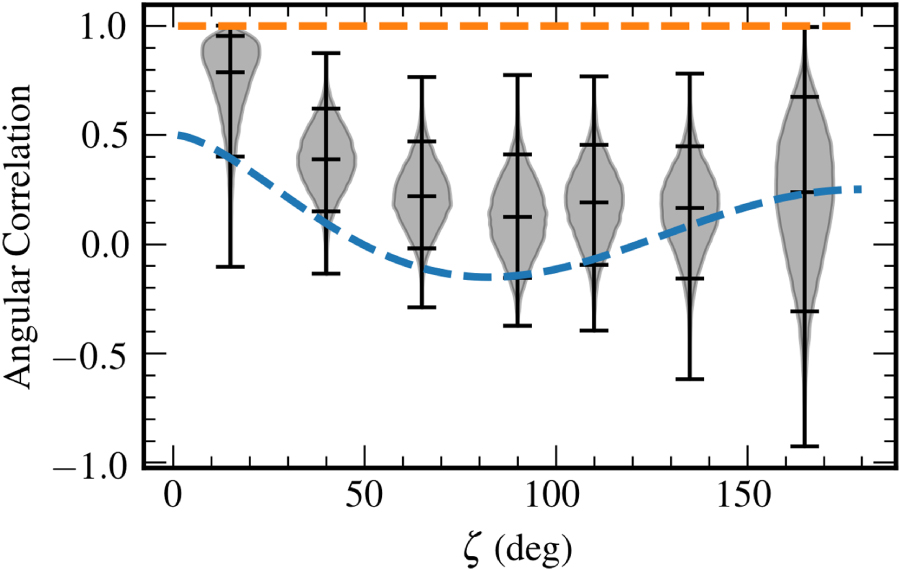} 
\caption{{\em Top:} Sky-averaged sensitivity (shown as the upper limit on the GW strain amplitude) as a function of the GW frequency from the NANOGrav 5-year data set (green),  9-year data set (orange), and 11-year data set (blue). The data is presently most sensitive at the frequency of 8\,nHz and least sensitive around $3\times 10^{-8}\,{\rm Hz} = 1 {\rm year}^{-1}$, corresponding to the orbital frequency of Earth around the Sun. Any MBHBs with GW strain above the 11-year sensitivity curve would at this point be detected by NANOGrav. Figure from \citet{aggarwal19}. {\em Bottom:} Angular correlation of common-spectrum pulsar timing noise versus angular separation of pulsars on the sky, from the NANOGrav 12.5 year dataset. The violin plots (gray/black) show the marginalized posteriors for the interpulsar spatial correlations. For comparison, the orange dashed line shows the flat correlation signature between pulsars expected as a signature of error in the timescale (e.g., drifts in clock standards), and the blue dashed curve shows the Hellings \& Downs correlation predicted to result from a stochastic background of GWs from MBHBs. Figure from \citet{arzoumanian20}. }
\label{fig1}      
\end{figure}

Detection of gravitational radiation emitted by a MBHB heading for coalescence constitutes the most direct evidence for the existence of such systems. At the time of this writing, GWs from MBHBs have not yet been definitively detected. However, expectations for such detections of gravitational radiation have been raised by the success of the LIGO-Virgo Collaboration \citep[e.g.,][]{abbott16a}, by the selection of the Laser Interferometer Space Antenna (LISA) for a large-class mission in the European Space Agency science program, and by the success of the LISA Pathfinder mission \citep{armano16} and the TianQin-1 experimental satellite \citep{luo20}. Recently, three PTA experiments, the North American Nanohertz Observatory for Gravitational Waves (NANOGrav), the Parkes Pulsar Timing Array (PPTA) and the European Pulsar Timing Array (EPTA) reported strong evidence for a signature of the common-spectrum, stochastic process \citep{arzoumanian20, goncharov21,chen21}. Even though the signal does not show sufficient evidence of quadrupolar correlations needed to claim detection of a stochastic background of GWs from MBHBs (see Fig.~\ref{fig1} and the text below), it provides a useful testing ground for a variety of theoretical models, and adds to the sense that an era of low-frequency GW astronomy is imminent.

PTAs seek to detect GWs by searching for correlations in the timing observations of a network of millisecond pulsars \citep{hellings83, foster90}. Specifically, the deviation of pulse arrival times from different pulsars due to an intervening nHz GW is predicted to exhibit a characteristic correlation with the angle between the pulsars on the sky; this is the so-called \citet{hellings83} curve. The three mentioned PTA experiments currently in operation, NANOGrav \citep{mclaughlin13}, EPTA \citep{desvignes16}, and PPTA \citep{hobbs13}, together form the International Pulsar Timing Array \citep[IPTA;][]{verbiest16}.  PTAs are sensitive to GWs with frequencies between a few and few hundred nHz, and their sensitivity increases for longer observational time baselines. This is illustrated in Fig.~\ref{fig1}, which shows the sensitivity as function of the GW frequency of NANOGrav based on the 5, 9 and 11 years of timing observations of about forty pulsars \citep{aggarwal19}. PTA sensitivity is also being steadily improved as more pulsars are identified and added to the network.

The expected signals for PTAs include (i) the aforementioned stochastic GW background, which is produced by an ensemble of unresolved MBHBs emitting GWs at different frequencies and distances throughout the Universe, and (ii)  continuous GWs from individual MBHBs, which would exhibit negligible frequency evolution on human timescales. At this point it is unclear whether the stochastic GW background will be detected first \citep{rosado15, arzoumanian20} or whether it will be detected contemporaneously with the individual MBHBs \citep{mingarelli17, kelley18}. In either case, a wide range of theoretical predictions agree that GWs from MBHBs are expected to be observed by the PTAs within the next decade. If so, their discovery and analysis will be well timed to inform expectations for GW detections with the space-based observatory LISA, which is planned to launch in the mid-2030s. Because PTAs are sensitive to the most massive MBHB systems out to $z\sim2$ \citep{sesana08, chen17}, they stand a chance of being associated with some of the most EM luminous AGNs in the local Universe. This bodes well for the prospects for multimessenger studies of MBHBs in the near future. 

PTAs are sensitive to individual MBHBs with high total mass, $\sim 10^8$--$10^{10}\,M_\odot$, and orbital periods on the order of months to years. For example, the NANOGrav collaboration performed a search in their 11-year dataset for GWs from individual MBHBs on circular orbits. While no statistically significant evidence for GWs has been found in the data, a physically interesting limit has been placed on the distances to, or equivalently masses of, individual MBHBs \citep{aggarwal19}. The 11-year data can rule out the existence of MBHBs within 120\,Mpc with chirp mass\footnote{The chirp mass is defined as $\mathcal{M}\equiv (M_1\,M_2)^{3/5}/(M_1+M_2)^{1/5} = q^{3/5}M/(1+q)^{6/5}$, where $M_1$ and $M_2$ are the masses of the primary and secondary MBH, $M=M_1 +M_2$ is the binary mass, and $q=M_2/M_1\leq 1$ is the mass ratio. Note that $\mathcal{M} \approx q^{3/5}M$ for $q\ll1$ and $\mathcal{M} \approx M/2$ for $q=1$.} $\mathcal{M} = 10^9\,M_\odot$ emitting GWs with frequency $8$\,nHz, which falls in the most sensitive part of the spectrum (see Fig.~\ref{fig1}). The existence of binaries with even larger chirp masses of $\mathcal{M} = 10^{10}\,M_\odot$, emitting at the same frequency, can be ruled out to even higher, cosmological distances of 5.5\,Gpc.


\begin{figure}[t]
\center
\includegraphics[trim=0 0 0 0, clip, scale=1.0]{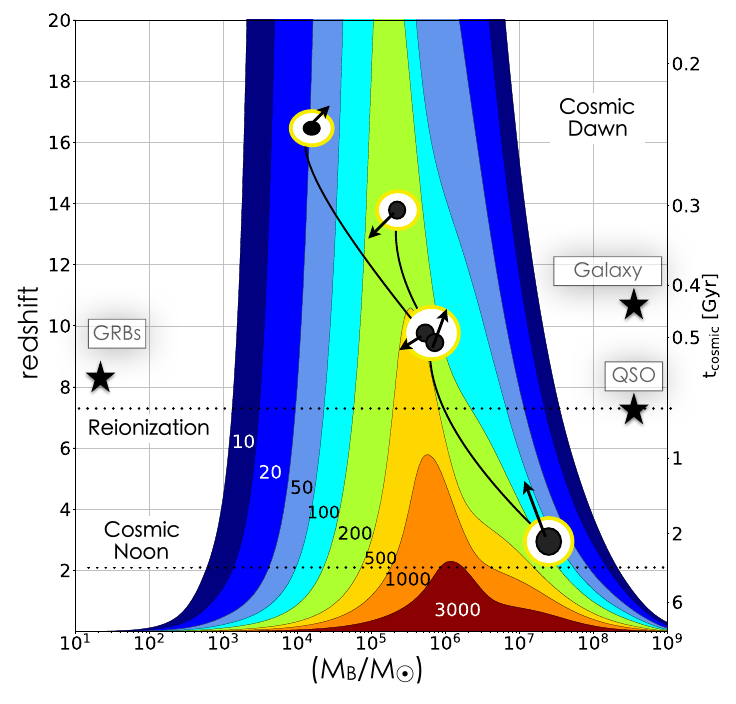} 
\caption{Contours of constant signal-to-noise ratio for the LISA observatory plotted as a function of redshift (or equivalently, cosmic time) and MBHB mass. The contours are calculated for systems of non-spinning MBHs with mass ratio $q=0.5$. Overlaid is the illustration of an evolutionary track, which ends with a formation of a $\sim 10^7\,M_\odot$ MBH merger remnant at $z\approx3$. Black dots and arrows represent the MBHs and their spins, respectively. MBHs are embedded in galactic halos (illustrated by white-yellow circles) and grow through episodes of accretion (black lines) and mergers. Black stars mark the redshifts of the most distant quasar, long gamma-ray burst and galaxy detected so far. Figure from \citet{colpi19}.}
\label{fig2}      
\end{figure}

In contrast to PTAs, the planned LISA and TianQin space-based missions will detect higher GW frequencies and MBHBs with lower total masses \citep{as17, mei21}. LISA will be sensitive to GWs with frequencies in the range from about $100\,\mu$Hz to 100\,mHz. The frequency of GW radiation emitted by MBHBs with total masses of $10^4$--$10^7\,M_\odot$ falls squarely within this bandwidth in the last stages of their evolution \citep{klein16}. Figure~\ref{fig2} illustrates that for many binary configurations in this mass range, high signal-to-noise merger detections will be possible out to redshift of $z\sim20$, which will in turn allow measurements of the masses and spins of coalescing MBHs to a few $\%$ accuracy. LISA will therefore be able to place the evolution of MBHs in the context of important cosmic epochs that coincide with the formation of MBH seeds (marked as Cosmic Dawn in Fig.~\ref{fig2}), the most distant quasars (Reionization), and the epoch of growth through accretion and mergers (Cosmic Noon). TianQin will operate over a similar (albeit not identical) parameter space and, like LISA, is expected to detect GWs from actual MBHB merger events \citep{feng19,wang19}.  Both in terms of the mass range and redshift distribution, the space-based GW observatories will therefore unearth a population of MBHBs that is different from and complementary to those discovered by PTAs.

\begin{figure}[t]
\center
\includegraphics[trim=0 0 0 0, clip, scale=0.5]{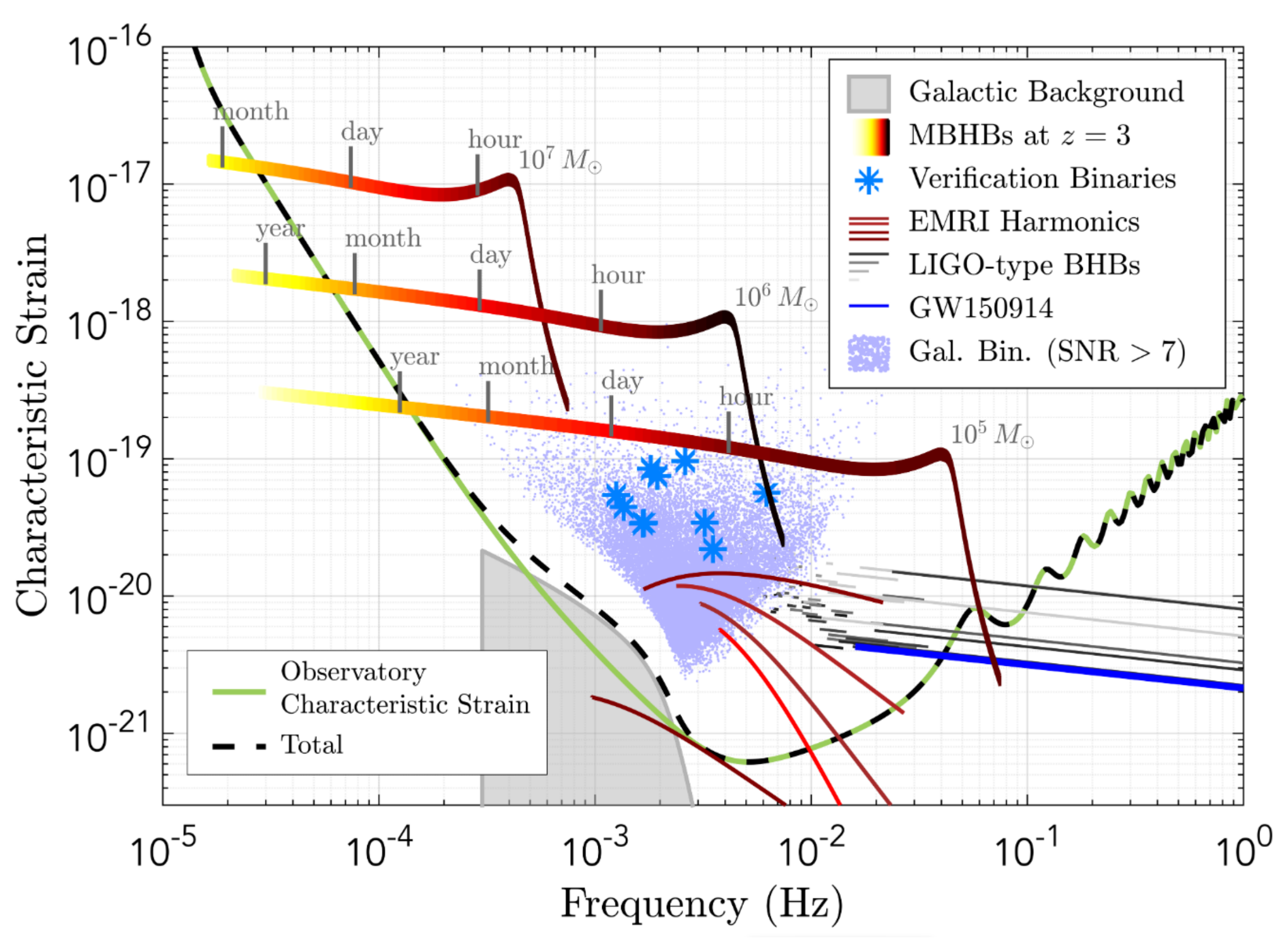} 
\caption{Illustration of the evolutionary tracks (multicolor) through the LISA frequency band for three equal mass MBHBs at redshift $z=3$ with total intrinsic masses $10^5$, $10^6$ and $10^7\,M_\odot$. The remaining time before coalescence is marked on the tracks. Also shown are the signals from the extreme mass ratio inspiral (EMRI), stellar origin black hole binaries of the type discovered by the LIGO-Virgo Collaboration, and from resolved galactic binaries (including the already known verification binaries). The solid green curve shows LISA's sensitivity and black dashed line is the sensitivity with an additional confusion signal at $\sim 1$\,mHz contributed by the unresolved galactic binaries. The sensitivity curves correspond to a configuration with six links, 2.5 million km arm length and a mission lifetime of 4 years. Figure from the LISA L3 mission proposal \citep{as17}.}
\label{fig3}      
\end{figure}

From all MBHBs detected with the space-based GW observatories, the population that stands the best chance of also being detected in EM observations includes systems at intermediate and low redshifts. Figure~\ref{fig3} shows the tracks for several representative MBHBs with equal mass ratios at redshift $z=3$ as they evolve through the LISA frequency band. MBHBs with mass $10^5 - 10^7\,M_\odot$ merging in galaxies at $z \leq 4$ are sufficiently loud GW sources to be detected by LISA weeks before coalescence. For systems in this mass and redshift range, the sky localization determined from the GW observations alone is expected to reach accuracy of about $10-100\,{\rm deg}^2$ in the weeks to hours before coalescence and $\approx 0.1\,{\rm deg}^2$ at merger \citep{haiman17, mangiagli20}. As long as these systems are also sufficiently EM luminous and can maintain emission regions until very close to the coalescence, this localization accuracy provides an opportunity for subsequent identification of an EM counterpart to GW detection. For example, if the culprit source (an AGN on the sky) exhibits characteristic EM variability that correlates with the GW chirp, this could lead to a convincing EM identification of the binary headed for coalescence.  Under such conditions, \citet{dalcanton19} find that given a fiducial LISA detection rate of 10 mergers per year at $z<3.5$, a few detections of modulated X-ray counterparts are possible over the nominal, 5 year duration of the LISA mission, with an X-ray telescope with a relatively large (1\,deg$^2$) field of view. Whether MBHBs have unique EM signatures in this or any other stages of their evolution that can be distinguished from regular AGNs is presently a subject of active investigation.

\subsection{Direct imaging of double nuclei}
\label{ssec:imaging}

A clear EM manifestation of a parsec-scale MBHB is an image of a binary AGN that forms a gravitationally bound system (as opposed to an accidental projection on the sky). A practical obstacle in the detection of such objects arises from their small angular separation: for example, a parsec-scale binary at a redshift of $z=0.1$ subtends an angle of only $\sim0.5$~mas on the sky (neglecting the projection effects). Such scales are below the angular resolution of most astronomical observatories operating at present time, except the radio and millimeter observatories using the Very Long Baseline Interferometry (VLBI) technique \citep[e.g.,][]{dorazio18}.

\begin{figure}[t]
\center
\includegraphics[trim=0 10 0 10, clip, scale=0.6]{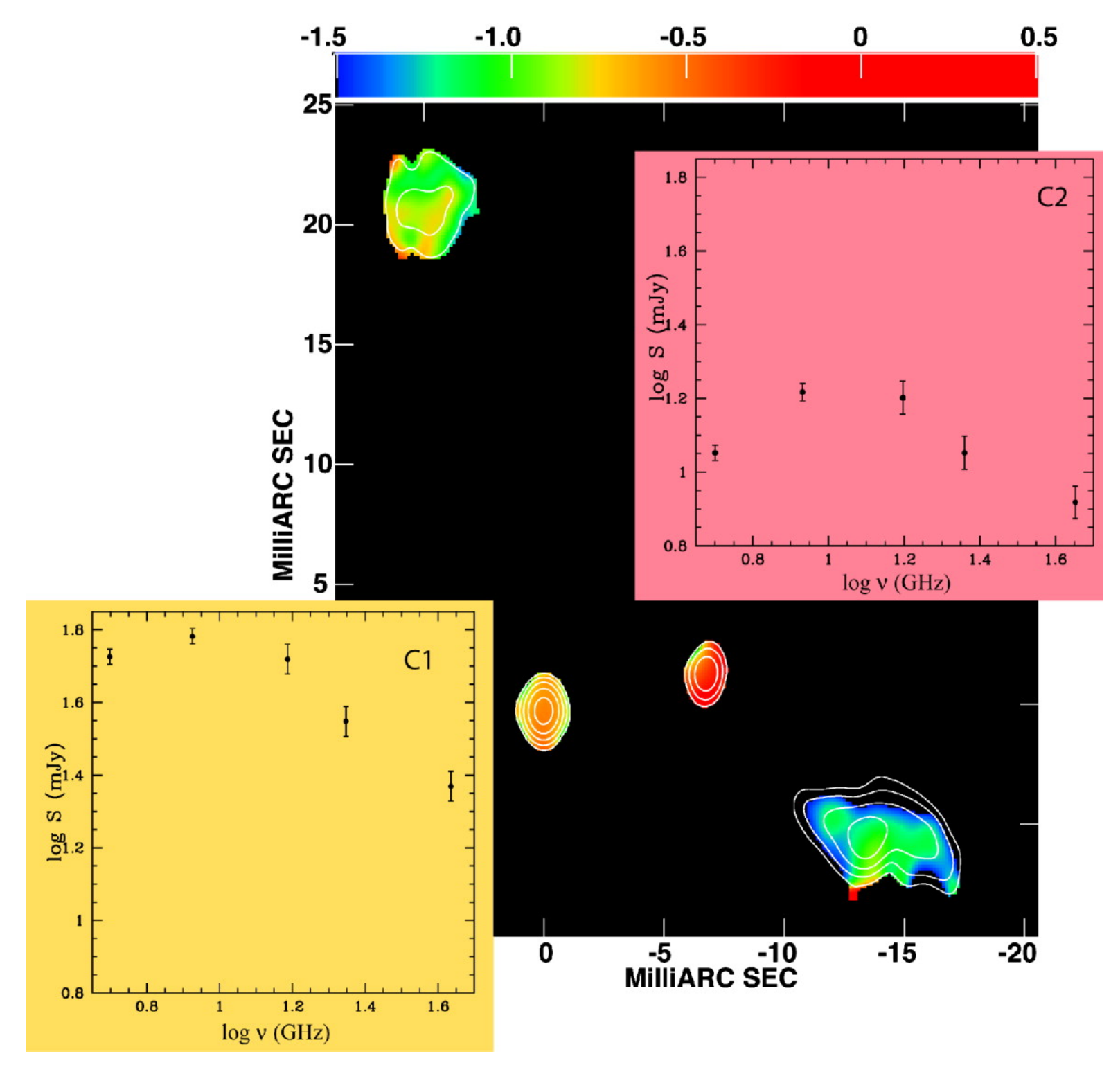} 
\caption{The MBHB candidate in the radio galaxy 0402+379. Two compact radio cores in the middle of the figure are separated by 7.3~pc on the sky and are thought to harbor MBHs. Both cores are characterized by flat radio spectra and have been identified as possible AGNs based on this signature. Radio lobes from a jet associated with one of the cores are also apparent. The color illustrates the spectral index distribution between 8 and 22\,GHz from the VLBA observations. The inset panels show flux density as a function of the logarithm of frequency for each core. Figure from \cite{rodriguez06}.}
\label{fig4}      
\end{figure}

This approach has been used to identify the most convincing MBHB candidate thus far, in the radio galaxy 0402+379 \citep{rodriguez06,rodriguez09,morganti09}. This system hosts a pair of two compact radio cores at a projected separation of 7.3~pc on the sky (Fig.~\ref{fig4}). Both cores are characterized by flat radio spectra and have been identified as possible AGNs based on this signature \citep{maness04,rodriguez06}. \citet{bansal17} subsequently reported that long term VLBI observations reveal relative motion of the two cores, consistent with orbital motion, lending further support to the MBHB hypothesis.  While the MBHB candidate in the galaxy 0402+379 was discovered serendipitously, it demonstrated the power of radio interferometry in imaging of small-separation MBHs.

In a subsequent investigation, \citet{burke11} searched for binaries in the archival VLBI data. The search targeted spatially resolved, double radio-emitting nuclei with a wide range of orbital separations ($\sim3\,{\rm pc} -5\,$kpc) among 3114 radio-luminous AGNs in the redshift range $0 < z \leq 4.715$. Another investigation, in addition to MBHBs, searched for the recoiling MBHs spatially offset from the centers of their host galaxies \citep{condon11}. The latter study is based on Very Long Baseline Array (VLBA) observations at a frequency of 8GHz of a sample of 834 nearby radio-luminous AGNs with typical distances of $\sim200$~Mpc. Radio sources brighter than 100~mJy were selected from the 1.4GHz NVSS catalogue data, under the hypothesis that the most massive galaxies have undergone a major merger in their history and may therefore contain a binary or recoiling MBH. More recently, \citet{tremblay16} performed multi-frequency imaging of a flux-limited sample of $\sim$1100 AGN included in the VLBA Imaging and Polarimetry Survey, all of which were pre-selected to be bright and flat-spectrum. None of these searches unearthed new instances of double radio nuclei, leaving several possible interpretations: (a) there is a true paucity of MBHBs, (b) MBHBs may be present but have low radio brightness, or (c) only one component of the binary is radio-bright and it may or may not show a detectable spatial offset relative to the center of the host galaxy. This highlights the difficulties in using radio imaging as the primary technique for selecting gravitationally bound MBHB candidates or their progenitors, given their unknown radio properties. 

\begin{figure}[t]
\center
\includegraphics[trim=0 0 0 0, clip, scale=0.7,]{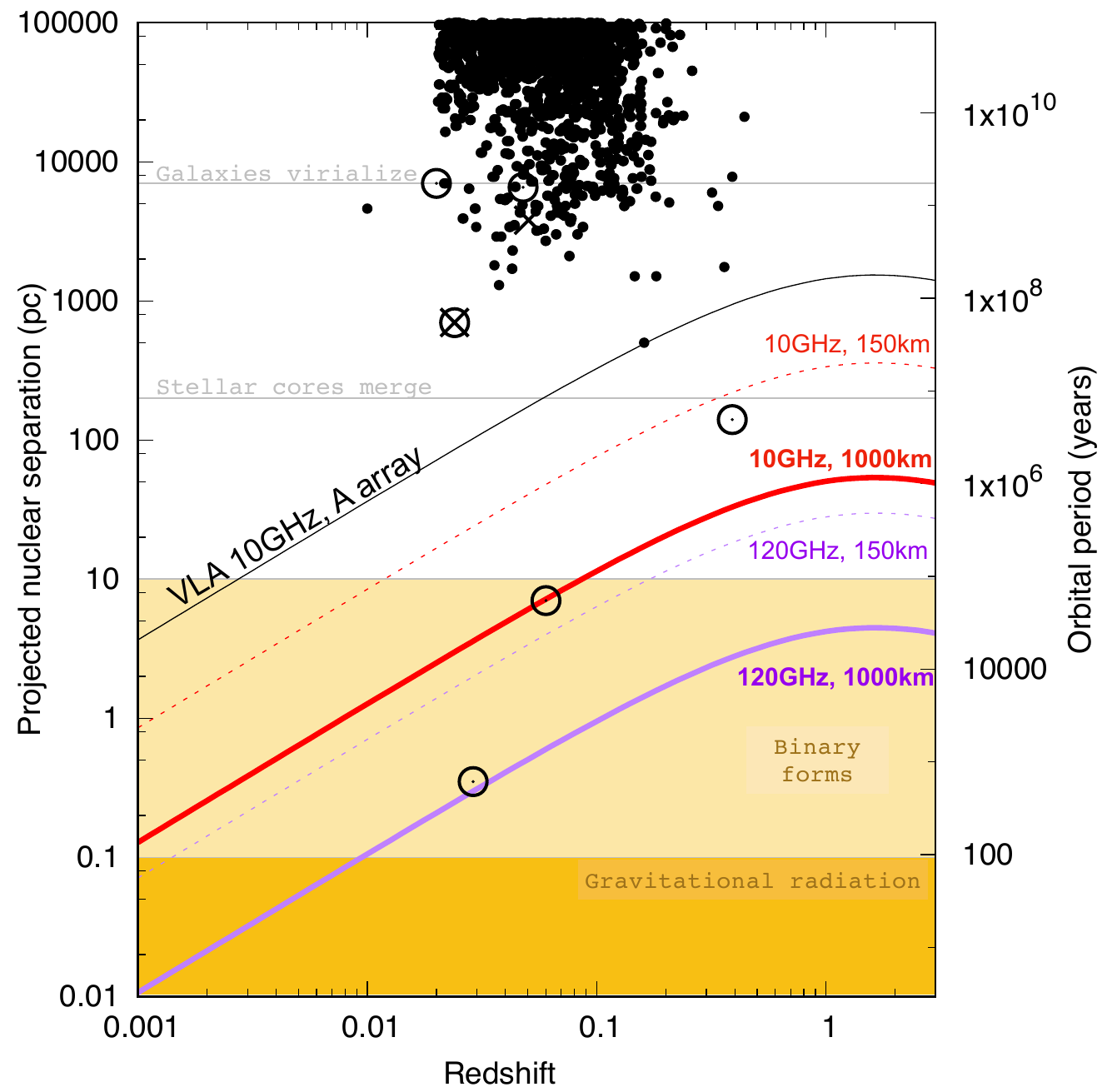} 
\caption{Projected nuclear separation as a function of redshift for the observed population of candidate AGN pairs and binaries, compared with the spatial resolution of the proposed ngVLA observatory. Orbital periods associated with each nuclear separation have been calculated neglecting the orbital projection effects. Each symbol represents a candidate discovered in observations in the radio (circles), optical / near-IR (dots), or X-ray ($\times$) part of the spectrum. Horizontal lines mark approximate evolutionary stages of MBH pairs and binaries. The black curve marks the spatial resolution limit at 10\,GHz of the VLA observatory in the A-configuration. The red and purple curves indicate the spatial resolution limit in the 10\,GHz and 120\,GHz bands, respectively, for the ngVLA 150\,km baseline array (dashed) and 1000\,km extended-baseline array with VLBI capabilities (solid). Longer baselines and higher frequencies have the potential to resolve MBHBs in the gravitational wave regime in the nearby universe (indicated by the curves that cross the dark yellow region). Figure from \citet{burke18_ngVLA}.}
\label{fig5}      
\end{figure}

Notwithstanding the challenges, VLBI imaging still represents a powerful technique. One promising avenue for VLBI may be to test candidate MBHBs selected by other techniques, especially if longer baseline arrays operating at higher frequencies become available in the future. Figure~\ref{fig5} summarizes the capabilities of one such array, proposed as an extension of the Very Large Array (VLA), and named the Next Generation VLA  \citep[ngVLA;][]{ngVLA18,burke18_ngVLA}.  The ngVLA could resolve gravitationally bound binaries at $z\lesssim0.1$ when operating at 10\,GHz, with the main array baseline of 1000\,km and subarrays extending to $\sim 9000$\,km used in VLBI mode. The same array configuration operating at 120\,GHz could in principle resolve MBHBs at even higher redshifts, as well as those that are inspiraling due to the emission of GWs at $z\lesssim0.01$. Therefore, VLBI searches with longer baselines and higher frequencies have the potential to enable both multi-wavelength and multimessenger searches for MBHBs at a variety of separations. If funded, the construction of the ngVLA could commence as early as 2024 with full operations starting in 2034, enabling it to operate contemporaneously with another next generation radio observatory currently under development in the Southern hemisphere, the Square Kilometer Array\footnote{\url{https://www.skatelescope.org/technical/info-sheets}}, as well as with the low-frequency GW detectors described in the previous section.

\subsection{Photometric measurements of quasi-periodic variability}
\label{ssec:photometry}

\begin{figure}[t]
\center{
\includegraphics[trim=0 0 0 0, clip, scale=0.8]{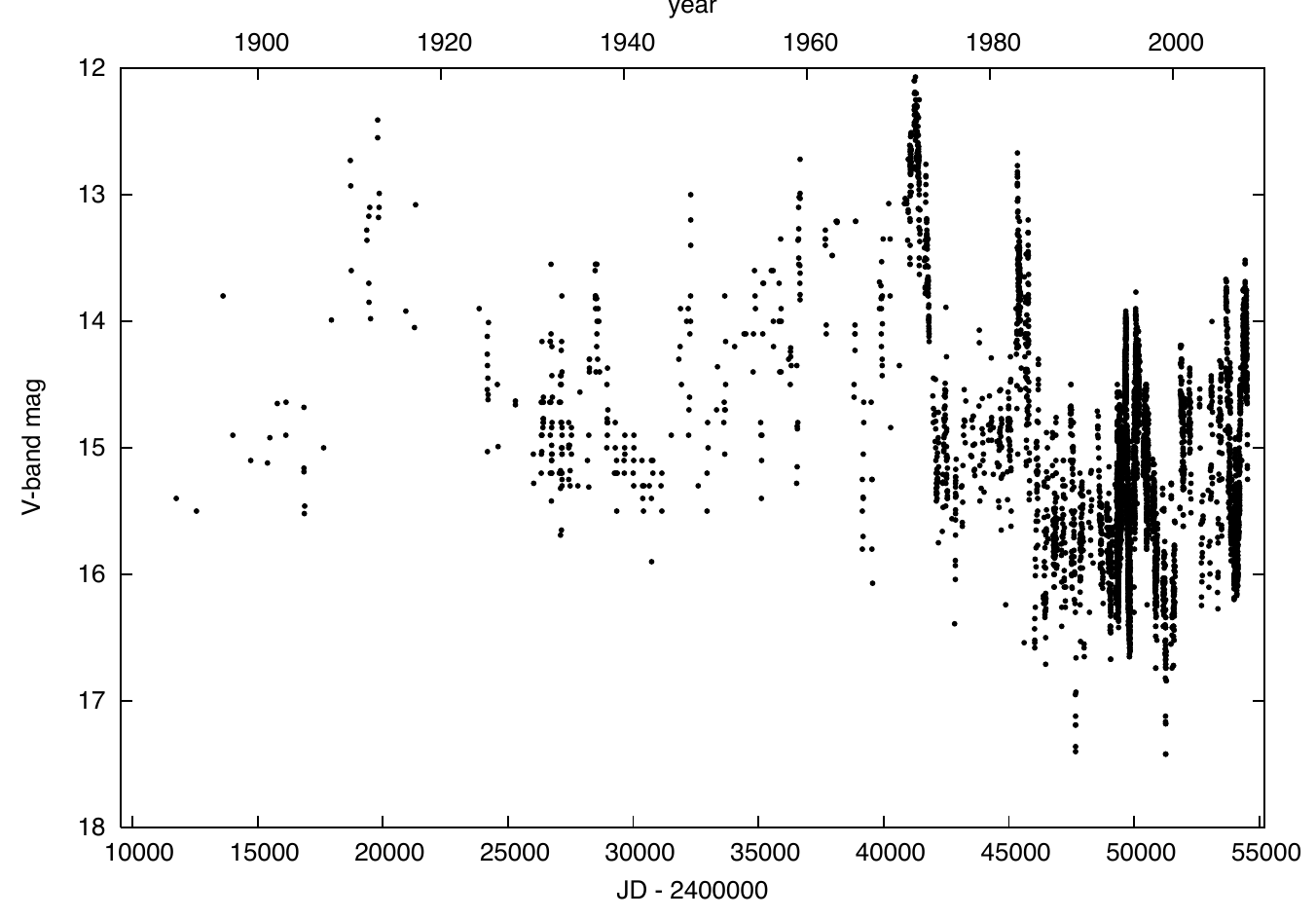}} 
\caption{Historical light curve of OJ~287 in V-magnitude recorded over more than 100 years. Quasi-periodic outbursts with an observed period of $\sim12$~yr in the light curve of this object have been interpreted as a signature of the MBHB orbital motion. Figure from  \citet{valtonen08}.}
\label{fig6}      
\end{figure}

MBHBs at $\sim$ mpc-scale separations have orbital periods of $\sim$ years, making any associated EM variability accessible on human timescales. Many MBHB candidates have been identified via signatures of periodic or nearly periodic variability in the lightcurves of quasars, which are interpreted as a manifestation of binary orbital motion. The possible physical origins of periodic EM variability from MBHBs are discussed in Sect.~\ref{sss_cbd}. A well-known example of a MBHB candidate in this category is the blazar OJ~287, which exhibits outburst activity in its optical light curve with an observed period close to 12 years \citep[see Fig.~\ref{fig6} from][]{valtonen08}. It is worth noting however that OJ~287 is unique among photometrically selected binary candidates because the first recorded data points in its light curve extend as far back as the 19th century. Similarly, OJ~287 has received an unprecedented level of observational coverage in modern times, yielding a high frequency of sampling (from 1970s onwards), in addition to long observational baseline. While the presence of quasi-periodicity has been claimed in other objects, it is usually less pronounced and recorded over much shorter time span than in the case of OJ~287, thus preventing an ironclad case for existence of MBHBs from being made \citep[for e.g.,][]{fan98,rieger00, depaolis02, liu14}. In 2019, a bright flare was observed from OJ~287 that occurred within 4 hours of its predicted arrival time, which provides further support for the MBHB hypothesis for this source \citep{laine20}.

Recently, there have been a number of systematic searches for MBHBs in photometric surveys, such as the Catalina Real-Time Transient Survey, the Palomar Transient Factory and others  \citep[e.g.,][]{graham15, charisi16, liu16}.  Although some candidates have been identified, there is an ongoing discussion about the level of evidence provided by the data.  For example, as pointed out by \citet{vaughan16} and later demonstrated by \citet{liu18}, the risk of false binary identifications in photometric surveys is especially high when the evidence is based on only a few apparent orbital cycles. If the current candidate fraction is treated as an upper limit, then the incidence of MBHBs among quasars is $< 5 \times 10^{-4}$, which is consistent with theoretical predictions for low redshift \citep[$<10^{-2}$;][]{volonteri09,kelley19}.


Because of the finite temporal extent of the surveys, which must record at least several orbital cycles of a candidate binary, most photometrically identified MBHB candidates have relatively short orbital periods, of the order a few years or less. While systematic searches like these signal more efficient approach to finding MBHB candidates in surveys, a confirmation of their identity requires additional evidence. This is because stochastically variable lightcurves of ``normal'' AGNs and quasars, powered by single MBHs, can be mistaken for periodic sources.  

Photometric searches for MBHBs will expand greatly in the next decade, as the capabilities of the optical time-domain surveys will be dramatically enhanced with the Zwicky Transient Facility (ZTF) and the advent of the Legacy Survey of Space and Time (LSST) with the Vera Rubin Observatory in 2023. Over time, photometric searches for MBHBs will also become increasingly sensitive to a wider range of orbital periods, as the existing surveys continue to collect data on longer timescales. For example, \citet{kelley19} find that if the host galaxy does not outshine the AGN, up to $\sim 100$ such MBHB candidates could be identifiable after 5 years of LSST observations with the Rubin Observatory. They also suggest existence of a ``sweet spot'' for binary mass ratios of $q \sim 0.1$, where differential accretion onto the secondary MBH could enhance variability. This could potentially improve the prospects for observing some MBHB systems with lower total mass. \citet{xin21} furthermore predict that from all quasi-periodic AGNs detected by LSST (of which there may be 20 to 100 million), there will be $10-150$ powered by MBHBs with short orbital periods ($\lesssim 1$\, day) that could in $\sim 5-15$ years be detected by LISA. Such sources would thus serve as ``LISA verification binaries", analogous to short-period Galactic compact-object binaries.

Already, some multimessenger constraints have been placed on the MBHB population by combining these photometrically variable candidates with upper limits from PTAs on their GW signals. PTAs are capable of probing the GW background at nanoHertz frequencies from sources such as MBHBs with orbital periods of a few years \citep{lentati15, shannon15, arzoumanian16, arzoumanian20}. Specifically, the upper limit placed by the sensitivity of PTAs largely rules out the amplitude of GW background resulting from the so far identified  $\sim150$ photometric binary candidates, implying that some fraction of them are unlikely to be MBHBs \citep{sesana18}. Subsequently, a similar approach was used to place limits on the presence of MBHBs in periodic blazars \citep{holgado18} and in ultraluminous infrared galaxies \citep{inayoshi18}. More recently, \citet{liu21} illustrated how detection and parameter estimation of individual MBHBs in PTA data could be improved by up to an order of magnitude with the addition of EM priors. These studies provide examples of the great potential of multimessenger techniques, even prior to a low-frequency GW detection.

\subsection{Spectroscopic measurements of offset broad emission lines}
\label{ssec:spectroscopy}

The principal assumption made by all spectroscopic searches is that some fraction of MBHBs at sub-parsec orbital separations reside in emission regions comparable in size to the broad line regions (BLRs) of regular AGNs\footnote{The term ``regular AGNs'' refers to the AGNs that do not host MBHBs.}. If so, the dynamical perturbation of a BLR by the binary gravitational potential can in principle be reflected in the low-ionization broad emission-line profiles \citep[e.g.,][]{gaskell83,gaskell96,bogdanovic08, montuori11, montuori12}. The broad emission lines of particular interest are H$\alpha~\lambda6563$, H$\beta\;\lambda4861$ and Mg\,\textsc{ii}$\;\lambda2798$, because they are prominent in AGN spectra and are commonly used as tracers of dense, low-ionization gas in BLRs at low (H$\alpha$ at $z < 0.4$) and high redshift (Mg\,\textsc{ii} at $z < 2.5$).

\begin{figure}[t]
\center{
\includegraphics[trim=0 0 0 0, clip, scale=0.6,angle=0]{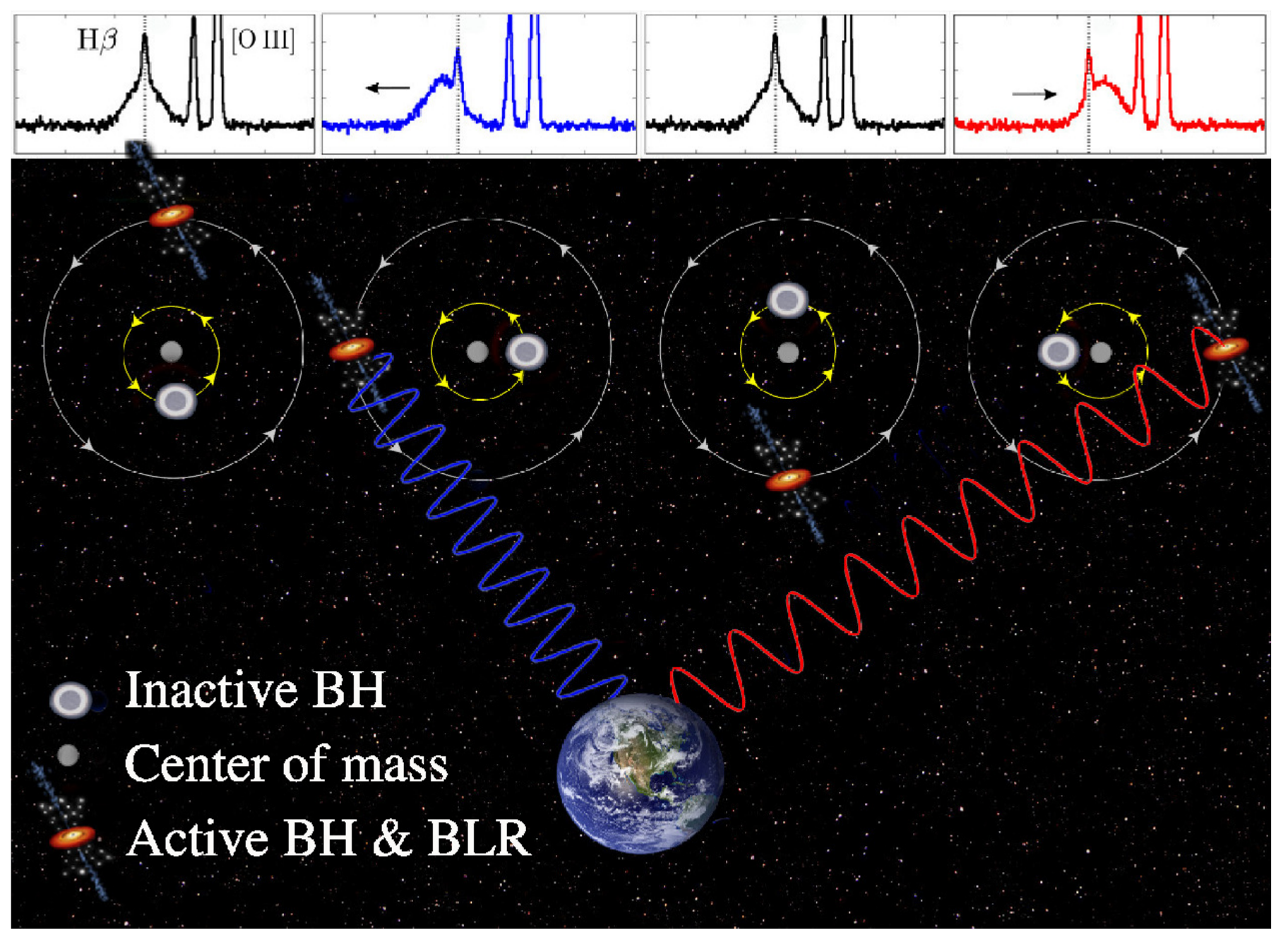}} 
\caption{Illustration of the spectroscopic radial velocity technique used to search for MBHBs. In the context of the MBHB model adopted by spectroscopic surveys, the broad profile (shown in the top panels as the H$\beta$ line) is attributed to the emission from the accretion disk of the smaller MBH. The offset of the broad emission-line profile with respect to its rest frame wavelength (marked by the vertical dotted line) is attributed to the binary orbital motion. The center of mass of the MBHB is assumed to be at rest with respect to the host galaxy, and the rest frame of the host galaxy is defined by the narrow H$\beta$ and [O\,$\textsc{iii}$] lines. Figure from \citet{guo19}.}
\label{fig7}      
\end{figure}

Spectroscopic searches rely on the detection of the Doppler shift of broad emission lines in the spectrum of a MBHB host caused by the binary orbital motion. This approach is reminiscent of a well established technique for detection of single- and double-line spectroscopic binary stars. In both classes of spectroscopic binaries, the lines are expected to oscillate about their local rest frame wavelength on the orbital time scale of a system. In the context of the MBHB model, the spectral emission lines are assumed to be associated with gas accretion disks that are gravitationally bound to the individual black holes \citep{gaskell83,gaskell96,bogdanovic09}. Given the high velocity of the bound gas, the emission-line profiles from the MBH disks are expected to be Doppler-broadened, similar to the emission lines originating in the BLRs of regular AGNs. Moreover, several theoretical studies have shown that in unequal-mass binaries, accretion occurs preferentially onto the lower mass object \citep{al94, gunther02, hayasaki07, roedig11, farris14}, rendering the secondary AGN potentially more luminous than the primary. If so, this indicates that some fraction of MBHBs may appear as single-line spectroscopic binaries. This scenario is illustrated in Fig.~\ref{fig7}, where the Doppler-shifted broad component of the H$\beta$ emission line traces the motion of the smaller MBH. \citep[See however][who suggest that the motion traced may actually be that of the primary]{nguyen20}.

This realization led to the discovery of a number of MBHB candidates based on the criterion that the culprit sources exhibit broad optical lines offset with respect to the rest frame of the host galaxy \citep{bogdanovic09a,dotti09a,bl09,tang09,decarli10, barrows11,tsal11, eracleous12, tsai13}\footnote{In an alternative approach, anomalous line ratios have been used to flag MBHB candidates with perturbed BLRs \citep{montuori11,montuori12}.}.
 Because this effect is also expected to arise in the case of a recoiling MBH receding from its host galaxy, the same approach has been used to flag candidates of that type \citep{komossa08a,shields09,civano10, robinson10,lusso14}. The key advantage of the method is its simplicity, as the spectra that exhibit emission lines shifted relative to the galaxy rest frame are fairly straightforward to select from large archival data sets, such as the Sloan Digital Sky Survey (SDSS). Its main complication, however, is that the Doppler-shift signature is not unique to these two physical scenarios; for example, broad line offsets can also be produced by AGN outflows. Complementary observations are therefore required in order to determine the nature of the observed candidates \citep[e.g.,][]{popovic12,barth15}.

To address this ambiguity, a new generation of time-domain spectroscopic searches has been designed to monitor the offset of the broad emission-line profiles over multiple epochs and identify sources in which modulations in the offset are consistent with binary orbital motion \citep{bon12, bon16, eracleous12, decarli13, ju13, liu13, shen13, runnoe15, runnoe17, li16, wang17,guo19}. For example, \citet{eracleous12} searched for $z < 0.7$ SDSS AGNs with broad H$\beta$ lines offset from the rest frame of the host galaxy by $\gtrsim 1000\,{\rm km\,s^{-1}}$. Based on this criterion, they selected 88 quasars for observational follow-up from an initial group of about 16,000. The follow-up observations of this campaign span a temporal baseline from a few weeks to 12 years in the observer's frame.  After multiple epochs of follow-up, statistically significant changes in the velocity offset have been measured for 29/88 candidates \citep{runnoe15, runnoe17}, all of which remain viable as MBHB candidates. This implies that $\lesssim 10^{-3}- 10^{-2}$ of quasars host MBHBs at $z < 0.7$, a limit that is so far consistent with theoretical predictions \citep{volonteri09,kelley19}.

\begin{figure}[t]
\center{
\includegraphics[trim=0 0 0 0, clip, scale=0.8,angle=0]{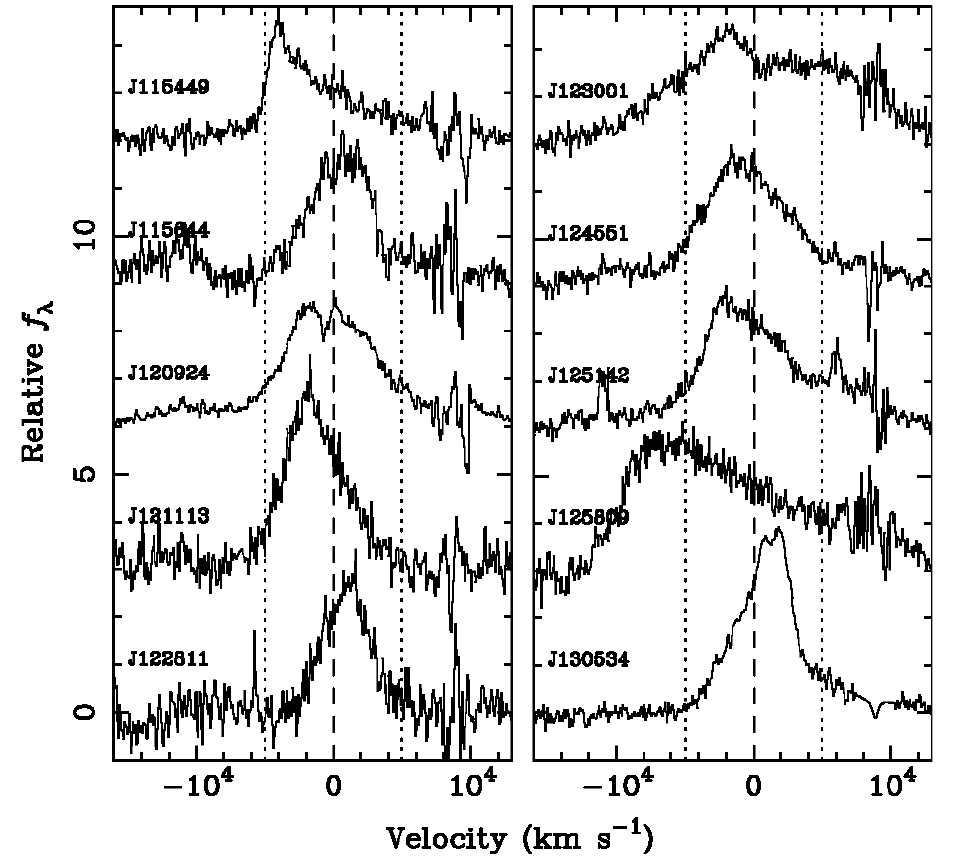}
}
\caption{H$\beta$ emission-line profiles of MBHB candidates selected in a spectroscopic survey by \citet{eracleous12}. Only the broad emission-line profiles are shown; the narrow line components were removed for clarity. Peaks of the asymmetric profiles are offset from the rest frame of the host galaxy (marked by the dashed vertical line) by $\sim{\rm few}\times 10^3\;{\rm km\,s^{-1}}$. If measured velocity offsets are associated with the orbital motion of the secondary MBH, they imply orbital periods in the range of $\sim {\rm few}\times10- {\rm few}\times 100$ yr for MBH mass ratios larger than 0.1. The dotted lines mark velocity offsets of $\pm 5000\;{\rm km\,s^{-1}}$.}
\label{fig8}      
\end{figure}

Figure~\ref{fig8} shows several representative broad $H\beta$ profiles selected in this search. The profiles are asymmetric and have peaks offset by $\sim{\rm few}\times 10^3\;{\rm km\,s^{-1}}$ from the rest frame of the galaxy, inferred from the wavelength of the narrow emission lines (the narrow emission-line components were subsequently removed from these profiles for clarity). If measured velocity offsets of this magnitude are associated with the orbital motion of the secondary MBH, as described above, then it is possible to show that searches of this type (with a yearly cadence of observations) are in principle sensitive to a subset of MBHBs with orbital separations $\lesssim {\rm few}\times 10^4$ Schwarzschild radii and orbital periods in the range of a few tens to a few hundreds of years \citep{pflueger18}. In comparison, current spectroscopic observational campaigns typically span a baseline about 10--15 years, shorter than the average period of MBHBs targeted by this technique. 

These relatively short monitoring baselines still allow one to reject MBHB candidates with velocity curves inconsistent with the MBHB model, even if they have been monitored for less than a full orbital cycle \citep{runnoe17,guo19}. At the same time, they preclude measurement of multiple orbital cycles, which have traditionally been used as a criterion for binarity in stellar systems. Therefore, this technique alone cannot uniquely confirm the identity of a MBHB; it must be supplemented by additional, independent observational evidence that can help to further elucidate their nature. Some of the more promising complementary approaches include direct VLBI imaging at millimeter and radio wavelengths of nearby MBHB candidates ($z\lesssim 0.1$) with orbital separations $a \gtrsim 0.01\,$pc \citep[e.g.,][]{dorazio18, burke18_ngVLA, breiding21}. 

As in the case of the photometric searches, a statistical test of the nature of spectroscopic MBHB candidates can be performed using GW constraints from PTAs. While hypothetical MBHBs detected spectroscopically do not emit significant amounts of GWs because they are at relatively large separations, they imply some number of MBHBs inspiraling toward coalescence, whose GW signal is reaching Earth at this very moment. For example, a comparison of this GW signal with the PTA sensitivity limit indicates that all the MBHB candidates in the spectroscopic sample of \citet{eracleous12} could be binaries without violating the observational constraints on the GW background \citep{nguyen20_pta}.

The risk of binary misidentifications in spectroscopic surveys can also be mitigated by continued monitoring and longer observational baselines. One example of a survey that can provide such opportunity in the next decade is SDSS-V, which is planned for the period 2020--2025\footnote{\url{https://www.sdss.org/future}}. As a part of SDSS-V, the Black Hole Mapper multi-object spectroscopic survey will provide multi-epoch optical spectra for more than 400,000 X-ray sources to probe the evolution of MBHs over cosmic time. This and similar surveys will therefore provide a considerable sample of active galaxies that are potential hosts to MBHBs.

\section{Physical processes and relevant timescales}
\label{section:timescales}

\subsection{Rates and typical redshifts}\label{ssection:rates_z}

In the standard model of structure formation in the Universe, bound halos of dark matter (along with their associated baryons) assemble in a hierarchical fashion, with small structures forming first and combining to make larger structures.  If halos contain central MBHs, then halo mergers can lead to eventual mergers of the black holes.  Such a coalescence will typically not happen if one halo is much less massive than the other, because in that case dynamical friction is inefficient and the lower-mass halo will either have an inspiral time longer than the Hubble time, or it will dissolve before it settles in the center.  Indeed, \citet{2003MNRAS.341..434T}  find that below mass ratios of 1:10, merging of dark matter halos is ineffective. As we discuss in the next section, a similar result is found in simulations of isolated (i.e., non-cosmological) galaxy mergers and cosmological simulations that follow pairing of MBHs.

\smallskip
\noindent\fcolorbox{white}{white}{%
    \minipage[!t]{\dimexpr0.50\linewidth-2\fboxsep-2\fboxrule\relax}    
Analyses of halo mergers \citep[][and others]{1974ApJ...187..425P, lacey93, sheth01, 2001ApJ...558..535M, 2002Ap&SS.281..501V, 2003ApJ...582..559V, wyithe03, sesana04, 2004ApJ...614L..25Y, 2004MNRAS.354..443I, 2004MNRAS.354..629I, 2005ApJ...620...59S, 2005ApJ...623...23S, 2007MNRAS.380.1533M} indicate that most of the halo mergers that might lead to MBH mergers occur at redshifts of several. One therefore finds that, if they follow promptly after their host halos merge, most MBH mergers would happen at redshifts $z>3$ and perhaps at  
 \endminipage}\hfill
\noindent\fcolorbox{black}{light-yellow}{%
    \minipage[!t]{\dimexpr0.47\linewidth-2\fboxsep-2\fboxrule\relax}
\begin{center} {\bf Physical mechanisms leading to orbital evolution of MBHBs} \end{center}
{ 
\noindent$\bullet$ Dynamical friction (during pairing of MBHBs).\\
\vspace{-2.0mm}

$\bullet$ Stellar ``loss-cone'' scattering.\\
\vspace{-2.0mm}

$\bullet$ Interactions with additional MBHs. \\
\vspace{-2.0mm}

$\bullet$ Torques from a circumbinary gas disk. \\
\vspace{-2.0mm}

$\bullet$ Emission of gravitational waves.\\
}
 \endminipage}\hfill\\
even higher redshifts. Such mergers would be difficult to detect electromagnetically because of their distance and the weakening of the signal due to redshift, and it will be even more difficult to identify their host galaxies. There is however evidence that delays due to orbital evolution of MBHs may shift the peak of the MBH coalescence rate to $z\sim 1-2$, a much more promising outcome for the EM followup \citep{volonteri20, li22a}. The space-based gravitational wave detectors (like LISA and TianQin) will nonetheless be able to detect mergers out to $z\approx20$ and will therefore provide unique insight into the era of structure assembly.

As discussed in Sect.~\ref{ssec:gw}, EM counterparts to MBH mergers would be much easier to detect at redshifts $z<1$.  The rate of such mergers, which may be accompanied by EM counterparts, is difficult to determine directly, but if we assume that major galaxy mergers (with mass ratios $>1/4$) inevitably lead to MBH mergers, then because galaxies comparable to or more massive than the Milky Way have typically had at least one major merger since $z\sim 1$ 
\citep{2006ApJ...652..270B,2008ApJ...672..177L,2009ApJ...694..643L,
2009ApJ...697.1369B,2010ApJ...715..202H}, and there are tens of billions of such galaxies in the universe, one might expect a few such mergers to happen per year. 

In the last few decades it has become clear that the mass of a central
MBH is correlated with properties of its host galaxy
\citep[e.g.,][]{kormendy95_araa,1998AJ....115.2285M,ferrarese00,gebhardt00,
2001ApJ...547..140M,tremaine02,2002ApJ...578...90F,
2006ApJ...644L..21F,2009ApJ...698..198G, mcconnell13}.  
Perhaps the tightest
correlation is between the black hole mass, $M$, and the stellar velocity
dispersion of the bulge, $\sigma$.  A recent assessment of this
correlation \citep{mcconnell13} gives 
\begin{equation}
\log(M/M_\odot)=\alpha+\beta\log(\sigma/200~{\rm km~s}^{-1}) \,,
\label{eq:M_sigma}
\end{equation}
with $\alpha=8.32\pm 0.05$, $\beta=5.64\pm 0.32$, and there
 is an intrinsic scatter in the logarithm

\smallskip
\noindent\fcolorbox{white}{white}{%
    \minipage[!t]{\dimexpr0.50\linewidth-2\fboxsep-2\fboxrule\relax}    
of the MBH mass of
$\epsilon_0=0.38$. This correlation implies that moderate
to massive galaxies typically contain black holes
with masses $M\sim 10^6$--$10^8\,M_\odot$. Mergers between such black
holes release energy in GWs that is comparable to
the total energy emitted by the stars in the galaxy over a Hubble time.  
As we discuss later in this section, some of that energy is emitted asymmetrically and hence, the
remnant can receive a kick sufficient to eject it from the merged
host galaxy.
 \endminipage}\hfill
\noindent\fcolorbox{black}{light-yellow}{%
    \minipage[!t]{\dimexpr0.47\linewidth-2\fboxsep-2\fboxrule\relax}
\begin{center} {\bf Characteristic timescales} \end{center}
{ 
\noindent$\bullet$ Galaxies merge in a few Gyr.\\
\vspace{-2.0mm}


$\bullet$ Bound MBHB forms in $\sim 10^{7-9}\,$yr due to dynamical friction. \\
\vspace{-2.0mm}

$\bullet$ MBHB shrinks in $\sim 10^{6-9}\,$yr due to interactions with stars and possibly circumbinary disk.\\
\vspace{-2.0mm}

$\bullet$ On the smallest scales MBHB coalesces rapidly due to the emission of GWs. \\
\vspace{-2.0mm}
}
 \endminipage}\hfill\\

\begin{figure}[t]
\center
\includegraphics[trim=0 0 0 0, clip, scale=1.2]{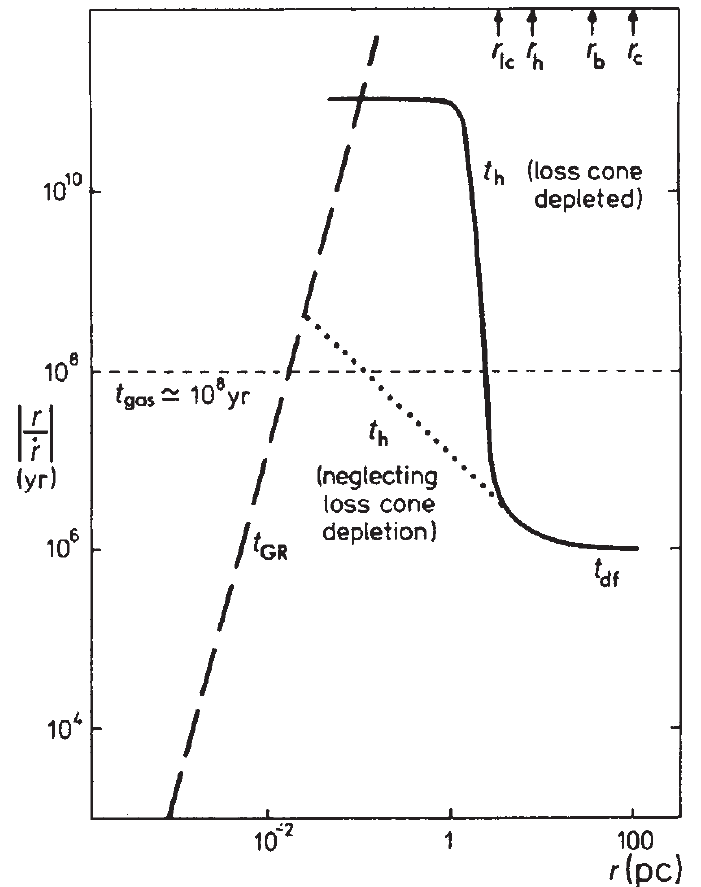} 
\caption{Diagram of timescales associated with different stages of MBHB evolution: dynamical friction ($t_{\rm df}$), loss-cone scattering ($t_{\rm h}$), gas processes ($t_{\rm gas}$), and GW emission ($t_{\rm GR}$). The timescales correspond to a relatively massive binary, $M \approx 10^8\,M_\odot$, residing in a massive elliptical galaxy with stellar velocity dispersion $\sigma = 300\,{\rm km\,s^{-1}}$. Unless the loss cone can be replenished on a timescale much less than the stellar relaxation time (dotted line), the MBHB orbital evolution takes longer than a Hubble time (solid). Gas dynamical processes can in principle lead to faster MBHB evolution (short dashed). For binaries that reach smaller separations, gravitational radiation takes over as the dominant mechanism at $\sim 10^{-2}\,$pc (long dashed). Figure from \citet{bbr80}.}
\label{fig12}      
\end{figure}

There are several important stages between a galaxy merger and a black hole merger, defined in terms of the physical mechanisms that determine the orbital evolution of the MBH pair. In brief:
\begin{enumerate}
\item The interacting galaxies settle together in a few orbital times. The bulges of the galaxies, which contain the MBHs, and are several hundred times more massive than the holes, are initially dense enough to act dynamically as single objects.  They therefore sink in the overall star-gas system of the remnant galaxy. This can take a few Gigayears.
\item When the cores get within several hundred parsecs of each other, they interact dynamically and the MBHs shed their stars. The holes now spiral separately in the potential of the remnant galaxy, and their orbit evolves due to dynamical friction against the stars and gas. This stage can take between $\sim 10^7$ to $10^9\,$yr.
\item When the MBHs get close enough so that their total mass significantly exceeds the mass of the gas and stars within their orbits (typically, $\sim {\rm few}\,$pc), their orbital evolution can slow, or even stall in principle, if much of the interacting stars and gas are ejected. As a consequence, the MBHB evolution in this stage is a sensitive function of the properties of its environment and spans a wide range of timescales ($\sim 10^6-10^9\,$yr).
\item If the holes can bridge this gap and reach separations of $10^{-2}-10^{-3}$~pc, then gravitational radiation by itself suffices to bring the MBHs to coalescence in less than a Hubble time. This phase can be very short, compared to the the other evolutionary stages mentioned above.
\end{enumerate}
The last three evolutionary stages are illustrated in Fig.~\ref{fig12}. We now discuss them in more detail, in the context of stellar and gas dynamical processes.

\subsection{Effects before MBHs are bound to each other}\label{ssection:effects_before}

\subsubsection{Stellar and gas dynamical friction}\label{sssection:df}

Dynamical friction is thought to be the principal mechanism responsible for orbital evolution of MBHs in the aftermath of galactic mergers and an important channel for formation of gravitationally bound MBH binaries. When two galaxies merge, their MBHs are initially tightly bound to their individual stellar bulges, so that the bulge-MBH system dynamically acts as one object. The parent bulges continue to sink towards each other until they reach $\sim1\,$kpc. Beyond this point the bulges merge and gradually dissipate until the MBHs are on their own. As long as the mass of the stars and gas between the MBHs is significantly larger than the total mass of the holes, the MBHs are dragged towards each other by dynamical friction exerted by the background stars and gas in the newly-formed remnant galaxy.  Thus, for a wide range of galactic merger scenarios, this process is expected to play an important role for MBHs with separations in the range from $\sim {\rm few}\times 100\,$pc  to $\sim {\rm few}\,$pc, when they become gravitationally bound. Dynamical friction is also expected to be important for MBHs that are ejected to comparable distances in dynamical interactions with other MBHBs, or in the aftermath of the GW recoil, the effects that we discuss in  Sect.~\ref{sssection:3body} and Sect.~\ref{ssection:GWphase}.

In this section we summarize the key aspects of dynamical friction in stellar and gaseous media, as applicable to MBHs in those scenarios.\footnote{We do not consider dynamical friction due to dark matter, which may play an important role for dynamics of intermediate mass black holes ($M\sim 10^{3-5}\,M_\odot$) in mergers of dwarf and dark-matter dominated galaxies \citep[see for example,][]{dicintio17,tamfal18}.} For a general introduction to this topic, we direct the reader to the seminal work of \citet{chandra43}, who developed the analytic theory for gravitational drag in collisionless systems, and to \citet{1999ApJ...513..252O}, who performed an equivalent treatment for a gaseous medium.

%
%

Dynamical friction arises when a massive perturber moving through some background (stellar or gaseous) medium creates in it a density wake. The wake trails the perturber and exerts on it gravitational force in the direction opposite to the motion, thus acting as a brake and earning this interaction a name: ``dynamical friction."  Consider, for example, an encounter of a MBH with mass $M$ moving at a velocity ${\bf v}$ relative to an isotropic background of stars with much lower mass, and a total mass density $\rho$. If the stars have a three-dimensional Maxwellian velocity dispersion $\sigma$, the drag experienced by the MBH is \citep[e.g.,][]{1987gady.book.....B}
\begin{equation}
{d{\bf v}\over{dt}}=-{4\pi\ln\Lambda\, G^2 M\rho \over{v^3}}
\left[{\rm erf}(X)-{2X\over{\sqrt{\pi}}}e^{-X^2}\right]{\bf v}
\label{eqn:accel_df}
\end{equation}
where $X\equiv v/(\sqrt{2}\sigma)$, $\ln \Lambda = \ln (b_{\rm max}/b_{\rm min}) \sim 10-20$ is the Coulomb logarithm, and $b_{\rm min}$ and $b_{\rm max}$ are the smallest and largest effective impact parameter, respectively. The characteristic timescale for the MBH to decelerate due to multiple encounters with stars is then
\begin{equation}
t_{\rm df} = v/(dv/dt)\sim v^3/(4\pi G^2 M \rho)\; .
\end{equation}
An important implication of this expression is that MBHs of larger mass, immersed in denser environments, can sink faster toward the center of the remnant galaxy. 

The exact timescale for decay of the MBH's orbit in the remnant stellar bulge will depend on the properties of the stellar bulge, such as the distribution of stars and the stellar velocity dispersion $\sigma$, which provides a characteristic speed of stars relative to the MBH. For example, following \citet[][see their Eq.~19]{antonini12}, who calculate the effect of stellar dynamical friction in nuclear star clusters around MBHs, we estimate the inspiral time for a $10^6\,M_\odot$ MBH initially located at $r=100\,$pc from the cluster center
\begin{equation}
t_{\rm df} \approx 45\,{\rm Myr} \, \left(\frac{r}{100\,{\rm pc}} \right)^2 
\left(\frac{\sigma}{100\,{\rm km\,s^{-1}}} \right)
\left(\frac{M}{10^6\,M_\odot} \right)^{-1}
\left(\frac{\ln \Lambda}{15} \right)^{-1} \,.
\label{eqn:t_df}
\end{equation}
Note that in the fiducial example considered in Eq.~\ref{eqn:t_df} inspiral from 1\,kpc due to stellar dynamical friction takes 4.5\,Gyr and is shorter for more massive black holes. In the estimate above, the stellar background is described by the mass density profile $\rho(r) =\rho_0(r/1\,{\rm pc})^{-1.8}$, where $\rho_0 = 1.5\times 10^5\,M_\odot\,{\rm pc^{-3}}$ is the normalization. This profile is an extrapolation of the observed distribution of stars in the central 10\,pc of the Galactic Center \citep[e.g.,][]{oh09}. Stellar density profiles that decrease less rapidly with radius would result in even shorter timescales. See for example an estimate in Eq.~(5) in \citet{souzalima17}, who instead assume a stellar distribution consistent with $\rho \propto r^{-1}$ \citep[i.e., the Hernquist bulge;][]{hernquist90}.

\begin{figure}[t]
\center
\includegraphics[trim=0 0 0 0, clip, scale=1.0]{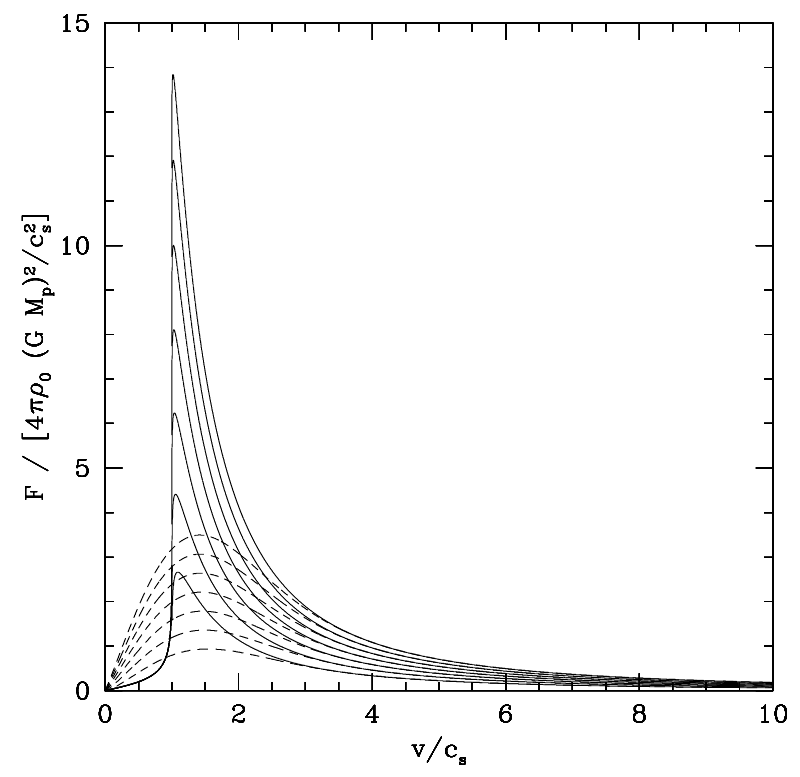} 
\caption{Magnitude of the dynamical friction force, as a function of the Mach number, $\mathcal{M} = v/c_s$. The force is exerted either by the gas (solid lines) or by the collisionless background (dashed) of the same density, and with a sound speed of the gas equal to the velocity dispersion of the collisionless system. From top to bottom, the lines correspond to the dynamical friction force contributed by regions of increasing size around the perturber. Figure from \citet{1999ApJ...513..252O}.}
\label{fig9}      
\end{figure}

If the remnant galaxy contains gas, the gas can also play an important role for pairing of MBHs. For example, \citet{1999ApJ...513..252O} evaluated the gravitational drag exerted by a uniform-density gas background on a massive perturber on a linear trajectory\footnote{See also \citet{kim07} for calculation with a massive perturber on a circular orbit.}.  Figure~\ref{fig9}, adapted from this work, shows a comparison of the magnitude of the dynamical friction force, exerted by the gas with the same density as a purely collisionless background, and with a sound speed $c_s$, equal to the velocity dispersion $\sigma$ of the collisionless system. \citet{1999ApJ...513..252O} found the following results as a function of the Mach number, ${\cal M}\equiv v/c_s$, where $v$ is the speed of the massive object relative to the local standard of rest of the gas:
\begin{enumerate}
\item For ${\cal M}\gg 1$, the gas drag is the same as the drag against a collisionless background, for the same average density $\rho$ and a velocity dispersion $\sigma$ of the collisionless background equal to $c_s$.
\item For ${\cal M}< 1$, the drag force is larger in a collisionless medium than in a gaseous medium. This is because the gas pressure forces result in a more symmetric distribution around the massive object.
\item For ${\cal M}\sim 1$, the dynamical friction force due to the gas can exceed the collisionless drag by a factor of four.
\end{enumerate}

\subsubsection{Implications for formation and properties of gravitationally bound MBHs}\label{sssection:implications_mbhb}

Because of its efficiency relative to the collisionless case, gaseous drag has been extensively investigated in simulations that follow pairing of MBHs in gas-rich mergers of galaxies \citep[see][for reviews]{mayer13,mayer17}. When pairing is successful, gaseous dynamical friction is capable of transporting the MBHs from galactic radii of few hundred pc to $\sim1$\,pc on timescales as short as $\sim10^7$\,yr \citep[for details see, e.g.,][]{escala04, escala05, dotti06, mayer07,vanwassen14}, and in many cases, tens of times faster than stellar dynamical friction \citep[e.g.,][]{berczik06, preto11, khan11, khb15, vasiliev15, pfister17}.  The propensity of gas to effectively couple with the orbiting MBHs can be understood by recalling that unlike collections of stars, gas can radiate energy.  Thus, again unlike stars, the same parcels of gas can interact multiple times with a black hole pair.  As a result, in the extreme case that only gas is present, the binary can in principle be induced to spiral inwards without severely depleting the gas reservoir.

Simulations of isolated galaxy mergers, however, show that MBH pairs with mass ratios $q< 0.1$ are less likely to form gravitationally bound binaries within a Hubble time at any redshift, primarily due to the inefficiency of dynamical friction on the lower-mass MBH \citep{callegari09,callegari11}. They find that in moderately gas rich galaxies MBH pairs with initially unequal masses tend to evolve toward equal masses, through preferential accretion onto a smaller MBH. This trend is also consistent with that found in cosmological simulations Illustris \citep{vogelsberger14, kelley17}, \texttt{ROMULUS}25 \citep{tremmel18, ricarte21}, and a suite of cosmological zoom-in simulations \citep{bellovary19, bellovary21}; moreover, evidence for wandering MBHs in nearby dwarf galaxies has recently been found \citep{reines20}. These works show that low-mass galaxies, which lack a dense stellar core, are more likely to become tidally disrupted early in the process of merging. As a consequence, they deposit their MBHs at large radii, resulting in a population of `wandering' MBHs. If so, this signals a preference for $q> 0.1$ mergers of MBHs. In the circumstance that the opposite is suggested by observations of GW mergers, this would point to an existence of a mechanism other than dynamical friction that can promote pairing of MBHs of disparate masses. 

Another important consideration for the pairing timescale of MBHs is the overall rotation of the stars and gas in the merged galaxy. For example, if a dominant component of overall speed of stars is in bulk motion, such as rotation, the velocity dispersion of the stars is then less than if their motion were random. Equation~\eqref{eqn:accel_df} indicates that the impact of dynamical friction (and more generally, dynamical relaxation) is stronger in systems with lower velocity dispersion $\sigma$. Studies of rotating systems confirm their fast evolution \citep{1999MNRAS.302...81E,2002MNRAS.334..310K,2004MNRAS.351..220K, 2009ApJ...695..455B,2010MNRAS.401.2268A}. Consequently, if the collision of galaxies that leads to the formation of dual MBHs also imparts significant rotation, as is likely in merger remnants, than the orbital evolution time of MBHs is much shorter than in non-rotating systems.  \citet{khb15} for example find that for all other things being the same, rotating stellar bulges drive MBHs more efficiently through the dynamical friction and three-body scattering phase, resulting in timescales that are between 3 (for corotating systems) and 30 times shorter (for counterrotating systems) than the non-rotating bulges. 

The latter timescale, obtained for systems in which a majority of stars are rotating in the opposite sense from the MBH pair, stems from the ability of counterrotating stellar orbits to effectively sap the orbital angular momentum of the MBHs. Therefore, the same effect should produce high eccentricity MBH pairs and bound binaries. This is important because if the eccentricity can truly increase arbitrarily close to unity, then the MBHB can get to a point where gravitational radiation is important and it will merge. This expectation is confirmed by studies that follow the evolution of binary eccentricity with respect to the host galaxy and find low eccentricities ($e\lesssim 0.1$) when the MBH binary and the galaxy are corotating, and high eccentricities ($e\lesssim 1$) and nearly radial orbits in the case of counterrotation \citep{2010MNRAS.401.2268A, sesana11, gualandris12, wang14, khb15, mirza17, rasskazov17}. Furthermore, these studies find that MBH pairs and binaries, whose orbital angular momentum is initially misaligned with respect to that of the stellar cusp, tend to realign their orbital planes with the angular momentum of the cusp. Similar outcomes have been observed for MBHs evolving in gaseous circumnuclear disks. Specifically, the corotation of MBHs with the disk results in efficient circularization of the black holes' orbits \citep{dotti07,li20a}. In the counterrotating case, however, gaseous dynamical friction drives the increase in MBH eccentricities until they experience ``orbital angular momentum flip" and start to corotate with the disk before a MBHB forms \citep{dotti09}. 

Simulations of isolated galaxy mergers have also revealed that pairing of MBHs may be inefficient in clumpy galaxies, such as the massive star-forming galaxies detected at redshifts $z\sim 1-3$. This is because gravitational scattering of MBHs by giant molecular clouds and stellar clusters can cause them to exhibit a random walk, rather than a smooth inspiral. Consequently, the orbital evolution of MBHs in such environments can be substantially slower (by one or two orders of magnitude relative to the timescale quoted above), or can stall permanently in some cases, at distances $\sim 10-10^3\,$pc from the center of the remnant galaxy \citep{fiacconi13,roskar15,tamburello17}. The long MBH pairing timescales obtained in these works leave room for the remnant galaxy to experience multiple mergers in this period of time. If so, the existence of triple, or even quadruple, MBHs in a single merger galaxy may be common, highlighting the need to understand their frequency of occurrence, and importance of dynamical interactions of multi-MBH systems. For example, using the Galaxy Zoo project catalog, \citet{darg11} find that a fraction of galaxy mergers involving more than two galaxies at a time is at least 2 orders of magnitude less than that of mergers of galaxy pairs. \citet{liu11} provide the incidence of multiple MBH in galaxy mergers, by investigating systems in which constituent MBHs accrete and shine as AGNs. They report the fraction of kpc scale AGNs pairs as $\gtrsim 3\times10^{-3}$, and kpc scale AGN triples as $\gtrsim 5\times10^{-5}$, both among optically selected AGNs at low redshift.

\begin{figure}[t]
\center
\includegraphics[trim=0 0 0 0, clip, scale=0.43]{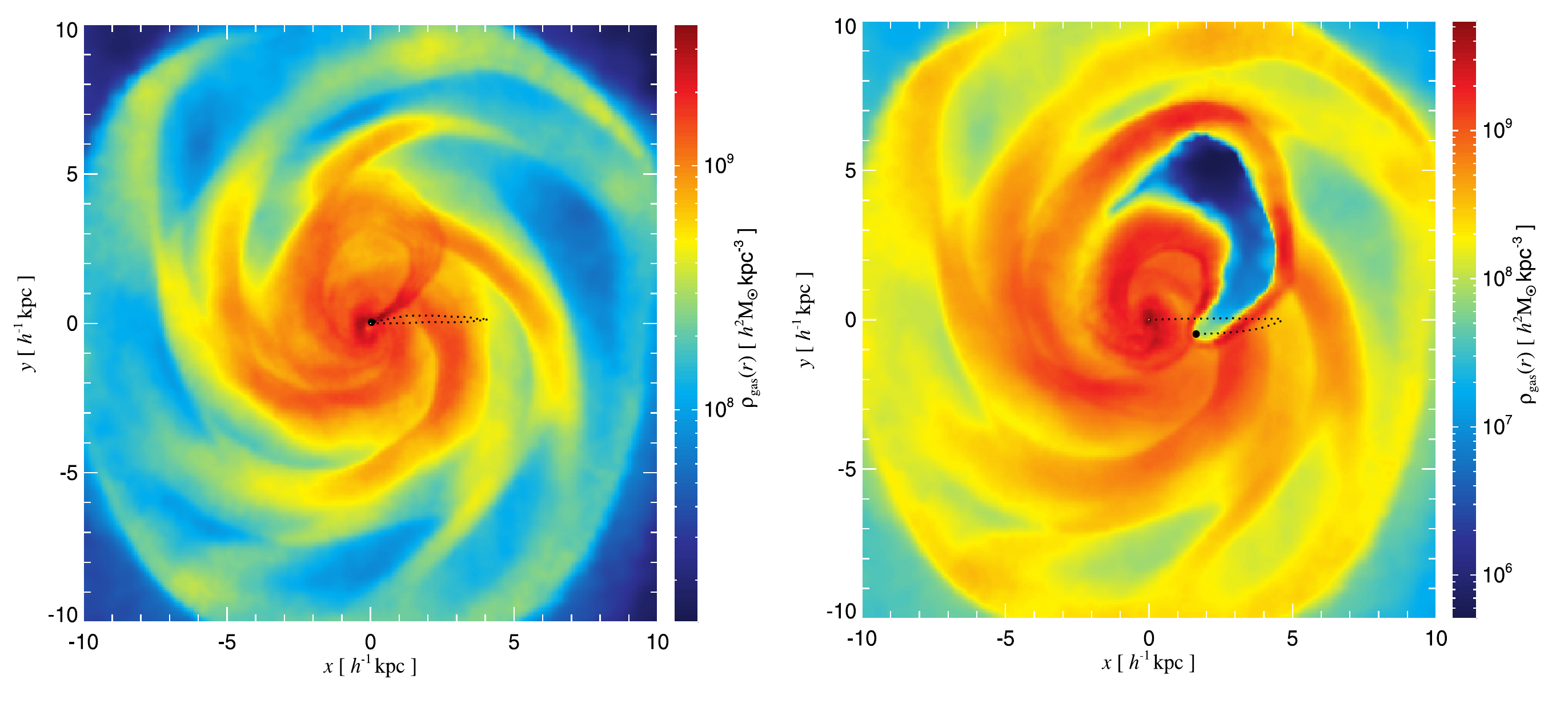} 
\caption{Mass-weighted gas density maps from hydrodynamic simulations of recoiling MBHs in massive gas-rich galaxies. Left (right) panel shows the results of the simulation without (with) feedback exerted by the accreting MBH. The scenario with feedback results in the evacuation of gas from the wake following the MBH. Both panels show the location of the MBHs approximately 40\,Myr after the recoil and the dotted lines trace their orbit from the start of each simulation. Figure adapted from \citet{sijacki11}.}
\label{fig10}      
\end{figure}

Interestingly, gravitational interaction of the MBH with the surrounding gas is quite local. For example, for a black hole with mass $\sim 10^6\,M_\odot$ most of the gravitational drag force is contributed by the gas that resides within only a few parsecs from the MBH
\citep{chapon13}. This proximity implies that the gas dynamical friction wake can be strongly affected, and possibly obliterated, by irradiation and feedback from an accreting MBH. Indeed, studies of dynamical evolution of MBHs find that dynamical friction can be significantly reduced due to  the ``wake evacuation" caused by purely thermal feedback from MBHs in simulations that follow gravitationally recoiled MBHs \citep[see Fig.~\ref{fig10};][]{sijacki11} and black hole pairing \citep{souzalima17,delvalle18}. These results bring into question the assumed efficacy of gaseous dynamical friction. In this context, \citet{park17},  \citet{toyouchi20} and \citet{gruzinov20} investigated the efficiency of gaseous dynamical friction in the presence of radiative feedback. The first two groups used local radiation hydrodynamic simulations and found that ionizing radiation, which emerges from the innermost parts of the MBH's accretion flow, strongly affects the dynamical friction wake and renders gas dynamical friction inefficient for a range of MBH masses and gas densities. They find that MBHs in this regime tend to experience positive net acceleration, meaning that they speed up, contrary to the expectations for gaseous dynamical friction in the absence of radiative feedback. This effect is more severe at the lower mass end of the MBH spectrum which, compounded with inefficiency of the gas drag for lower mass objects in general, suggests that $\lesssim 10^7\,M_\odot$ MBHs have fewer means to reach the centers of merged galaxies and form bound pairs with their counterpart MBHs. Indeed, \citet{li20b,li22a} show a significant suppression of pairing probability of such MBHs in merger galaxies with gas, in the presence of radiative feedback.  This is of importance because systems of MBHs in this mass range are direct progenitors of merging binaries targeted by the space-based GW observatory LISA.

Another important effect for the gaseous media is that when the gas accreting onto a MBH has much greater angular momentum than the hole itself, the MBH tends to align its spin axis with the angular momentum of the gas.  This effect was originally proposed by \citet{1975ApJ...195L..65B},  and several papers have investigated important aspects of this process \citep{1983MNRAS.202.1181P,
1998ApJ...506L..97N,1999MNRAS.309..961N,sorathia13b, sorathia13a,tremaine14,morales14,zhuravlev14,nealon15,polko17,hawley18, liska19}. Given that the radius of the accretion disk around an individual MBH can be thousands of times, or more, greater than the radius of the black hole, it follows that the angular momentum of the disk can dominate if it has more than a few percent of the mass of the MBH. An important implication of this statement is that it is, in principle, easy for the accretion flow to reorient the spin of the MBH. Note that this is different than the amount of  gas required to significantly change the spin \emph{magnitude} of the MBH, as conservation of angular momentum in this case requires accretion of the amount of gas comparable the original black hole mass.  

\begin{figure}[t]
\center
\includegraphics[trim=0 0 0 0, clip, scale=1.2]{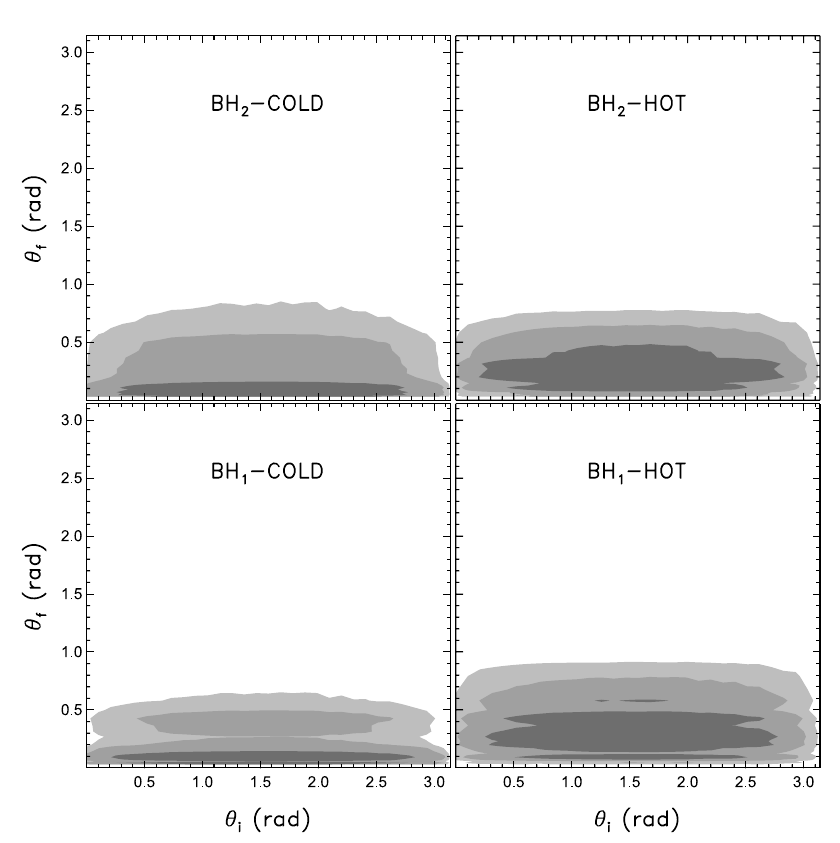} 
\caption{Density distribution of the initial ($\theta_i$) and final ($\theta_f$) angles of MBH spins with the vector of the binary orbital angular momentum. Left (right) panels refer to MBHs embedded in cold (hot) circumnuclear gas disks, described with a polytropic equation of state with $\gamma = 7/5$ ($\gamma = 5/3$).  Upper (lower) panels refer to the spin of the secondary (primary) MBH,  both with mass $M=4\times 10^6\,M_\odot$. Dark, medium and light-grey surfaces refer to high density regions encompassing 68.3, 95.5 and 99.7 per cent of the realizations. $\theta_i$ has been sampled isotropically, and the magnitudes of the dimensionless spin parameters (not shown) have been sampled from a constant probability distribution, between 0 and 1. Figure from \citet{dotti10}.} 
\label{fig11}      
\end{figure}

A major reason that this matters astrophysically is that, as we discuss in \S\ref{ssection:GWphase}, mergers between holes with coaligned spins lead to moderate GW kicks, whereas mergers between comparable-mass black holes with randomly oriented spins can produce kicks of thousands of kilometers per second, sufficient to eject the remnant from any galaxy.  This led \citet{bogdanovic07} to propose that gas alignment was effective in reorienting MBH prior to their becoming a gravitationally bound binary. Calculations by \citet{miller13} further find that gas torques are effective in aligning the MBH spins and their orbit with the large scale gas flow not only at large orbital separations but also well into the final parsec, where the two form a gravitationally bound binary. Simulations by \citet{dotti09,dotti10} and \citet{perego09} support the spin-alignment hypothesis, with the refinement that because the large gas disk in the inner few hundred parsecs of a galaxy will be somewhat turbulent, accretion is not always in the same direction and hence, rather than being exactly aligned, the spins of MBHs will be mostly aligned.  This effect is illustrated in Fig.~\ref{fig11}, which shows the efficiency of alignment in circumnuclear disks that correspond to more turbulent (hot) or less turbulent (cold) gas flows. This level of alignment reduces the probability of a large kick but does not eliminate it, and depending on specific assumptions one every few thousand or few tens of thousands of galaxy mergers might result in a kick exceeding 2,000 km~s$^{-1}$ \citep{lousto12}.

It is worth noting that there are also works proposing that the packets of gas that
arrive at MBHs are insufficient to reorient their spin axes. In this picture, the 
black holes accrete small quantities of gas stochastically (as opposed to from a disk 
with a well defined angular momentum) and as a consequence, the majority of MBHs 
are expected to spin slowly 
\citep{2005MNRAS.363...49K,king06,2007MNRAS.377L..25K, 2008MNRAS.385.1621K,2009ApJ...697L.141W,2012ApJ...753...15N, lodato13}.
The stochastic accretion hypothesis so far appears in tension with the spins of MBHs inferred from the X-ray observations of AGNs with relativistically broadened Fe~K$\alpha$ lines \citep{1996MNRAS.282.1038I,2002MNRAS.335L...1F,
2003PhR...377..389R,2006ApJ...652.1028B,2009ApJ...702.1367B,
2009ApJ...703.2171S,2011ApJ...736..103B,2012ApJ...755...88R}. 
If the properties of these observed MBHs are representative of the broader population in the Universe, these measurements imply that the dimensionless spin magnitudes of most MBHs exceed $s=0.6$. If not by the EM measurements, this hypothesis will ultimately be tested by the LISA GW observatory, which is expected to measure the distribution of MBH spins with exquisite precision.


\subsection{Effects after MBH are gravitationally bound to each other}\label{ssection:bound} 

If MBH inspiral due to the stellar and gas dynamical interactions described in previous section is efficient, it eventually results in formation of a gravitationally bound binary. Qualitatively, a gravitationally bound MBHB forms when the amount of gas and stars enclosed within its orbit becomes comparable to its mass, $M$. This orbital separation is comparable to the radius of gravitational influence of a single MBH, where the circular velocity around the black hole equals the stellar velocity dispersion, $r_{\rm inf} \approx GM/\sigma^2$. Combining this with the $M-\sigma$ relationship in Eq.~\eqref{eq:M_sigma} \citep{mcconnell13} gives 
\begin{equation}
r_{\rm inf} \approx 3.2\,{\rm pc}\,\, M_7^{0.645} = 6.6\times10^{6} r_g \,\, M_7^{-0.355}\; .
\label{eq:rinf}
\end{equation}
where $M_7 = M/10^7M_\odot$  and $r_g = GM/c^2\approx 0.1~{\rm au}~(M/10^7~M_\odot)$ is the gravitational radius. Therefore, bound binaries form at separations of $\sim1-10$\,pc for for a wide range of MBHB masses and host galaxy properties.

\subsubsection{Three-body scattering of stars and interactions with additional MBHs}\label{sssection:3body}

The simplest picture one can consider for a MBHB in this stage is that of a gravitationally bound binary that evolves due to interactions with individual stars in a completely spherically symmetric potential. For stars that come close enough to the binary,  three-body interactions lead to ejection of stars with greater energy than they had initially, a process that leads to gradual hardening of  the MBHB. In the limit where the MBHB masses are much larger than the masses of the interacting stars, the work of \citet{1996NewA....1...35Q} is especially relevant.  He showed that, given an unlimited supply of stars, the binary would harden at a rate given by the dimensionless parameter
\begin{equation}
H={\sigma\over{G\rho}}{d\over{dt}}\left(1\over a\right) \,,
\label{eq_Hparam}
\end{equation}
where he found $H\approx 10-20$ in the cases most relevant to us, in which the stellar speeds at infinity are much less than the binary orbital speed. In this case, the parameter $H$ provides a  dimensionless characterization of the rate of change of the binding energy of the binary. The associated hardening timescale can then be estimated as
\begin{equation}
t_{\rm h} = \left| \frac{a}{\dot{a}}  \right| = \frac{\sigma}{G\rho aH} =
1.5\times 10^6\,{\rm yr}\,\left(\frac{\sigma}{100\,{\rm km\,s^{-1}}} \right)
\left(\frac{10^4\,M_\odot\,{\rm pc^{-3}}}{\rho} \right)
\left(\frac{0.1\,{\rm pc}}{a} \right)
\left(\frac{15}{H} \right) \,,
\label{eq:t_3body}
\end{equation}
and it corresponds to the time that a MBHB of mass $\sim 10^7\,M_\odot$ spends at orbital separation of $a = 0.1\,$pc in a nuclear stellar cluster with velocity dispersion of $\sigma = 100\,{\rm km\,s^{-1}}$ and stellar mass density of $\rho = 10^4\,M_\odot\,{\rm pc^{-3}}$. Note also that the MBHB spends increasingly  longer times at smaller orbital separations and thus, its evolution slows down progressively.

The implication of Eq.~\eqref{eq:t_3body} is that three-body interactions with stars can in principle lead to very efficient orbital evolution of MBHBs, with a wide range of masses, given an unlimited supply of stars. This assumption is reasonable in the stages of evolution when the MBHB orbital speed is comparable to the velocity dispersion of the stars, because stars are not completely ejected and can return for another interaction. When the MBHB orbital speed is significantly greater than $\sigma$ however, interacting stars are ejected. In that scenario, the supply of stars that can interact with the MBHB is depleted and must, ultimately, be replenished if the binary is to continue its evolution.  The timescale on which the stars are replenished therefore determines the rate of the binary orbital evolution.  

This timescale is ordinarily determined by gravitational two-body interactions between stars in the nuclear star cluster around the MBHB, which can change the energy of a star by the order of itself. This is referred to as the \emph{energy relaxation time} \citep{1987gady.book.....B,1987degc.book.....S} and is directly related to the dynamical friction time shown in Eq.~\eqref{eqn:t_df}, as it describes the same type of gravitational interaction between the two point masses. For example, using Eq.~\eqref{eqn:t_df} one can estimate a relaxation time of about 41\,Gyr for a $M=1\,M_\odot$ star residing in the nuclear star cluster with $\sigma = 100\,{\rm km\,s^{-1}}$, from an initial radius $r \approx 3\,$pc, corresponding to the influence radius of a $10^7\,M_\odot$ binary. Such a MBHB would thus be unable to shrink its orbit much below the influence radius by scattering of stars and would in principle have to rely on other processes, like emission of GWs, to continue its evolution. At this point, however, gravitational radiation driven inspiral would also take longer than a Hubble time for such a binary, leading to a hang-up in the evolution of the MBHB. This phenomenon, first described by \citet{bbr80}, has been dubbed the ``final parsec problem'' \citep{2003AIPC..686..201M}.

Thus, an important question for MBHBs evolving due to three-body interactions with stars is: are there physical scenarios in which the stars are supplied at a sufficiently high rate, so as to guarantee the delivery of the MBHBs to the GW regime and coalescence in less than a Hubble time? The answer depends on the orbital properties of the stars. If the interacting star is on an orbit with eccentricity $e$, then its specific angular momentum is $\ell=\sqrt{GMa(1-e^2)}$, where $M$ is the mass of the central object (either a single MBH or a MBHB).  Two-body relaxation is a diffusive process, so the time required for the angular momentum to change by the order of itself is $(1-e^2)$ times the energy relaxation time. Thus, orbits with high eccentricities can change their angular momenta rapidly.  The rapidity with which stars can change their angular momenta leads to the distinction between full and empty loss cones \citep[for discussion of stellar dynamics in the centers of galaxies, see, e.g.,][]{1972ApJ...178..371P,1976MNRAS.176..633F,1976ApJ...209..214B, 1977ApJ...216..883B,1977ApJ...211..244L,2004ApJ...600..149W,
2005ApJ...629..362H}.  

In general, the ``loss cone" is the solid angle inside of which a particular star can have an important interaction with a target object. For example, for tidal disruption around a single MBH, the loss cone at a given distance from the black hole is the set of angles such that if a star's orbit is inside that solid angle, it will be disrupted. In the case of interactions with a MBHB, the loss cone is the solid angle such that the star will pass within $\sim 2$ binary semimajor axes of the binary, where the star can have a significant interaction with it. If the angular momentum can change by of order itself or more in a single orbit, the orbit is in the \emph{full} loss cone regime, so-called because any star that is ejected or destroyed from such an orbit will be replaced immediately by another star.  In contrast, if the typical change in the angular momentum in one orbit is less than the magnitude of the angular momentum, then the orbit is in the \emph{empty} loss cone regime, and loss of stars must be replaced diffusively, on the timescale discussed in the previous paragraph. In such scenarios, there is a decrease in the phase space density inside empty loss cones, leading to significant slow down or complete hang-up in the orbital evolution of the MBHB.  These two regimes are illustrated in Fig.~\ref{fig12}, which shows different evolutionary tracks for MBHBs that find themselves in full and empty loss cones.

Given the importance of the final parsec problem for plausibility of coalescences of MBHBs, there have been many papers devoted to its solution. One approach has been to consider the evolution of MBHBs in non-spherical potentials. The motivation for this has been twofold. Firstly, galaxies that are remnants of major mergers are not expected to be spherically symmetric. This is because the relaxation timescales in their nuclei (discussed above) are expected to exceed a Hubble time in many relevant physical scenarios. Secondly, if the underlying stellar potential is triaxial then, because it lacks the azimuthal or spherical symmetry which would guarantees conservation of angular momentum for individual orbits, individual stars can replenish the loss cone much more rapidly than would be possible via two-body relaxation.  Indeed, for galaxies that can be described as triaxial ellipsoids, there are a substantial number of stars on ``centrophilic" orbits that can plunge arbitrarily close to their center \citep{2002ApJ...567..817H,2004ApJ...606..788M}. Studies of these orbits have demonstrated that even moderate triaxiality suffices to allow a MBHB to merge within a Hubble time \citep[e.g.,][]{khb06,berczik06, preto11, khan11, khb15, vasiliev15, pfister17}.

For example, recent studies show that in all cases, a MBHB forms and hardens on timescales of at most 100\,Myr, coalescing on another few-hundred-megayear timescale, depending on the binary orbital eccentricity and the characteristic density of stars \citep{khan18}. Higher central densities provide a larger supply of stars that can efficiently extract energy from the MBHB orbit, and shrink it to the phase where GW emission becomes dominant, leading to the coalescence of the MBHs. Overall, the sinking of the MBHB in such studies takes no more than $\sim0.5\,$Gyr after the merger of the two galaxies is complete, but this can in principle be much faster for plunging binary orbits. 

In addition to non-spherical potentials, a number of studies have considered the evolution of MBHBs in rotating galactic nuclei \citep[see the discussion in Sect.~\ref{sssection:implications_mbhb} and][]{2010MNRAS.401.2268A, sesana11, gualandris12, wang14, khb15, mirza17, rasskazov17}. An important result of these studies is that in general, rotation has been found to increase the hardening rate of MBHB even more effectively than galaxy geometry alone. In this case too, the corotation (counterrotation) of a MBHB with stars in the galactic nucleus results in relatively low (high) binary orbital eccentricity. For MBHBs that end up on relatively eccentric orbits, a residual eccentricity may be preserved all the way into the GW regime, potentially allowing to distinguish different evolutionary scenarios from GW observations.

It is worth noting that the same types of interactions that lead to efficient binary hardening also result in some number of stellar tidal disruptions. As a result, the evolution of the tidal disruption rates happens in tandem with the binary hardening rates, since both are driven by the depletion of loss-cone orbits \citep{chen11, lezhnin19}. Specifically, the stellar tidal disruptions of unequal mass MBHBs can reach a rate $\sim 10^{-1}\,{\rm yr^{-1}}$ (three orders of magnitude higher than the corresponding value for a single MBH) soon after the gravitationally bound binary is formed \citep{chen09}. Depletion of loss-cone orbits leads to a precipitous drop in the disruption rate after $\sim 10^6$\,yr, which asymptotes to the value typical of single MBHs ($\sim 10^{-5}$ to $\sim 10^{-4}\,{\rm yr^{-1}}$) for binaries in the GW regime. While the MBHB hardening rate is only weakly influenced by tidal disruptions \citep{chen11}, they can nevertheless give rise to bright EM flares \citep{coughlin17, coughlin18}, that may allow identification of a galaxy that is the host to a gravitationally bound binary, as long as this scenario can be distinguished from a disruption by a single MBH and variability of regular AGNs.


While stellar dynamical processes seem capable of delivering many MBHBs into the GW regime, it is prudent to consider what happens to those that experience ineffective stellar scattering and stall as a result at sub-parsec orbital separations. Given enough time, the remnant galaxy could undergo subsequent mergers with other galaxies and acquire additional (i.e., third, fourth, etc.) MBHs in the process \citep{mikkola90, heinamaki01, blaes02}. As noted before, studies of galaxy pairs lead to the conclusion that, on average, all massive galaxies have experienced a merger event in the last 10 billion years \citep{2006ApJ...652..270B,2008ApJ...672..177L}. Assuming an average binary lifetime of 1 billion years implies that about 10 percent of MBHBs may form a triplet. Indeed, \citet{kelley17} and \citet{bonetti18} find that about 20--30 percent of all MBHBs could involve the presence of a third MBH. 

Some of the consequences of the intrusion of another MBH into the MBHB system are: ($i$) the enhanced rate of loss cone refilling due to perturbation of stellar orbits by the intruding MBH \citep{perets08}, ($ii$) chaotic, non-hierarchical three-body interactions, which tend to shrink the binary semimajor axis and to increase its orbital eccentricity  \citep{valtonen91}, and ($iii$) if they form a hierarchical triple, the merger time of the inner binary can be dramatically reduced due to eccentricity oscillations induced by the Kozai-Lidov mechanism \citep{blaes02}. The latter two effectively increase the binary eccentricity to values close to unity, resulting in the GW losses and eventual MBHB coalescence on a timescale of $\sim {\rm few}\times 10^8\,$yr. All of them are likely to accelerate the MBHB coalescence and increase the ejection rate of (typically less massive) MBHs \citep{hoffman07, as10, ryu18, bonetti19}. This bodes well for GW searches for MBHBs, as it provides multiple means for them to reach this evolutionary stage.

\subsubsection{Interactions with the circumbinary disk}\label{sssection:circumbinary} 

The evolution of sub-parsec MBHBs in predominantly gaseous environments has also been a topic of a number of theoretical studies in the past 15 years \citep[][etc.]{an05, macfadyen08, bogdanovic08, bogdanovic11, cuadra09, haiman09, hayasaki09, roedig12,shi12,shi15,noble12,kocsis12b, kocsis12a,dorazio13,farris14,rafikov16,schnittman16, tang17, miranda17, bowen17, bowen18,bowen19, tang18, dascoli18,moody19}. Most have focused on a particular physical scenario, commonly referred to as the \emph{binary in circumbinary disk}, in which a MBHB is embedded within a geometrically thin and optically thick gas disk. This setup is of interest because a dense and relatively cold circumbinary disk can, in principle, act as an effective conveyor belt for angular momentum. Such disks are also thought to be radiatively efficient and similar to disks that power AGNs -- thus, they can, as an added bonus, provide a luminous EM tracer for inspiraling MBHBs.

\begin{figure*}[t]
\centering{
\includegraphics[trim=130 120 120 130, clip, scale=0.35,angle=90]{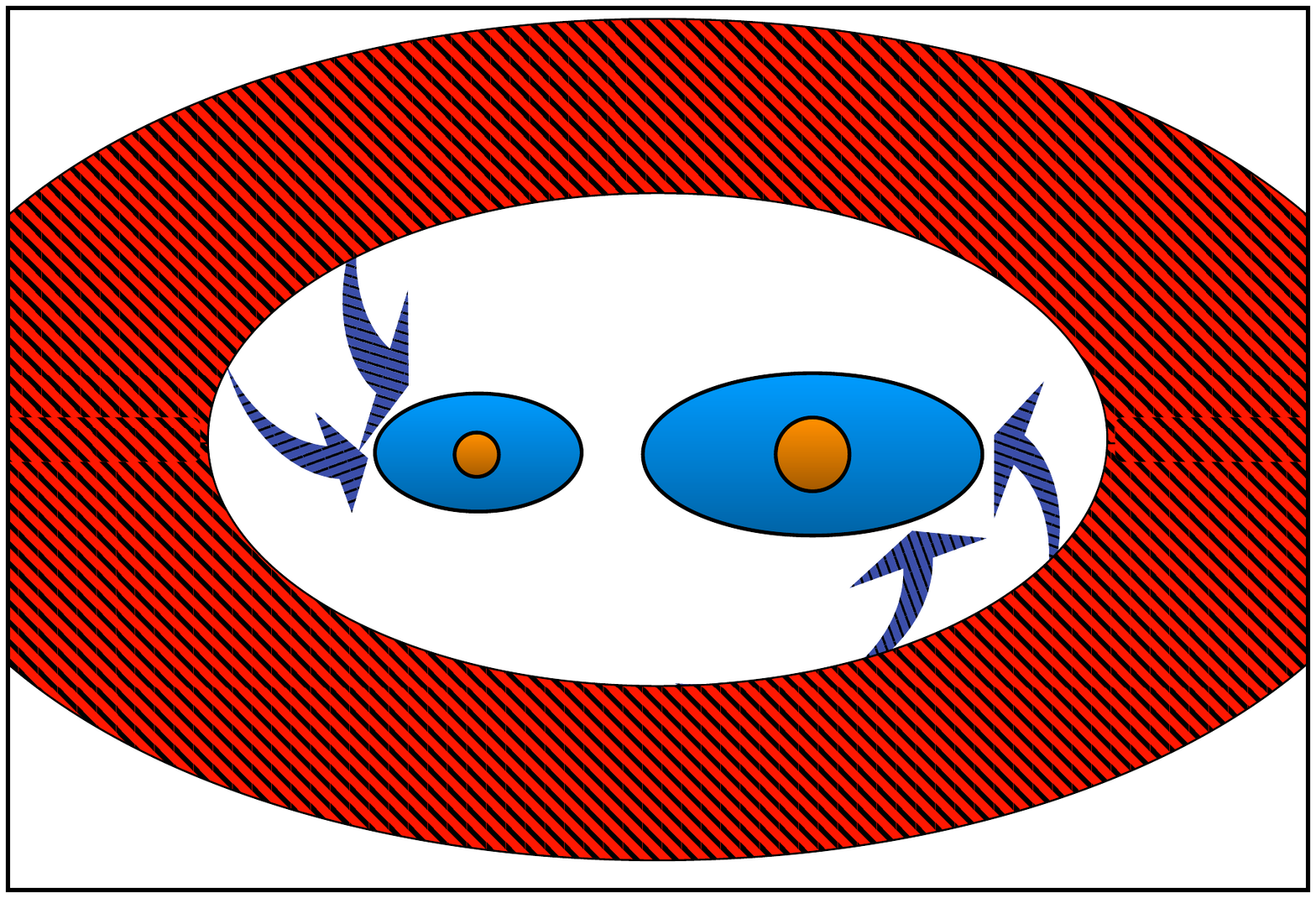} 
\includegraphics[trim=260 355 260 350, clip, scale=1.55,angle=0]{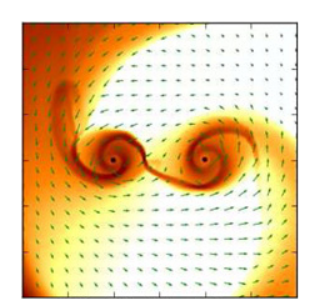} 
}
\caption{Illustration of the geometry of circumbinary region after the binary with comparable mass MBHs has created a low density region in the disk. {\it Left:} Simulations indicate that MBHs in this evolutionary phase can accrete by capturing gas from the inner rim of the circumbinary disk and can maintain mini-disks bound to individual holes. Adapted from \citet{bogdanovic15}. {\it Right:} Snapshot of gas surface density from a 2D hydrodynamic simulation of circumbinary disk around a MBHB with mass ratio 1:4. Orbital motion is in the counterclockwise direction. Superposed green arrows trace the fluid velocity. Adapted from \citet{farris14}.}
\label{fig13}      
\end{figure*}

As noted in Sect.~\ref{sssection:implications_mbhb}, simulations of isolated galaxy mergers indicate that MBH pairs with mass ratios considerably lower that $q \approx 0.1$ are less likely to form gravitationally bound binaries within a Hubble time, primarily due to the inefficiency of dynamical friction on the lower-mass MBHs. If, at the point of formation of the gravitationally bound binary, the two MBHs have a mass ratio larger than a few percent, theoretical studies indicate that the binary torques can truncate a sufficiently cold circumbinary disk and create an inner low-density cavity with a radius corresponding to approximately twice the binary semimajor axis \citep[see][and references above]{lp79,dorazio16}. MBHs in this phase can accrete by capturing gas from the inner rim of the circumbinary disk and can in this way maintain mini-disks bound to individual holes. The left panel of Fig.~\ref{fig13} illustrates the geometry of the circumbinary region for MBHBs in this evolutionary stage, found in simulations that have investigated this scenario. 

Simulations also indicate that despite strong binary torques, accretion into the central cavity continues unhindered relative to the single MBH case \citep[e.g.,][]{dorazio13,farris14,shi15}. If so, this has two important implications: (1) that necessary conditions exist for efficient transport of angular momentum through the circumbinary disk and potentially, orbital evolution of the MBHBs (see the discussion below, however) and (2) by analogy with single MBHs, binary MBHs powered by accretion can also shine as AGNs. Another property that many simulations of MBHBs in circumbinary disks identify robustly is that in unequal-mass binaries, accretion occurs preferentially onto the smaller of the two objects, because it orbits closer to the inner edge of the circumbinary disk \citep[e.g.,][]{al94,gunther02,hayasaki07,roedig11,nixon11, farris14}. This indicates that the lower-mass secondary MBH may grow at a rate higher than the primary and that, given enough time, the binary may evolve toward an equal-mass configuration. 

Thus, the simplest scenario that one can consider in this case is that of a MBHB orbit that decays under the influence of accretion torques. In this case the inner rim of the circumbinary disk follows the MBHB inward until the timescale for orbital decay by gravitational radiation becomes shorter than the viscous timescale\footnote{The timescale on which angular momentum is transported outward through the disk.} of the disk \citep{milosavljevic05,an05}. At that point, the rapid loss of orbital energy and angular momentum through gravitational radiation can cause the binary to detach from the circumbinary disk and to accelerate towards coalescence. If the orbital evolution of the binary is driven by the circumbinary disk, the rate of shrinking of the MBHB orbit can be described in terms of the viscous inflow rate at the disk inner edge as 
\begin{eqnarray}
\label{eq:dotavisc}
\dot{a}_{\rm visc} &=& - \frac{3}{2}\,\alpha \left(\frac{h}{r} \right)^2 v_{\rm Kep}\\ \nonumber
&=& -1.1\times 10^{-8} c 
\left(\frac{\alpha}{0.1} \right)^{4/5}
\left(\frac{\dot{M}}{0.1\dot{M}_{\rm E}} \right)^{2/5}
\left(\frac{10^4 r_g}{a} \right)^{2/5}
\left(\frac{10^7 M_\odot}{M} \right)^{1/5} ,
\end{eqnarray}
where $\alpha$ is the dimensionless viscosity parameter and $h/r$ is the geometric aspect ratio of the disk in the \citet{ss73} accretion disk model. $v_{\rm Kep} = (GM/2a)^{1/2}$ is the circular orbital speed of the circumbinary disk with the inner edge at $r=2a$, around a central object, here assumed to be a binary with semimajor axis $a$ and mass $M$. Here $\dot{M}_{\rm E} = L_{\rm E}/\eta c^2$ is the Eddington mass accretion rate, $\eta$ is the radiative efficiency, $L_{\rm E} = 4\pi GM m_p c/\sigma_{\rm T}$ is the Eddington luminosity, $\sigma_{\rm T}$ is the Thomson cross section, and other constants have their usual meaning. We obtain the second line of Eq.~\eqref{eq:dotavisc} by using the expression for $h/r$ for a gas pressure dominated disk \citep[see Eq.~2.16 in][]{ss73}. 

The associated timescale for the inner edge of the disk to move inward is then
\begin{equation}
\label{eq:tvisc}
t_{\rm visc} = - \frac{a}{\dot{a}_{\rm visc}} \approx
 1.7\times10^6\,{\rm yr} \,
\left(\frac{0.1}{\alpha}\right)^{4/5}
\left(\frac{0.1\dot{M}_{\rm E}}{\dot{M}} \right)^{2/5}
\left(\frac{a}{10^4 r_g} \right)^{7/5}
\left(\frac{M}{10^7 M_\odot} \right)^{6/5} ,
\end{equation}
%
%
%
where we assumed a binary with orbital separation $a=10^4\,r_g$ surrounded by a disk with a steady state accretion rate of 10\% of the Eddington limit. It is worth noting that the Shakura-Sunyaev solution for an accretion disk around a single MBH implies that for each $M$ and $\dot{M}$ there is a critical radius beyond which the disk is gravitationally unstable to fragmentation and star formation. This truncation radius in general arises somewhere between $10^2\,r_g$ and $10^6\,r_g$, depending on the properties of the disk and the mass of the central object \citep[see Eqs.~15 and 16 in][for example]{haiman09}. This is commonly considered as the outer edge of the accretion disk, beyond which the transport of angular momentum transitions from accretion torques to some other mechanism, either stellar or gaseous. Given that the structure and stability of circumbinary accretion disks is an area of active research, the existence and location of such an outer truncation radius are still open questions. 

Equation~\eqref{eq:tvisc} implies that if accretion torques are responsible for evolution of the MBHB orbit, its semimajor axis should monotonically decrease with time until the binary reaches the GW regime. Since the MBHB-circumbinary-disk system has more angular momentum than the nominal Shakura-Sunyaev MBH-disk system, one may also expect the timescale for orbital evolution to be longer than estimated above by some factor that depends on the properties of the MBHB and circumbinary disk. The expectation of systematic shrinking of binary orbit has been confirmed by many theoretical works leading to 2017 (see the references above), based on hydrodynamic and magnetohydrodynamic simulations of MBHBs in circumbinary disks. Works after 2017, however, show that significant uncertainties in the MBHB migration still remain. 

The basic challenge with determining whether and how fast the MBHB orbit shrinks arises from the fact that angular momentum loss by gravitational torques is nearly canceled by angular momentum gain in accretion \citep{shi12}, making quantitative determination of the net sensitive to details of the calculation \citep{dittmann21}. For example, using 2D hydrodynamic simulations \citet{tang17} found that the torques that are responsible for MBHB orbital evolution emerge from the gas very close to the MBHs, and are contributed by the small asymmetries in the shape of the gas distribution of the mini-disks. The implication of this result is that the timescale for orbital evolution is sensitive to both numerical effects (numerical resolution and dimensionality of the model) and numerical implementation of the physical processes (such as accretion, thermodynamic properties of the disk, etc.). Surprisingly, the simulations by \citet{tang17} showed that when MBHs are modeled as rapid sinks\footnote{These simulations do not resolve the spatial scales close to the event horizon of the MBHs and instead model them as sinks of larger size.} and accrete gas on a timescale comparable to or shorter than the local viscous time, the binary experiences positive torques, leading to expansion of its orbit. A few other groups, which performed close studies of the angular momentum transfer to the MBHB and the resulting binary orbital evolution, also report positive orbital torques: (a) \citet{miranda17}, based on 2D simulations of circular binaries with different mass ratios and eccentricities, (b) \citet{munoz19} for equal-mass binaries with different eccentricities, modeled in 2D, and (c) \citet{moody19} based on their 2D and 3D simulations of circular binaries with equal-mass ratios, coplanar and misaligned with the circumbinary disk. 

More recently, \citet{tiede20} modeled equal-mass MBHBs on circular orbits and found that  binary orbital evolution switches from outspiralling for warm, thicker disks (with aspect ratios $h/r \sim  0.1$), to inspiralling for cooler, thinner disks at a critical value of $h/r  \sim 0.04$. They explain that the reason why a group of earlier studies find positive orbital torques is because they model disks with $h/r =  0.1$. \citet{dittmann22} arrive to a similar conclusion and find an additional dependence of the orbital torque on the assumed value of kinematic viscosity. These results underscore that the evolution of MBHBs is sensitive to the viscous and thermodynamic properties of the disk (reflected in its temperature and $h/r$). They also raise additional questions regarding the degree to which the calculation of the MBHB orbital torques may be affected by other approximations (like the description of the accretion flow equation of state) or by effects not captured by most present-day simulations (such as loss of angular momentum due to magnetically or radiatively driven outflows). At the time of this writing, some insight is provided by a study by \citet{williamson22}, the first one to investigate orbital evolution of binary AGNs at parsec-scale separations affected by radiation pressure from the AGNs themselves. The study shows that radiation pressure cyclically destroys MBH mini-disks, leading to reduced gravitational torques and accretion, and significantly increased coalescence time scale.

The uncertainties mentioned above should also affect the evolution of the MBHB orbital eccentricity, which is a function of both the orbital angular momentum and energy. The works mentioned above \citep{tang17,miranda17,munoz19,moody19, tiede20, dittmann21, dittmann22} assume a MBHB with fixed orbital elements, and thus make no predictions about the evolution of eccentricity. Earlier works that reported the inspiral of MBHBs found that in prograde binaries, the exchange of angular momentum with the circumbinary disk drives a steady increase in binary eccentricity \citep{an05,cuadra09}, which saturates in the range of $0.45-0.6$ \citep{roedig11,zrake21}. This finding seemed intuitive, as the comparable mass MBHBs were found to produce circumbinary disks with non-axisymmetric, lopsided cavities. These were in turn expected to induce the increase of binary orbital eccentricity through gravitational interactions. For retrograde MBHBs, the cancelation in angular momentum with the circumbinary disk was found to lead to growth in the orbital eccentricity, which is slow for initially circular binaries, and fairly rapid and leading to $e\approx 1$ for initially eccentric binaries \citep{nixon11,roedig14,bankert15, schnittman15, as16}. The question of MBHB orbital eccentricity is important for the final outcome of the binary evolution (stalling or coalescence), as very eccentric MBHBs that reach small enough pericentric distances may evolve directly into the GW regime and coalesce rapidly (see the next section). Similarly, the orbital eccentricity of the MBHB may also be imprinted in its EM and GW signatures at coalescence, as we discuss in Sect.~\ref{section:signatures}.

As noted in \S\ref{sssection:implications_mbhb}, the MBHs are expected to achieve an effective alignment of their spin axes with the rotation axis of the circumnuclear disk (and thus, with each other) by the time they form a gravitationally bound binary \citep{bogdanovic07}.  This alignment is expected to be further enforced after they form a binary, if their evolution continues in a circumbinary disk with a well defined orbital angular momentum \citep{miller13}. Alternatively, there is no a priori reason to expect the alignment of the MBH spins or high spins in scenarios characterized by the absence of coherent accretion or absence of gas altogether  \citep{2005MNRAS.363...49K,king06,2007MNRAS.377L..25K, 2008MNRAS.385.1621K,2009ApJ...697L.141W,2012ApJ...753...15N,lodato13}.

\subsection{The gravitational-wave phase}\label{ssection:GWphase}

\begin{figure*}[t]
\centering{
\includegraphics[trim=0 0 0 0, clip, scale=0.5,angle=0]{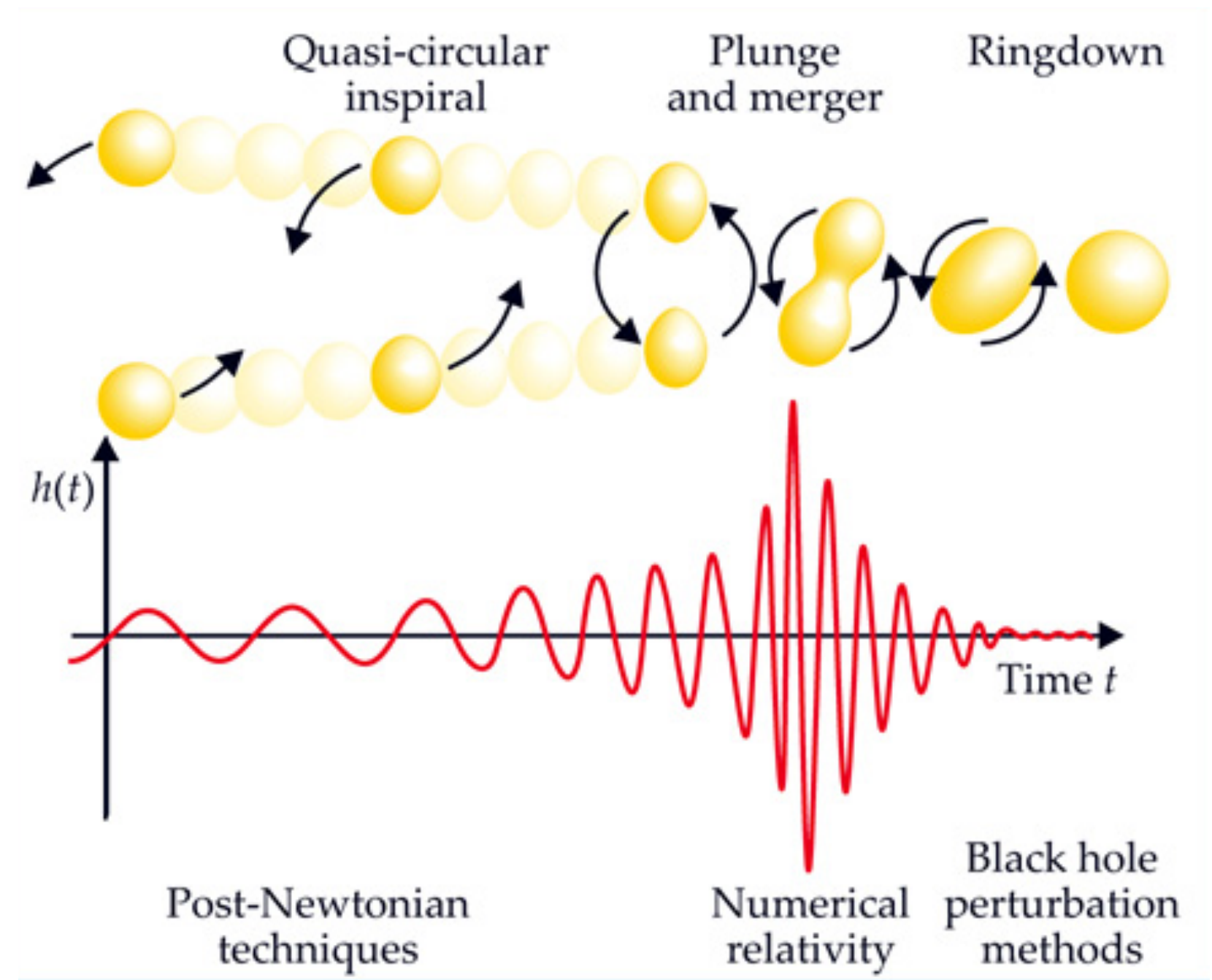} 
}
\caption{The loss of energy and angular momentum via the emission of gravitational radiation drives a binary to coalescence, which proceeds in three different phases, illustrated here in terms of the GW strain amplitude as a function of time. Figure from \citet{baumgarte11}.}
\label{fig14}      
\end{figure*}

When the binary separation is small enough, the evolution of the MBHB is driven primarily by gravitational radiation, rather than by interactions with stars or gas. The subsequent evolution and coalescence of the black holes is commonly divided into three approximately defined phases, illustrated in Fig.~\ref{fig14}.  The most protracted phase is \emph{inspiral}, in which the black holes are sufficiently separated that their inward radial migration is slow compared to their orbital motion. The second phase is \emph{merger}, in which the black holes plunge together and merge. The third phase is \emph{ringdown}, in which the initially asymmetric combined horizon radiates away modes and settles into a Kerr configuration.  The motivation for this classification originates from different techniques used to describe these evolutionary stages. Specifically, inspiral and ringdown phases can be treated semi-analytically and are modeled using the post-Newtonian technique and perturbation method, respectively. The merger phase, however, involves full nonlinear gravity and must usually be treated numerically \citep[in some cases the Effective One Body approach can also be used to compute the full GW signal;][]{buonanno99,buonanno00,damour00}.

Emitted gravitational radiation carries away some fraction of the MBHB energy and angular momentum, and it is illustrative to examine how this affects the properties of the holes in the three evolutionary stages mentioned above. Consider a binary with component masses $m_1$ and $m_2$, total mass $M=m_1+m_2$, and reduced mass ratio $\mu=m_1m_2/M$.
Since the inspiral phase includes the evolution of the MBHB from infinity to the innermost stable circular orbit\footnote{The smallest circular orbit in which a test particle can stably orbit a massive object in general relativity.} (ISCO), the energy released is $\mu$ times the specific binding energy at the ISCO, so that $E_{\rm inspiral}\sim \mu$. The total energy emitted in the merger and ringdown phases may be  estimated using the order of magnitude rule that the strain amplitude from a binary with semimajor axis $a$,
at a luminosity distance $d_L$ from us, is $h\sim (G\mu/ac^2)(GM/d_L\, c^2)$.  For the merger and ringdown phases, the semimajor axis is comparable to the size of the event horizon of the binary\footnote{Here we adopt $GM/c^2 \equiv M$ as a natural unit of length, expressed in geometrized units when $G=c=1$, and refer to it as the \emph{gravitational radius}.}, $a \sim M$, so $h\sim \mu/d_L$.  Since the GW luminosity is $L\sim d_L^2\, h^2 f^2$, where $f$ is the GW frequency, then $L\sim \mu^2 f^2$. If the merger and ringdown phase last a time $\tau$, then the total energy released is $E\sim \mu^2 f^2\tau$.  But the characteristic GW frequency at merger is $f \sim 1/M$ and the characteristic time is $\tau\sim M$, resulting in $E_{\rm merger}\sim \mu^2/M$, or a factor $\sim \mu/M$ times the energy released in the inspiral. As a consequence, for a comparable-mass binary the energy emitted in the merger and ringdown could be a few percent of the rest mass energy of the binary, and of the same order as the energy emitted in the inspiral. For systems with extreme mass ratios however, the inspiral is likely to be the only detectable phase.

In most of the inspiral phase, down to the separation $\sim 10-20\,M$, the pure gravitational radiation driven evolution is well described by quadrupolar radiation \citep{1964PhRv..136.1224P}. In this approximation, the orbital semimajor axis, $a$, and eccentricity, $e$, of the binary evolve as
\begin{equation}
\biggl\langle{da\over{dt}}\biggr\rangle=-{64\over 5}{G^3\mu M^2\over{
c^5a^3(1-e^2)^{7/2}}}\left(1+{73\over{24}}e^2+{37\over{96}}e^4\right)
\end{equation}
and
\begin{equation}
\biggl\langle{de\over{dt}}\biggr\rangle=-{304\over{15}}e{G^3\mu M^2\over{
c^5a^4(1-e^2)^{5/2}}}\left(1+{121\over{304}}e^2\right)
\end{equation}
where the angle brackets indicate an average over an orbit.  One can show that these rates imply that the quantity
\begin{equation}
ae^{-12/19}(1-e^2)\left(1+{121\over{304}}e^2\right)^{-870/2299}
\label{eq_ae}
\end{equation}
is constant throughout the inspiral, meaning that for moderate to 
small eccentricities we have $e\propto a^{19/12}$.
These equations imply that the characteristic time to inspiral for
high-eccentricity orbits is
\begin{equation}
\label{eq:tgw}
t_{\rm gw}\approx 10^9~{\rm yr}
\left(\frac{10^{18}~M_\odot^3}{\mu M^2}\right)
\left(\frac{a}{0.001~{\rm pc}}\right)^4
(1-e^2)^{7/2}\; ,
\end{equation}
and that the eccentricity also decreases on that timescale. The dependence of the characteristic inspiral time on $e$, as well as circularization due to emission of gravitational radiation can be understood by recalling that the GW luminosity depends very strongly on the orbital frequency (and thus speed) of two masses. For high eccentricities the orbital speed is significantly larger at pericenter than it is at most places in the orbit. If one simplifies by assuming that GWs are \emph{only} emitted at pericenter (as in the impulse approximation), the loss of energy and angular momentum at pericenter means that the apocenter distance decreases with time, and consequently, so does the eccentricity.

The strong dependence of the inspiral time, $t_{\rm gw}\propto a^4$, means that at large distances the binary inspirals slowly. This timescale should be compared to the characteristic timescale for the inner edge of the circumbinary disk to move inward, $t_{\rm visc} \propto a^{7/5}$, given by Eq.~\eqref{eq:tvisc}. This implies that for MBHBs surrounded by circumbinary disks, the radial diffusion of the circumbinary disk initially keeps the inner edge of the disk at a radius $r\approx 2a$ as the binary shrinks. When $a$ becomes small enough however, the GW inspiral time can become shorter than the disk radial diffusion time, so that the MBHB may decouple from the disk  
\citep{pringle91, al94, 1996ApJ...467L..77A,liu03,milosavljevic05}. Comparison of the two timescales indicates that the radius of decoupling can be $\sim 100\,M$ for comparable-mass binaries and accretion at rates comparable to Eddington. It is reduced if one member of the MBHB is much less massive than the other, because gravitational
radiation is weaker in that case. The implied decoupling radius is however larger at lower accretion rates, as long as the disks are in the radiatively efficient regime 
\citep[with luminosity above $\sim 0.01\,L_{\rm E}$; see, e.g.,][]{ichimaru77, rees82,ny94,1999MNRAS.303L...1B}. 
This is because the disks then have smaller vertical thickness and hence smaller inward radial speeds (see Eq.~\eqref{eq:dotavisc}). 

The question of decoupling of the MBHB from the circumbinary gas disk before coalescence is important, because coalescences that occur in gas poor environments may not be accompanied by detectable EM counterparts. For that reason, this question has been examined in hydrodynamic simulations of inspiraling MBHBs in circumbinary disks.  For example, \cite{noble12} pointed out that magnetic fields in the circumbinary disk are stretched
as the binary shrinks, and this can lead to much larger stresses at its inner edge than previously realized. They find that the disk can follow the binary down to $\sim 20M$,
which has encouraging implications for the detectability of EM counterparts. Similarly, \citet{tang18} find that the MBHs capture gas from the circumbinary disk and continue to accrete efficiently via their own mini-disks, well after their inspiral outpaces the viscous evolution of the disk.


In addition to effects on the semimajor axis and eccentricity, \citet{2004PhRvD..70l4020S} showed that the directions of the spins of the two black holes can change systematically during GW inspiral. In particular, if the more massive black hole is at least partially
aligned with the orbital axis and the mass ratio is close to unity, then the spins of both holes tend to align with the orbital axis. This has been proposed by \citet{2010ApJ...715.1006K} and \citet{2012PhRvD..85l4049B} as an important mechanism to further align already partially aligned spins, which (as we discuss below) in turn reduces the kicks produced by the MBH
merger. Table~1 of \citet{2010ApJ...715.1006K} shows that alignment is only effective if the mass ratio is very close to unity and that even for mass ratios of 1/3, alignment
is ineffective. Thus, this alignment mechanism might play a role in a limited set of mergers. Interestingly, however, \citet{gerosa15b} show that even initially aligned spin configurations can be brought out of alignment as a consequence of an instability that exists for a wide range of spin magnitudes and mass ratios, and can occur in the strong-field regime near the merger. They show that when the spin of the higher-mass black hole is aligned with the orbital angular momentum and the spin of the lower-mass black hole is antialigned, binaries in these configurations are unstable to precession to large spin misalignment. If prevalent in real MBHBs, this instability provides a way to circumvent gas-driven astrophysical spin alignment that may be acquired at large binary separations (see discussion in Sect.~\ref{sssection:implications_mbhb} and \ref{sssection:circumbinary}), allowing significant spin precession prior to merger.

The orientation of spins, as well as the asymmetry in masses of the two black holes, play an important role for the resulting magnitude of the ``GW kick", received by the daughter black hole that forms in the coalescence.  The merger phase is the origin of the majority of this kick, relative to the original center of mass of the binary. The small pitch angle of the orbit during MBHB inspiral means that net emission of momentum in GWs nearly self-cancels
over an orbit.  Analytical calculations of the kick during inspiral
\citep{1962PhRv..128.2471P,1973ApJ...183..657B,1983MNRAS.203.1049F,
1984MNRAS.211..933F,1989ComAp..14..165R,1995PhRvD..52..821K,
2004ApJ...607L...5F,2005ApJ...635..508B,2006PhRvD..73l4006D,
2007CQGra..24.5307K,2007ApJ...662L..63S,2008PhRvD..77d4031S}
 suggest that prior to the final orbit the net speed relative to the original center of mass is at most tens of km~s$^{-1}$. However, the momentum radiated in the last half-orbit before merger is \emph{not} self-cancelled, coinciding with moments before coalescence when the emission of gravitational radiation is strongest.  

The breakthroughs that ushered in stable numerical simulations of black hole coalescence through ringdown 
\citep{2005PhRvL..95l1101P,campanelli06,2006PhRvD..73j4002B}
have allowed GW kicks to be computed for many configurations, from nonspinning black holes to fully generic mergers of comparable-mass black holes without special spin orientations. They showed that if the spins are aligned with each other (as opposed to antialigned or at some angle) and the orbital angular momentum, then the maximum kick is moderate: less than 200\,km~s$^{-1}$ \citep{2007PhRvL..98i1101G}. However, starting in 2007 other configurations were found that can lead to much greater kicks. \citet{campanelli07a, campanelli07b} showed that for equal-mass maximally spinning black holes with oppositely-directed spins in the orbital plane (configuration also known as the ``super kick"), the kick can reach almost 4,000~km~s$^{-1}$. A different, ``hangup kick" configuration, in which the spins are partway between the orbital plane and orbital axis, was introduced in \citet{2011PhRvL.107w1102L} and \citet{lousto12}.  The maximum velocity predicted for this configuration was shown to be close to 5,000\,km~s$^{-1}$.

As numerical simulations of black hole mergers are extremely computationally expensive, several groups have developed approximate analytic formulae, calibrated on simulations, that predict kick velocity as a function of mass ratio and spin magnitudes and directions 
\citep{2007ApJ...668.1140B,baker08,2008PhRvD..77d4028L,
2009PhRvD..79f4018L,2010ApJ...719.1427V,2010CQGra..27k4006L,lousto12}. We show the one by \citet{lousto12} that includes the super kick configurations 
\begin{eqnarray}
{\bf v_{\rm kick}}&= &v_m\,\hat{e}_1 + v_\perp (\cos{\xi}\,\hat{e}_1 + \sin{\xi}\,\hat{e}_2) + v_\parallel\,\hat{L}  \,, \\
v_m&=&A_m\, \frac{\eta^2(1-q)}{1+q}(1+B_m\eta)  \,,\\
v_\perp&=& H\frac{\eta^2}{1+q}(s_1^\parallel-qs_2^\parallel)  \,, \\
v_\parallel&= &\frac{16\eta^2}{1+q}(V_{11}+V_AS_z+V_BS_z^2+V_CS_z^3) 
| {\bf s_1}^\perp - q\,{\bf s_2}^\perp | \cos{(\phi_\Delta - \phi_1)} \,.
%
\end{eqnarray}
%
%
The velocity component $v_m$ refers to the kick velocity contribution from unequal masses and $v_\perp$, $v_\parallel$, are contributions from the spins. The indices $\perp$ and $\parallel$ refer to the components perpendicular and parallel to the orbital angular momentum, respectively.  Here, $\hat{e}_1$, $\hat{e}_2$ are the orthogonal unit vectors in the orbital plane, $\xi$ is the angle that determines the projection of the spin contribution to the kick velocity in the orbital plane ($v_\perp$) on $\hat{e}_1$ and $\hat{e}_2$, and $\hat{L}$ is the unit vector in the direction of the orbital angular momentum.  $s_1=(cS_1/Gm_1^2)$ is the dimensionless spin angular momentum of the more massive, primary black hole, and $s_2$ is that of the secondary, with similar definition. The parameter $q =m_2/m_1 \leq 1$ is the mass ratio, $\eta = q/(1+q)^2$ is the symmetric mass ratio, and $S_z= 2(s_1^\parallel+q^2 s_2^\parallel)/(1+q)^2$. $\phi_1$ is the phase angle associated with the primary black hole and $\phi_\Delta$ is the angle between the in-plane vector, defined in \citet{lousto12} as a linear combination of the two spins, and a fiducial direction at merger. The best fit coefficients are $A_m=1.2\times 10^4$\,km\,s$^{-1}$, $B_m=-0.93$, $H=6.9\times 10^3$\,km\,s$^{-1}$, $V_{11}=3677.76$\,km\,s$^{-1}$, $V_A=2481.21$\,km\,s$^{-1}$, $V_B=1792.45$\,km\,s$^{-1}$, $V_C=1506.52$\,km\,s$^{-1}$, and $\xi = 145^\circ$.
 
It is worth noting that improved, albeit more complex versions of the analytic formulae are now available in the literature \citep[e.g.,][]{healy18}. In addition to the analytic formulae, a family of non-analytic, so called ``surrogate models'', has been developed more recently. Surrogate models are trained against numerical simulations and were shown to be almost indistinguishable from simulations in the portion of parameter space where $s_1$ , $s_2 \leq 0.8$ and $q \leq 1/4$ \citep{varma19a, varma19b}.

\begin{figure*}[t]
\centering{
\includegraphics[trim=0 0 0 0, clip, scale=1.0,angle=0]{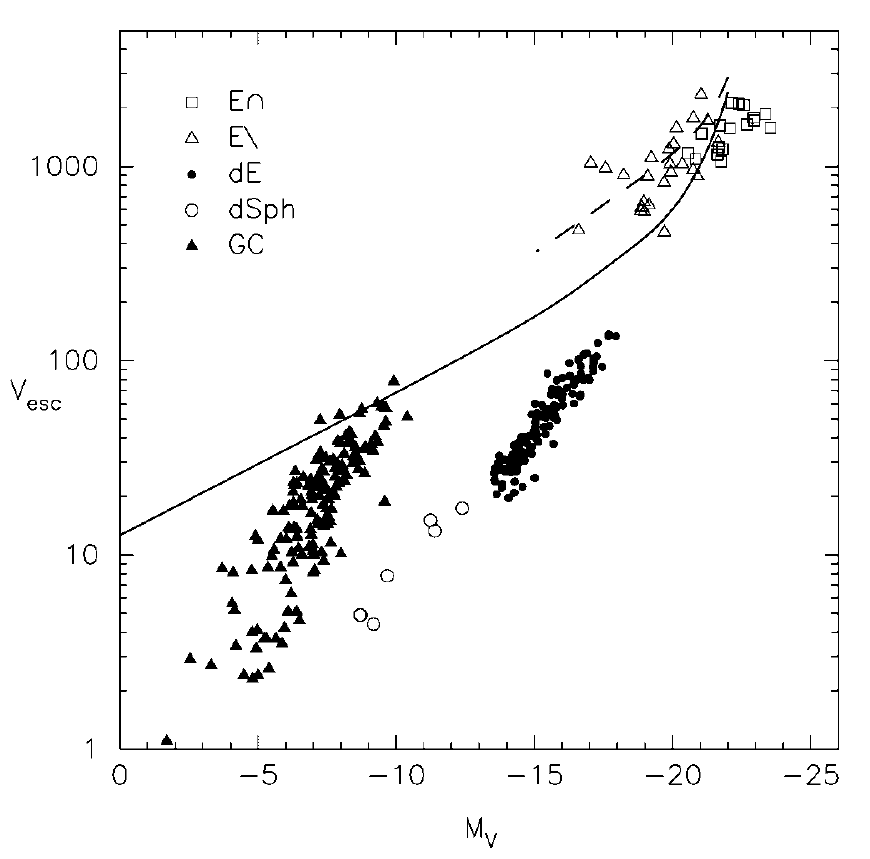} 
}
\caption{Central escape speeds in units of ${\rm km\,s^{-1}}$ from different types of stellar systems: core ellipticals (E$\cap$), power-law ellipticals (E$\backslash$), dwarf ellipticals (dE),  dwarf spheroidals (dSph), and globular clusters (GC). The solid line is the mean escape speed from the DM halos associated with the luminous matter. The dashed line is the escape speed from the combined luminous+mean DM potentials for E galaxies. Figure from \citet{2004ApJ...607L...9M}.}
\label{fig15}      
\end{figure*}

As discussed by \citet{2004ApJ...607L...9M} and illustrated in Fig.~\ref{fig15}, GW kicks exceeding $\sim 2,000$~km~s$^{-1}$ suffice to eject the remnant MBH from any galaxy.  Kicks exceeding several hundred km~s$^{-1}$ are enough to eject remnant MBHs from galaxies comparable to or smaller than the Milky Way. Thus, if such kicks are produced in some
galactic mergers, there may exist some number of ``empty nest'' galaxies without a central MBH or with the MBH offset from the galaxy center. We discuss possible observational signatures associated with this phenomenon further in Sect.~\ref{section:postmerger}.

\section{Emission signatures before, during, and after coalescence}
\label{section:signatures}

In this section we discuss the emission signatures that arise before, during and after coalescence.  Because our goal is to review the EM counterparts to merging MBHs that are sources of GWs, we focus on stages when the binary evolution is driven by emission of gravitational radiation (inspiral, merger and ringdown) rather than interactions with stars or gas. As discussed in \ref{section:timescales}, this may happen when the MBHs are within few hundred to a thousand gravitational radii of each other, or $\sim 10^{-3} (M/10^8\,M_\odot)$\,pc, and hence their orbital period is a few years or less. At such separations, the EM signatures are most likely to be associated with gas accretion flows surrounding the MBHBs. As noted in previous section, stellar tidal disruptions are not likely to play an important role for MBHBs in the GW-dominated phase of their inspiral, since their enhanced rate would have long since asymptoted to the value typical of single MBHs. We therefore focus attention on the MBHBs in circumbinary disks and assess pathways to EM signatures by considering several questions:
\begin{itemize}
\item What is the semimajor axis at which MBHB evolution driven by stellar and gas interactions transitions to binary evolution driven by gravitational radiation? 
\smallskip
\noindent\fcolorbox{white}{white}{%
    \minipage[!t]{\dimexpr0.50\linewidth-2\fboxsep-2\fboxrule\relax}    
\item What is the orbital eccentricity of the MBHB once it transitions to GW-driven evolution? Because emission of GWs leads to circularization, this transition determines the eccentricity of the MBHB at smaller separations and thus affects possibilities for time-variability of its EM signatures.

\item What fraction of the matter in the circumbinary disk eventually accretes
 \endminipage}\hfill
\noindent\fcolorbox{black}{light-yellow}{%
    \minipage[!t]{\dimexpr0.47\linewidth-2\fboxsep-2\fboxrule\relax}
\begin{center} {\bf EM counterparts of merging MBHBs can originate from} \end{center}
{ 
\noindent$\bullet$ The circumbinary disk.\\
\vspace{-2.0mm}

$\bullet$ Mini-disks and accretion streams. \\
\vspace{-2.0mm}

$\bullet$ Magnetized jets.\\
\vspace{-2.0mm}


$\bullet$ If all else fails: emission from the host galaxy. \\
\vspace{-2.0mm}
}
 \endminipage}\hfill\\
onto either of the MBHs?  Even if this fraction is small, the high efficiency of energy production in the mini-disks around each BH could mean that they dominate the overall luminosity.
\item Down to what spatial scale can the circumbinary disk follow the GW-driven shrinking of the MBHB?  
This could have a large impact on the simultaneity of EM and GW emission during the merger, as well as the EM luminosity of any merger-related event.
\item What is the role of periodic or quasi-periodic modulation of EM emission, as opposed to secular variability? Nearly periodic modulation, if observed over enough cycles, will be more easily distinguished from background noise than a gradual increase in intensity. Another important question is whether the fractional variation in intensity for periodic MBHB sources will be large enough to be detected, especially when combined with emission from the host galaxy.
\end{itemize}
We will keep these questions in mind as we lay out analytic estimates and discuss possibilities for production of characteristic EM signatures from merging MBHBs presented in the literature.

\subsection{Electromagnetic radiation before the merger}
\label{section:precursors}

\subsubsection{Emission from the circumbinary disk}
\label{sss_cbd}

The possible locations of EM emission before the MBHB merger are the circumbinary disk, the  mini-disks around the individual black holes, and the streams feeding the mini-disks (see Fig.~\ref{fig13}). Let us focus first on the circumbinary disk, where we will present some simple analytic estimates to guide the intuition and compare them with the results from simulations, paying attention to situations where they differ. The analytic estimates are obtained
under the assumption that the circumbinary disk luminosity is produced by gravitational energy and angular momentum transport rather than by perturbations induced by the binary torques.  We will also assume that the mass accretion rate through the disk is constant, and that it does not change as the MBHB inspirals and the inner rim of the circumbinary disk shrinks. As the binary inspirals, the disk then becomes more luminous and its spectrum peaks at higher photon energies.  

This picture is similar to the scenario envisioned by many researchers \citep{pringle91, al94,1996ApJ...467L..77A,liu03,milosavljevic05}.  
They noted that  the GW-driven inspiral time decreases more rapidly with semimajor axis ($t_{\rm gw}\propto a^4$, Eq.~\eqref{eq:tgw}) than does the viscous inflow time ($t_{\rm visc}\propto a^{7/5}$, Eq.~\eqref{eq:tvisc}).  Therefore, when the binary semimajor axis decreases sufficiently, the binary may decouple from the disk. As noted in  Sect.~\ref{ssection:GWphase}, comparison of the two timescales indicates that the decoupling may happen when $a \sim 100\,M$ for comparable-mass binaries and  accretion rates comparable to Eddington. If so, this implies that the gradual refilling of the hole after MBHB merger could provide a distinct signature of the merger \citep[as suggested by][]{milosavljevic05}. 
It also implies that the circumbinary disk will not be very luminous. The reason is that if the inner edge of the circumbinary disk is at $r_{\rm in} \approx 2a \sim 200\,M$ when the binary decouples, then the energy is released with only $M/r_{\rm in} \sim 0.5$\% efficiency. In comparison, the radiative efficiency of any matter accreting all the way to ISCO of either MBH is about 10\%, so the luminosity of the circumbinary disk is at least a factor of twenty less. Based on these simple arguments, the emission luminosity from the MBH mini-disks is expected to dominate that from the circumbinary disk in a steady-state accretion flow some time before merger.

\begin{figure*}[t]
\centering{
\includegraphics[trim=0 0 0 0, clip, scale=0.48,angle=0]{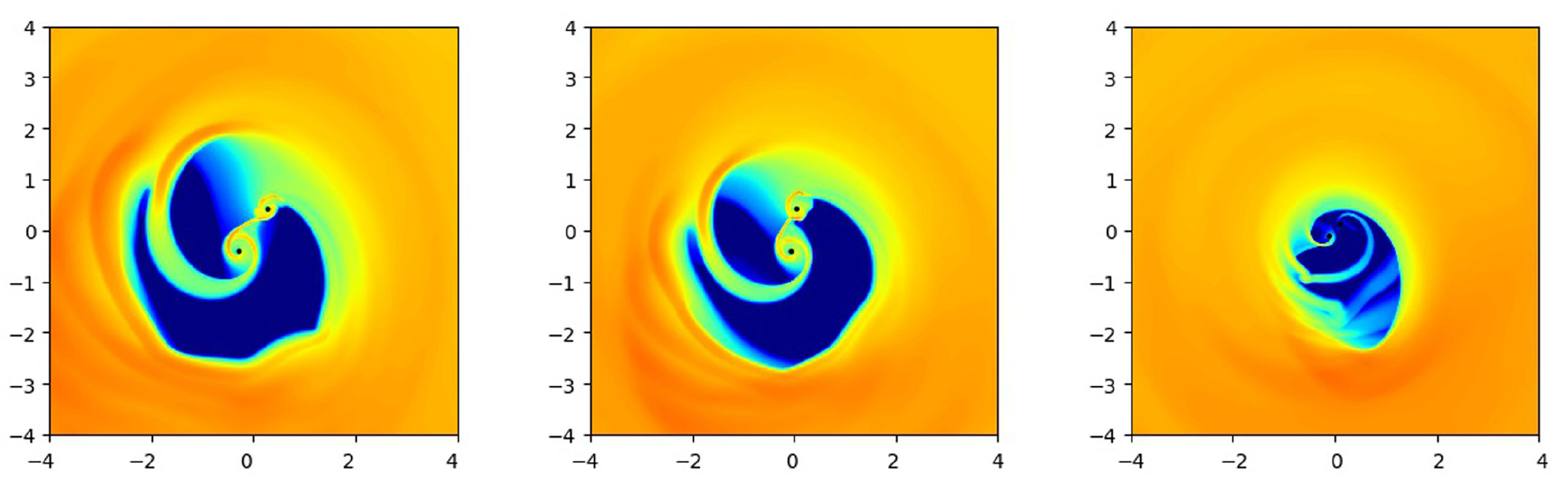} 
}
\caption{Surface density of the gas around an equal mass $10^6\,M_\odot$ MBHB evolving from the initial separation of $60\,M$ (left panel) to merger (right), on a timescale of about a week. Despite the presence of an asymmetric low density cavity, the two MBHs continue to accrete gas carried by the narrow gas streams all the way through coalescence. The blue and green (orange and red) colors indicate low (high) surface densities. Figure adapted from \citet{tang18}.}
\label{fig16}      
\end{figure*}

These considerations also imply that when the MBHBs merge, the circumbinary disk may be left behind at some appreciable distance. If so, there may be little matter around the MBHB at the moment of coalescence and hence, few opportunities for direct interaction of that matter with the dynamically changing spacetime near the MBHs that could produce a luminous EM counterpart. This early concern was mitigated by simulations \citep{noble12, farris15b, bowen17, tang18}, which indicate that the inner rim of the circumbinary disk is able to follow the binary to radii of $\sim 10M$. This is also true for the spinning MBHBs, as the spins do not strongly affect the bulk properties of the circumbinary disk \citep{lopez_armengol21}. For the circumbinary disk with an inner rim radius of $\sim 20M$, the energy release could then be $\sim 5$\% and substantial enough to produce an EM luminous signature. Similarly, the inspiral time of the gas is then much shorter than previously assumed, thus allowing the MBHs to accrete all the way to coalescence. This is illustrated in Fig.~\ref{fig16}, where the two MBHs are fed by the narrow gas streams till coalescence.

The prime difficulty with gravitationally-powered circumbinary disk emission as a signature of binary MBHB coalescence is that the gradual increase in luminosity and change in spectrum is smooth and thus, might not be easily distinguishable from other gradual changes that exist naturally in disks around single AGN.  As a reminder, we expect at most a few MBHB mergers per year out to a redshift $z\sim 1$, compared to the millions of AGNs that exist to that redshift.  Thus without a
contemporaneous GW signal and decent sky localization to guide a search, one needs to identify a signature that exists in fewer than one AGN per million. It is not clear that smooth changes in the luminosity and spectral properties of the circumbinary disk would provide such an opportunity (see however the discussion of the ``notches" in the total spectrum contributed by the circumbinary disk, the streams, and the mini-disks at the end of Sect.~\ref{sss_minidisks}).

To explore other possibilities related to circumbinary disks, we return to the prospect of torques exerted by the MBHB on the disk. Consider first a circular binary.  The magnitude of the torque it exerts on the circumbinary disk is likely to be time-independent, once we factor out the slow inward motion of the binary and disk. The torque is not, however, axisymmetric because the binary potential is not. Thus, there could be azimuthal modulation of the emission that appears periodic, depending on the viewing direction of the observer.  How strong this modulation is depends on multiple factors. For example, if the disk is sufficiently optically thick that the radiation diffusion time is longer than an orbital period, one would expect that visible modulation of the emission would be strongly suppressed. If in contrast the radiation diffusion time is much less than an orbital period, the modulation might be visible, but the mass of the disk that is torqued is decreased as a result.

We can quantify these considerations by imagining a disk with a fixed aspect ratio $h/r$ and surface density $\Sigma$.  Let the inner radius of the disk be $r_{\rm in}=2a$ and the area $f\pi r_{\rm in}^2$, where $f<1$ is a fraction of the disk experiencing strong torques. Then the optical depth of the disk from the midplane is $\tau_{\rm T}=\Sigma\kappa_{\rm T}$, where $\kappa_{\rm T}$ is the opacity to Thomson scattering. For a MBHB with a total mass $M$, the orbital period at the inner edge of the disk is $P_{\rm in}=2\pi(r_{\rm in}^3/GM)^{1/2}$. The radiation diffusion time from the midplane is 
$t_{\rm diff}=(h/r)\tau_{\rm T} (r_{\rm in}/c)$. The condition  $t_{\rm diff}<P_{\rm in}$ thus implies
\begin{equation}
\Sigma< 2\pi \left( \frac{h}{r}\,\kappa_{\rm T} \right)^{-1} \left(\frac{r_{\rm in}c^2}{GM}\right)^{1/2}\; .
\end{equation}
This corresponds to a quantitative value of the surface density
\begin{equation}
\Sigma< 2\times 10^3~{\rm g~cm}^{-2}
\left(\frac{h/r}{0.1} \right)^{-1}
\left(\frac{\kappa_{\rm T}}{0.4~{\rm cm}^2~{\rm g}^{-1}} \right)^{-1}
\left(\frac{r_{\rm in}}{200\,M}\right)^{1/2} \,,
\end{equation}
and total mass
\begin{equation}
m < f\pi r_{\rm in}^2\,\Sigma \approx
30\,M_\odot\,f   
\left(\frac{h/r}{0.1} \right)^{-1}
\left(\frac{\kappa_{\rm T}}{0.4~{\rm cm}^2~{\rm g}^{-1}} \right)^{-1}
\left(\frac{r_{\rm in}}{200\,M}\right)^{5/2}
\left(\frac{M}{10^8~M_\odot}\right)^2 \,.
\end{equation}

\noindent Thus, the portion of the circumbinary disk experiencing strong torques is likely to be small compared to the mass of the MBHB. Note also that the torque luminosity is unlikely to exceed the emission luminosity of the circumbinary disk produced by gravitational energy and angular momentum transport. If it did, such energy injection would increase significantly the thickness of the disk, facilitating fast radial inflow of the gas toward the MBHB. With that in mind, for the purposes of this section we consider the torque luminosity limited by the Eddington luminosity of the flow, $L_t \lesssim L_{\rm E} \approx 1.3 \times 10^{46}\,{\rm erg\,s}^{-1} (M/10^8\,M_\odot)$.

For a circular MBHB the modulation of luminosity emitted by the circumbinary disk is likely be small, because even if all the energy is released at a single orbital phase, the modulation would rely on Doppler boosting, which is of order $v_{\rm orb}/c\sim 0.1$ for $a=100\,M$. Greater modulation is possible if the MBHB has significant eccentricity, because then the total torque also has a time modulation. This implies that the viability of this mechanism as a signature relies on the eccentricity of the binary. Several studies have shown that when the binary evolution is driven by
torques from the circumbinary gas disk, then the MBHB orbital eccentricity tends toward higher values \citep[][]{1992PASP..104..769A, an05, cuadra09}. More specifically, binaries that start their evolution with low to moderate eccentricities reach a limiting value of $e\approx 0.4-0.6$ \citep{roedig11,zrake21}. Once the binary's evolution is dominated by gravitational radiation however, the eccentricity drops roughly as $e\sim f_{\rm orb}^{-19/18}$ for small to moderate eccentricities \citep{1964PhRv..136.1224P}.  Thus, to determine the eccentricity when MBHB torques are relevant in this phase, we need to estimate the orbital separation at which the binary's evolution is driven by GWs, and the largest separation at which we might be able to identify periodic fluctuations.

We estimate the first radius by comparing the torque luminosity to the luminosity of gravitational radiation. The GW luminosity for a binary with total mass $M$, symmetric mass ratio $\eta$, semimajor axis $a$, and eccentricity $e$ is \citep{1964PhRv..136.1224P}
\begin{equation}
L_{\rm GW}=-{32\over 5}{G^4\eta^4M^5\over{c^5a^5(1-e^2)^{7/2}}}
\left(1+{73\over{24}}e^2+{37\over{96}}e^4\right)\; .
\end{equation}
Setting $e_{\rm GW}=0.6$ in this stage of evolution, we rewrite this as
\begin{equation}
L_{\rm GW}=65\left(\frac{\eta}{0.25} \right)^4
\left(\frac{GM}{ac^2}\right)^5
\left(\frac{c^5}{G} \right)\; .
\end{equation}
Equating this to $(L_{\rm t}/L_{\rm E}) L_{\rm E}$ gives the transition semimajor axis 
\begin{equation}
\label{eq_agw}
a_{\rm GW}\approx 370\,M 
\left(\frac{M}{10^8\,M_\odot} \right)^{-1/5}
\left(\frac{L_{\rm t}}{L_{\rm E}} \right)^{-1/5}\; ,
\end{equation}  
where we assumed an equal mass binary with $\eta = 0.25$.

When it comes to the consideration of the maximum separation at which the periodic modulation in EM luminosity of the circumbinary disk can be reliably detected, we note that $\sim 10$ cycles must be seen to be confident in the periodic nature of the signal and rule out an AGN with unusual stochastic variability. Thus, the longest MBHB orbital period we consider for the purposes of this estimate is $P\sim 1$~yr. The semimajor axis that satisfies $2\pi\sqrt{a^3/GM}=1$~yr is
\begin{equation}
\label{eq_ayr}
a_{\rm yr}\approx 460\,M
\left(\frac{M}{10^8\,M_\odot} \right)^{-2/3}\;. 
\end{equation}
Equations~\eqref{eq_agw} and \eqref{eq_ayr} indicate that $a_{\rm yr} \sim a_{\rm GW}$ when $L_{\rm t}/L_{\rm E} \lesssim 1$. According to Eq.~\eqref{eq_ae}, $e\propto a^{19/12}$ for MBHBs in the GW regime, and so
the eccentricity at $a_{\rm yr}$ can be estimated 
\begin{equation}
e_{\rm yr}\approx e_{\rm GW} \left(\frac{a_{\rm yr}}{a_{\rm GW}} \right)^{19/12}\approx
0.6 \left(\frac{M}{10^8~M_\odot} \right)^{-133/180} \left(\frac{L_{\rm t}}{L_E} \right)^{19/60}\; ,
\end{equation}
or $e_{\rm yr}\approx 0.6$ if the expression above would suggest a higher eccentricity. Thus, if the torque luminosity is close to Eddington, the MBHB eccentricity and associated EM variability could be substantial at a one year period. To estimate the length of time during which such variability could be sustained we determine the characteristic GW inspiral time from a one year orbital period by substituting an expression for $a_{\rm yr}$ in Eq.~\eqref{eq:tgw} to obtain
\begin{equation}
\label{eq_tgw_yr}
t_{\rm GW} \approx 2\times 10^4\,{\rm yr}\,
\left(\frac{M}{10^8~M_\odot} \right)^{-5/3} \,,
\end{equation}
where we evaluated $t_{\rm GW}$ for an equal-mass MBHB and $e_{\rm yr} = 0.6$. 

It is illustrative to consider how many variable MBHB systems could be visible on the sky at any given time. Suppose that there is an instrument that can survey the entire sky over a $\sim 1$~year period with sensitivity corresponding to a bolometric flux of $F=10^{-12}\,{\rm erg\,cm}^{-2}{\rm s}^{-1}$. Such an instrument would be able to see a source with a luminosity $L$ to a luminosity distance
\begin{equation}
d_{\rm L}\approx 10\,{\rm Gpc}\,
\left(\frac{L}{L_E}\right)^{1/2}
\left(\frac{M}{10^8~M_\odot} \right)^{1/2}
\left(\frac{F}{10^{-12}\,{\rm erg\,cm}^{-2}{\rm s}^{-1}}\right)^{-1/2}\; .
\end{equation}
Thus, a $10^8~M_\odot$ MBHB with a circumbinary disk that emits at nearly the Eddington luminosity would be visible to a redshift of $z\sim 1.5$.  The number density of galaxies that harbor $\sim 10^8~M_\odot$ black holes is approximately $10^{-3}~{\rm Mpc}^{-3}$
\citep{2004MNRAS.351..169M}. Since the comoving volume at $z\sim 1.5$ is $\sim150$~Gpc$^3$,  this implies $\sim 10^8$ such galaxies in that volume. Some estimates suggest that tens of percent of galaxies have had at least one, and possibly a few major mergers since $z\sim 1$ \citep{2006ApJ...652..270B,2008ApJ...672..177L, 2006ApJ...647..763M, 2010ApJ...715..202H}.
Taking into account that the light travel time from $z\sim 1.5$ is about 10\,Gyr, we calculate the approximate merger rate of $(10^8)(0.1)(1)/10^{10}\,{\rm yr}\sim 10^{-3}$\,yr$^{-1}$. We estimated in Eq.~\eqref{eq_tgw_yr} that the maximum extent of time over which the EM variability could be sustained is ${\rm few}\times 10^4\,$yr, implying that there are potentially few 10s of such systems that are visible during the circumbinary disk phase, if they radiate at the Eddington luminosity. If instead the luminosity is a percent of Eddington, one can show that such sources would be visible to a redshift $z\sim 0.2$. As a result, their number drops to a few systems at a time that are active and have a period of a year. 

It is worth noting that the estimate above applies to the more massive end of MBHB population, targeted by the PTAs. It is consistent with findings by \citet{zrake21}, based on high-resolution hydrodynamic simulations, who predict that comparable-mass PTA binaries should be detected with $e \approx 0.4-0.5$. The LISA sources on the other hand undergo GW circularization and are likely to enter the LISA band with a measurable eccentricity of $10^{-2} - 10^{-3}$. The mergers of lower mass MBHB systems, targeted by LISA, Tian-Qin and similar space-based observatories, will be more numerous but less EM luminous in absolute terms, even if they radiate at the Eddington luminosity. For sub-Eddington systems in this mass range, stars in the host galaxy may outshine the emission from the circumbinary disk, particularly in the rest-frame infrared and optical band. This indicates that such systems may be comparatively more difficult to identify, at least based {\it solely} on their circumbinary disk emission (see Sect.~\ref{ssec:photometry} for predicted detections rates of MBHBs exhibiting quasi-periodic EM variability).

\subsubsection{Emission from the mini-disks and accretion streams}\label{sss_minidisks}

As discussed in the previous section, a possible limitation of circumbinary disk emission is that its luminosity is restricted by its low efficiency, because the inner rim of the disk will remain at a large radius compared to the ISCO (located at $6M$ for non-spinning black holes) for most of the inspiral. We therefore turn our attention to emission from mini-disks that may exist around the individual MBHs. In these disks, the matter spirals all the way to the ISCO and has the opportunity to release $\sim$10\% of its mass-energy. If, for example, the inner edge of the circumbinary disk is at $\sim 1000M$, it releases $\sim 0.1$\% of the mass-energy of the matter. The hundred times higher efficiency of the mini-disks implies that only 1\% of the circumbinary disk gas needs to make its way to the mini-disks for their luminosity to be comparable to that of the circumbinary disk. If a significantly greater fraction of the circumbinary gas spills over, the mini-disks can dominate the luminosity and might provide better opportunities for the EM signatures.

Hence, in order to tackle the question of the luminosity of the MBH mini-disks, one needs to understand how much gas is channelled to them from the circumbinary disk. Early models that investigated this question assumed the circumbinary disk to be axisymmetric as well as being vertically averaged. In such models there can be no accretion onto the individual MBHs, and there is significant pileup of matter at the inner edge of the circumbinary disk \citep{2002ApJ...567L...9A,2009MNRAS.398.1392L,2010MNRAS.407.2007C,
kocsis12b,kocsis12a,2012arXiv1205.5017R}. This led to the proposal of signatures of MBHBs evolution that rely on this pileup. Simulations that subsequently relaxed the axisymmetry condition yielded qualitatively different results. They find in particular that the nonaxisymmetric and time-dependent gravitational accelerations tend to fling matter out of the system or cause accretion, rather than acting as a simple barrier \citep[][and others]{2012MNRAS.423L..65B, dorazio13,farris14,shi15}. They established that despite strong binary torques, accretion into the central cavity continues unhindered and is comparable to the single MBH case. They also found that the portion of the stream that is flung by the MBHB toward the inner rim of the circumbinary disk produces a non-axisymmetric density enhancement at its inner edge, often called a ``lump" \citep{noble12, shi12, farris14, gold14a}. An interesting consequence of this density distribution is that the lump quasi-periodically modulates the accretion flux into the central cavity and the mini-disks, even when the orbital eccentricity of the MBHB is modest. The relative amplitude of the lump (and the associated EM periodic signal) were however found to diminish with greater magnetization of the accretion flow and the decreasing MBH mass ratio, vanishing completely between $0.2<q<0.5$ \citep{noble21}.

We can use simple considerations to explore what could happen to the matter that follows a path that allows it to be captured by either MBH. For example, a key question related to the observability of modulation is whether the {\it inspiral time through the mini-disks} is significantly less than, comparable to, or significantly greater than the MBHB orbital time. We will see in the following few paragraphs that this depends on the orbital separation of the MBHB as well as the thermodynamic properties of the disk. If the inspiral time is long, then mini-disks can provide effective buffering of the incoming modulations in the accretion flux driven by the lump. In this case the lump period can still be imprinted in the low-energy (optical and infrared) emission associated with the streams and the cavity wall, but would be absent in accretion rates of the two MBHs \citep{ws21}. If the inspiral time is short, then major modulation is possible, because the feeding rate to the binary mini-disks (which is modulated at the orbital period of the binary and the lump) would be reflected in the accretion rate onto the black holes. We next estimate the inspiral time through a mini-disk to an individual MBH and compare it with the orbital period of the binary. Suppose that the MBHB has mass ratio  $q$, a semimajor axis $a$ and an eccentricity $e$. The maximum extent of either mini-disk is set by its Roche lobe at the pericenter of the binary orbit, since the mini-disk that expands beyond that radius gets truncated by tidal forces from the other MBH. 

Using the formula of \cite{1983ApJ...268..368E}, the Roche lobe radius around the lower-mass black hole is
\begin{equation}
\frac{r_2}{a(1-e)}={0.49q^{2/3}\over{0.6q^{2/3}+\ln(1+q^{1/3})}}\; .
\end{equation}
For example, in the equal-mass case $q=1$ and $e=0.6$, $r_1=r_2=r = 0.38a(1-e) \approx 0.15a$. In the standard disk solution of \cite{ss73}, the inspiral time of the gas is given by the viscous timescale $t_{\rm visc}=2r/[3\alpha(h/r)^2(GM/2r)^{1/2}]$ (see Eq.~\eqref{eq:tvisc}), where we accounted for the fact that the mass of a single MBH is $M/2$. Comparison with the MBHB orbital period, $P_{\rm orb}=2\pi(a^3/GM)^{1/2}$, yields
\begin{equation}
\frac{t_{\rm visc}}{P_{\rm orb}}\approx 
\left(\frac{\alpha}{0.1} \right)^{-1}
\left(\frac{h/r}{0.3} \right)^{-2}\; .
\end{equation}
It follows that, if the mini-disk has sufficient geometrical thickness, it is possible that the inspiral time will be shorter than the orbital time. According to the \cite{ss73} solution, the disks are this geometrically thick only at high accretion rates and at small radii, where they are supported by radiation pressure gradients. For example, at the Eddington accretion rate, $h/r>0.3$ only when $r<23\,M$. Thus, if the accretion rate to the individual MBHs is large, and the holes are fairly close together at pericenter (so that $a \lesssim 150\,M$), a modulation in the accretion rate from the streams may appear as a modulation in the accretion rate onto the holes. 

When the MBHs are at such close separations, additional modulation effects arise as a consequence of the relativistic dynamics and the shape of the gravitational potential between the two MBHs. One is the mass exchange between the MBH mini-disks, which happens because the potential between the two MBHs becomes shallower than in the Newtonian regime, causing the quasi-periodic ``sloshing'' of gas at $\sim2-3$ times the MBHB orbital frequency \citep{bowen17}. The second effect arises when the radius of the Hill sphere (the sphere of gravitational dominance) of an individual MBH becomes comparable to the radius of its innermost stable circular orbit. In this case, the absence of stable orbits for the gas in the mini-disks precludes them from maintaining the inflow equilibrium. As a consequence, the mini-disk masses show significant quasi-periodic fluctuations with time \citep{gold14b, bowen18, bowen19}, potentially providing another time-dependent feature in the MBHB's EM emission. 

Because the size of the Hill sphere and ISCO depend on the black hole mass ratio and spins, one would expect that the 
total mass of the mini-disks and thus, the brightness of their EM signatures depend on these parameters. This dependence was studied in a series of recent works based on GRMHD simulations of mini-disks associated with inspiraling MBHBs with $a \lesssim 20\,M$. \citet{combi21} for example find that whether the MBHs are spinning or not, the mass and accretion rate of mini-disks maintain periodicities set by the beat frequency between the orbital frequencies of the MBHB and an overdense lump in the circumbinary disk, corresponding to approximately 0.72 of the MBHB orbital frequency, $\Omega_{\rm B}$. \citet{gutierrez21} subsequently performed a post-processing analysis of the EM emission from simulations resented in \citet{combi21} and report that it also exhibits periodicity driven by the lump dynamics, albeit at different characteristic frequencies corresponding to the radial oscillations of the lump ($\sim 0.2\,\Omega_{\rm B}$) and at twice the beat frequency ($\sim 1.44\,\Omega_{\rm B}$). They also show that for spinning MBHs with dimensionless spins $s_1 = s_2 = 0.6$ aligned with the orbital angular momentum of the binary, the mini-disks are more massive, and consequently $\sim 3-5$ times brighter relative to the non-spinning binary configurations. Regardless of the spin properties however, mini-disks in MBHBs are characterized by 25-70\% lower radiative efficiencies than ``standard" disks around single MBHs, and are thus expected to be dimmer. In another study of inspiraling MBHBs with $a < 20\,M$, \citet{paschalidis21} show that the exact timing of the mini-disks disappearance and EM dimming, as they become smaller than the ISCO, also depends on the MBH spins. They propose that in inspiraling MBHB systems where an early GW detection allows a prompt EM follow-up, the timing of the mini-disk fading can provide a new probe of the MBH spins, different from the GW measurement.

However, another possibility exists. The accretion streams, which fall nearly ballistically from the inner edge of the circumbinary disk, will hit the outer edges of the mini-disks, producing shocks and emission that might have a different spectrum from the disk spectra. \citet{roedig14b} find that the hotspots created by shocks should radiate Wien spectra with temperatures $\sim 100\,$keV and that their cooling time is smaller than the MBHB orbital time. If so, modulation in the stream rates may be directly reflected in modulation of the shock properties. The total energy release in such shocks might not be large however. Instead of releasing the $\sim 10$\% of the mass-energy available at the ISCO, the impacts typically release only the binding energy at the Roche radius. In our equal-mass MBHB example, this amounts to the binding energy at $r=0.15a$ around a mass $M/2$ black hole, so the specific energy release is only a few times higher than the specific energy released in the circumbinary disk (see Sect.~\ref{sss_cbd}). This may render this signature challenging to detect, given the current limitations of hard X-ray detectors.

\begin{figure*}[t]
\centering{
\includegraphics[trim=0 0 0 0, clip, scale=0.50,angle=0]{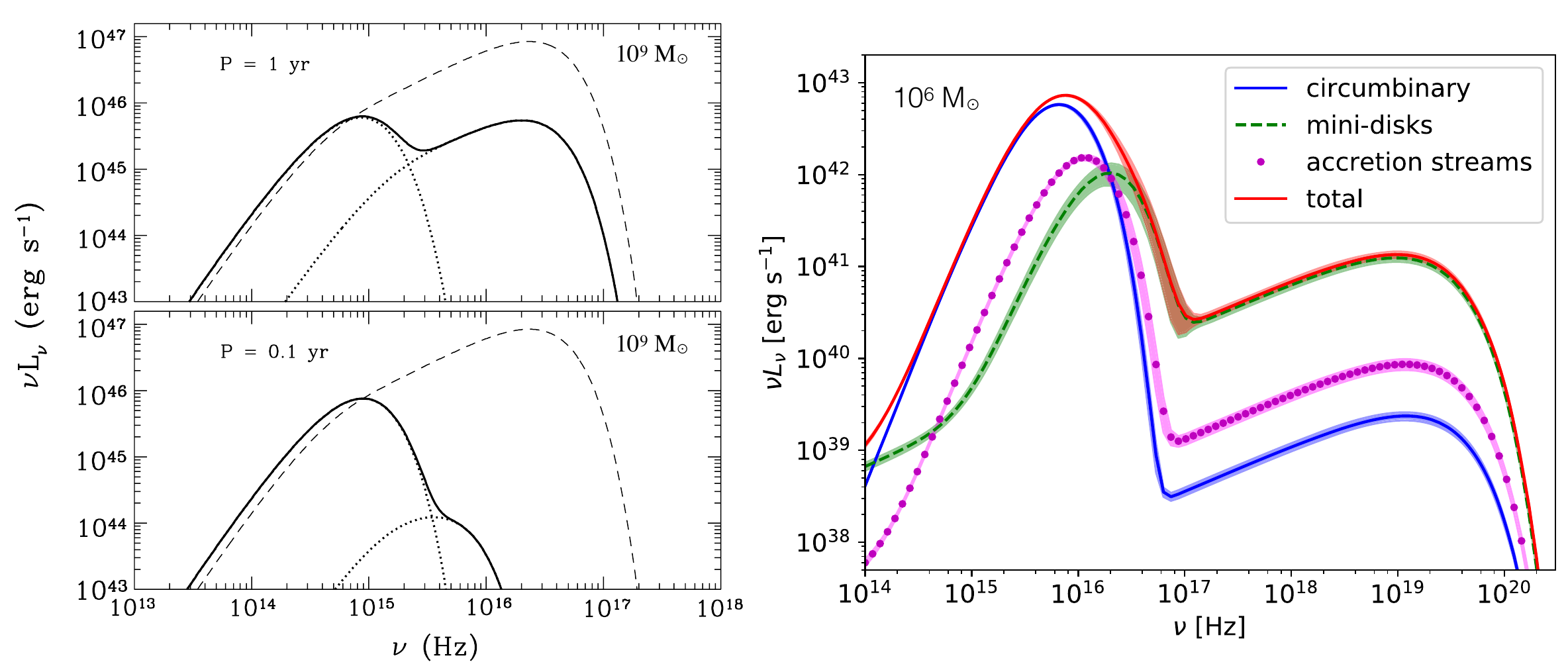} 
}
\caption{{\it Left:} Thermal luminosity spectrum of a non-spinning $10^9\,M_\odot$ MBHB with a mass accretion rate $3\, \dot{M}_E$ through the circumbinary disk. The top (bottom) panel corresponds to a binary with an orbital period $P = 1\,$yr (0.1\,yr) or equivalently $a = 100\,M$ ($22\,M$). Dotted lines show the spectrum from the circumbinary disk (peaking at lower frequencies) and the secondary mini-disk (higher frequencies). Both panels show the combined spectrum (solid thick lines) and the corresponding AGN spectrum for an Eddington-limited thin disk around a single MBH with the same mass (thin dashed). Figure adapted from \citet{tanaka12}. {\it Right:} Luminosity spectrum of a non-spinning $10^6\,M_\odot$ MBHB with mass accretion rate through the circumbinary disk $0.5\, \dot{M}_E$. The spectrum is calculated for a binary inspiraling from $a=20\,M$ and is time-averaged over one orbit. It includes contributions from the mini-disk regions, accretion streams and the circumbinary disk. Figure adapted from \citet{dascoli18}.}
\label{fig17}      
\end{figure*}

Given the anticipated properties of emission from the circumbinary disk, the mini-disks and accretion streams, it is interesting to examine their relative contribution to the spectral energy distribution of the MBHB. A number of works in the literature considered the thermal spectrum from the components of this gas flow and the imprint in it created by the low density gap \citep{gultekin12,kocsis12b,tanaka12,tanaka13,roedig14b,farris15a,ryan17,tang18}. The left panel of Fig.~\ref{fig17} shows two examples of a thermal luminosity spectrum from \citet{tanaka12}, calculated using a model of a thin, viscous disk for a $10^9\,M_\odot$ MBHB and mass accretion rate $\dot{M} = 3\, \dot{M}_E$ through the circumbinary disk. These spectra include contributions from the circumbinary disk and the mini-disk of the lower mass secondary MBH, which is expected to capture a larger fraction of the mass accretion from the circumbinary disk. Figure~\ref{fig17} illustrates that in the system with the orbital period $P = 1\,$yr  (or equivalently, $a = 100\,M$) the circumbinary and mini-disk emit a comparable amount of power, albeit at different frequencies. In the more compact MBHB system with $P = 0.1\,$yr ($a = 22\,M$), the luminosity of the mini-disk corresponds to only $\sim 1\%$ of that of the circumbinary disk. This is contrary to simple expectations based on the efficiency of dissipation and to those laid out at the beginning of this section. More specifically, it indicates that at small enough binary separations ($a < 100\,M$), the reduced mass of the mini-disk leads to a diminished luminosity of its thermal emission, despite a higher efficiency of dissipation relative to the circumbinary disk.

Because the time-variable, quasi-periodic emission associated with accreting MBHBs is more likely to be associated with the mini-disks than the circumbinary disk, the relative dimness of the mini-disks brings into question the ability to identify it in observations. For this reason, it is important to account for all anticipated components of emission, in addition to the thermal spectrum. For example, outside of the thermalized regions the mini-disks are also expected to be significant sources of coronal emission, where inverse Compton scattering between photons and high-energy electrons gives rise to the hard X-ray emission. The right panel of Fig.~\ref{fig17} shows an example of the luminosity spectrum calculated from a general relativistic MHD simulation where the mini-disks, the accretion streams and the circumbinary disk are sources of both thermal and coronal emission \citep{dascoli18}. In this case, the emission is associated with a  $10^6\,M_\odot$ binary, whose mass accretion rate through the circumbinary disk is $\dot{M} = 0.5\, \dot{M}_E$, evolving from a separation of $a=20\,M$. 

The total spectrum is reminiscent of classical AGNs powered by single MBHs and exhibits two peaks: one produced by the thermal emission at the UV/soft X-ray frequencies and the other produced by coronal emission in hard X-rays. In the UV/soft X-ray band, the luminosity of the circumbinary disk is comparable to or larger than that of the two mini-disks. The hard X-ray luminosity is on the other hand dominated by emission from the mini-disks by more than an order of magnitude. The emission from the accretion streams is subdominant across the entire range of frequencies. \citet{dascoli18} find that in their simulated scenario, $\sim 65\%$ of the emission luminosity comes from the circumbinary disk and $\sim 25\%$ from the mini-disks. While the exact percentages may be subject to change in the future, this result nevertheless illustrates an important point: the bolometric luminosity and variability associated with the mini-disks may not dominate throughout the MBHB inspiral but their emission may be distinguished from that of the circumbinary disk in the X-ray band (at $\geq 10^{17}\,$Hz, corresponding to the rest frame energy of $\geq 0.4\,$keV). 

We now return to an important characteristic feature of these spectra, visible in the left panel of Fig.~\ref{fig17} as an inflection or a ``notch" that is caused by a deficiency of emission from the gap in the circumbinary disk. The diagnostic powers of the notch, which could potentially be used to identify MBHBs and place constraints on their geometry, were extensively discussed in the literature (see the references above, following the mention of a low density gap). This feature is visible in the modeled thermal spectra of accreting MBHBs with $a \geq 100\,M$ and becomes weaker with a diminishing contribution of the mini-disks to the overall luminosity of the gas accretion flow in MBHBs with smaller separations. Indeed, in the right panel of Fig.~\ref{fig17}, where the MBHB separation is $a < 20\,M$, there is no apparent notch in the spectrum at $\sim 2\times 10^{16}\,$Hz. If this is a general trend in MBHBs surrounded by circumbinary flows, it suggests that the spectral notch may be a more effective diagnostic tool for more widely separated binaries, where it would be visible in the IR/optical/UV part of the spectrum. On the flip side, if either the notch or the periodically varying X-ray emission are distinct enough to be detected in the spectra of many AGNs, they can in principle be used as a smoking gun to identify hundreds of MBHBs with mass $>10^7\,M_\odot$ in the redshift range $0.5 \lesssim z \lesssim 1$ \citep{krolik19}.

\subsubsection{Other modulation possibilities before the merger}\label{sss_other}


In addition to the broadband X-ray emission discussed above, the two mini-disks may produce variable relativistically broadened Fe\,K$\alpha$ emission lines with rest-frame energy 6.4\,keV. This line commonly originates from the central region of MBH accretion flows (within $\sim 10^3\,M$). It is prominent in the X-ray spectrum due to the high abundance and high fluorescence yield of iron, making it easy to identify even if it is strongly distorted by relativistic Doppler shifts and gravitational redshift \citep{fabian89}. The Fe\,K$\alpha$ line has already been used to probe the innermost regions of accretion disks and infer the spin magnitudes of single MBHs in a subset of AGNs \citep{reynolds13}. In the context of the MBHBs this technique can in principle be applied to binaries with orbital separations of $\lesssim 10^3\,M$ and used to probe the circumbinary flow and the spin magnitudes of both MBHs. 

\begin{figure*}[t]
\centering{
\includegraphics[trim=0 0 0 0, clip, scale=0.50,angle=0]{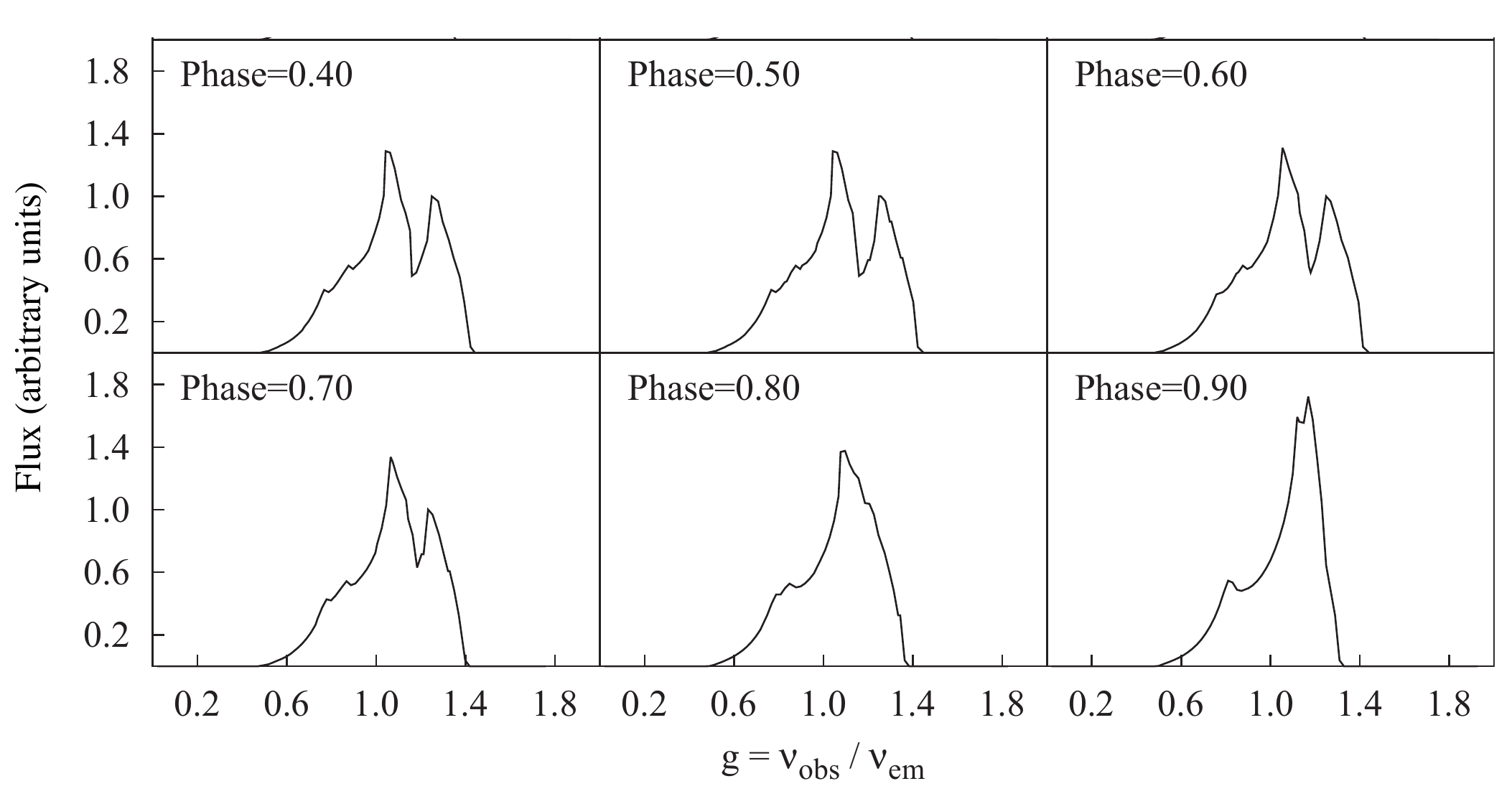} 
}
\caption{Evolution of a composite Fe\,K$\alpha$ profile from an equal-mass MBHB, showing several different orbital phases. The profiles were calculated for two slowly spinning MBHs with orbital separation $a\approx 2000\,M$, eccentricity $e=0.75$ and mini-disks aligned with the binary orbital plane. The horizontal axis shows the ratio of the observed and rest frame frequency. Figure adapted from \citet{jovanovic14}.}
\label{fig18}      
\end{figure*}

The Fe\,K$\alpha$ emission properties of close MBHB systems have been investigated in a handful of theoretical models that predict the shape of the composite emission lines from circumbinary accretion flows \citep{yu01, sesana12, mckernan13, jovanovic14}. Figure~\ref{fig18} shows a sequence of the Fe\,K$\alpha$ line profiles as a function of the binary orbital phase calculated for an equal-mass binary with two mini-disks of equal luminosity. This and other models illustrate that the Fe\,K$\alpha$ emission-line profiles can vary with the orbital phase of the binary and be distinct from those observed in the single MBH systems. If so, they warrant further investigation as a potentially useful MBHB diagnostic. For example, in the case of the MBHBs targeted by the PTAs, the combination of the Fe\,K$\alpha$ and GW signatures could provide a unique way to learn about the properties of MBHBs, since GW alone will not place strong constraints on the binary parameters \citep{arzu14, shannon15, lentati15}. In the case of the inspiraling MBHBs detectable by the LISA observatory \citep{klein16}, the constraints on the orbital and spin parameters obtained from the two messengers could provide two independent measurements that can be combined to increase the precision of the result. The prospects for identifying Fe\,K$\alpha$ profile signatures of MBHBs will be greatly enhanced by future observatories such as \textit{Athena} \citep{nandra13} and \textit{XRISM} \citep{xrism20}, which will be equipped with X-ray micro-calorimeters to enable very high resolution X-ray spectroscopy.

Special and general relativistic effects could also lead to the modulation of the observed EM emission, even in MBHB systems without any intrinsic variability (like the oscillating mini-disks and sloshing streams discussed earlier). For example, \citet{bode10} modeled inspiraling MBHBs immersed in hot accretion flows with a smooth and continuous density distribution of the gas. They find that quasi-periodic EM signatures can still arise as a consequence of shocks produced by the MBHB combined with the effect of relativistic beaming and Doppler boosting. 
An object for which the signature of relativistic Doppler boosting has been modeled in some detail is a quasar PG\,1302-102, whose periodic light curve led to a suggestion that it may harbor a MBHB with the rest-frame orbital period of about 4 years \citep{graham15_pg1302}. In this context \citet{dorazio15} showed that the amplitude and the sinusoid-like shape of its light curve can be explained by relativistic Doppler boosting of emission from a compact, unequal-mass binary with separation 0.007--0.017\,pc. While signatures of relativistic beaming and Doppler boosting remain of interest for inspiraling MBHBs in general \citep[see][]{charisi21}, the binary candidacy of PG\,1302-102 was called into question by \citet{liu18}, who found that the evidence for periodicity decreases when new data points are added to the light curve of this object. They note that for genuine periodicity one expects that additional data would strengthen the evidence, and that the decrease in significance may therefore be an indication that the binary model is disfavored. 

It is worth noting that relativistic beaming and Doppler boosting could also lead to non-axisymmetric irradiation of the accretion flow outside of the MBHB orbit. The impact of this effect is not entirely obvious: for example, it could simply involve reprocessing of the beamed radiation, or it could have structural effects, if photo-heating and radiation pressure inflate the accretion flow and produce outflows. In either case, the natural frequencies of modulation in the EM emission of the gas that is ``anchored'' to the orbiting MBHs (i.e., the mini-disks) would be the binary orbital frequency and its overtones.


\begin{figure*}[t]
\centering{
\includegraphics[trim=0 0 0 0, clip, scale=0.55,angle=0]{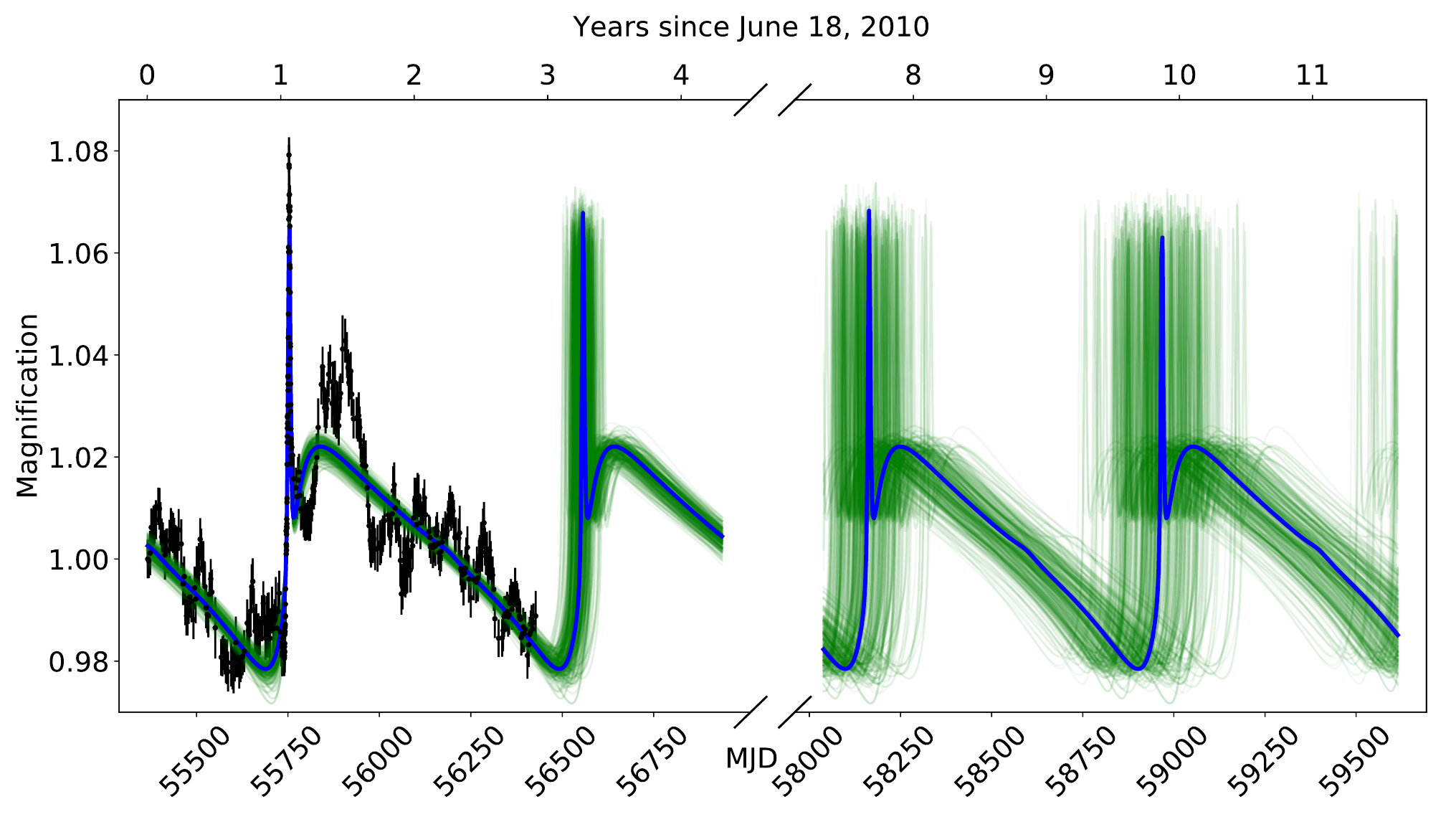} 
}
\caption{Optical light curve for Spikey, a candidate for MBHB with gravitational lensing signature, overlaid with a maximum likelihood model accounting for Doppler boosting and self-lensing. The model uncertainty is extended to show predicted flares in 2013, 2018 and 2020. Figure from \citet{hu20}.}
\label{fig19}      
\end{figure*}

Because relativistic beaming and Doppler boosting are strongest in MBHB systems where the binary orbital plane is close to the line of sight, the same population could also be subject to periodic self-lensing. Namely, if at least one of the MBHs is accreting, the light emitted from its accretion disk can be lensed by the other black hole. For example, \citet{dorazio18_lensing} find that for $10^6$--$10^{10}\,M_\odot$ binaries with orbital periods $<10\,$yr, strong lensing events should occur in $\sim 1-10$\% of MBHB systems that are monitored over their entire orbital period. A similar fraction ($1-3\%$) may also show a distinct feature (a dip) in their self-lensing flares, imprinted by the black hole shadow from the lensed hole  \citep{davelaar21a, davelaar21b}. If so, this may provide an opportunity to extract MBH shadows that are spatially unresolved by VLBI.  Using similar assumptions, \citet{kelley21} find that the Rubin Observatory's LSST could detect tens to hundreds of self-lensing binaries. A light curve of one such binary candidate, nicknamed Spikey, is shown in Fig.~\ref{fig19} \citep{hu20}. In this case the authors identify a model that is a combination of Doppler modulation and a narrow, symmetric lensing spike, consistent with an eccentric MBHB with a mass $3 \times 10^7\,M_\odot$ and rest-frame orbital period of 418 days, seen at nearly edge-on inclination. 

Another modulation possibility was proposed by \citet{2008PhRvL.101d1101K}, who suggested that the GWs generated by the inspiraling MBHB would ripple through the disk and induce in it viscous dissipation. Under the assumption that the ripples would be dissipated efficiently, they calculated that the energy released in this way would dominate the locally generated energy at large radii.  However, the high frequency of the GWs compared to the local dynamical timescales farther
out in the disk means that dissipation will be extremely inefficient. Using this more realistic assumption, \cite{2012MNRAS.425.2407L} found that the energy release from dissipated ripples is insignificant in all plausible circumstances.  


\subsection{Signatures during or immediately after the merger}\label{section:coalescence}

\subsubsection{Reaction of matter to dynamic gravity}\label{SS_matter}

Most of the gravitational wave energy produced during MBH coalescence is emitted during the merger and ringdown phases. If significant matter and/or magnetic fields are present close enough to the binary, where they can interact with a dynamically changing spacetime, there are various avenues for EM emission during these stages \citep[these topics were also discussed in a comprehensive review of relativistic aspects of accreting MBHBs in the GW-driven regime by][]{gold19}. For example, \citet{2010ApJ...709..774K} showed that if the gas density in the immediate vicinity of the MBHB is high enough to make it optically thick, its characteristic luminosity is the Eddington luminosity, independent of the gas mass. It is worth noting, however, that such predictions are subject to an important assumption: that gas in the vicinity of the MBHBs can cool relatively efficiently all the way to the merger, such that it can remain sufficiently dense and optically thick. If so, the bulk of the gas will tend to reside in a rotationally-supported, optically thick but geometrically thin (or slim, $h/r <1$) accretion disk around the MBHB. The necessity for this assumption stems from practical reasons, as the thermodynamics of circumbinary disks (determined by the heating and cooling processes) is computationally expensive to model from first principles and is unconstrained by observations.

In reality, the balance of heating and cooling in a circumbinary flow can be significantly altered close to merger, when the gas is expected to be permeated by energetic radiation and heated by MBHB shocks. This leaves room for an additional possibility: that radiative cooling of a flow around the MBHB is inefficient. If this is the case, such a flow would resemble hot and tenuous, radiatively inefficient accretion flows \citep[RIAFs;][]{ichimaru77, rees82, ny94}. These tend to be less luminous than their radiatively efficient counterparts, as well as optically thin and geometrically thick ($h/r \gtrsim 1$). Therefore, the radiatively inefficient and radiatively efficient gas flows represent idealized scenarios that bracket a range of physical situations in which pre-coalescence MBHBs may be found. We review them both here for completeness and note when works in the literature adopt one or the other assumption. To provide continuity with the discussion of the EM precursors to merger in previous section (Sect.~\ref{section:precursors}), here we consider the properties of MBHBs that evolve from orbital separations of about $10\,M$ through merger and ringdown. 

(a) {\it Mergers in radiatively inefficient gas flows.} In RIAFs, most of the energy generated by accretion and turbulent stresses is stored as thermal energy in the gas, and the accretion flow is hot and geometrically thick \citep{ichimaru77, ny94}. The electron and ion plasmas in RIAFs can form a two-temperature flow in which the thermal energy is stored in the ion plasma while the electron plasma cools more efficiently (i.e., $T_p >  T_e$). In such cases, the temperature of the plasma is represented by the ion temperature, while the characteristics of emitted radiation depend on the properties of electrons. The temperature ceiling reached by the ion plasma is determined by cooling processes such as thermal bremsstrahlung, synchrotron, and inverse Compton emission, as well as the electron-positron pair production and the pion decay resulting from energetic proton-proton collisions. Which process dominates the energy loss of the plasma depends  sensitively on its density, temperature, and magnetic field strength, as well as the efficiency of coupling between ions and electrons. The latter process determines the rate with which energy can be transferred from hot ions to electrons, and consequently the ratio of their temperatures, $\epsilon = T_e / T_p$. In order to illustrate the emission properties of these flows, we will make a simplifying assumption that $\epsilon = 10^{-2}$ everywhere in the accretion flow.
 This is an idealization as $T_e / T_p$ is expected to vary in both space and time and can have a range of values between $\sim 10^{-2}$ and $0.1$ depending on the dominant plasma processes \citep[e.g.,][]{sharma07}.

To understand the dependence of the gas luminosity on the properties of the system, we first consider the bremsstrahlung luminosity emitted from the Bondi radius of gravitational influence of a single MBH with mass $M$, $R_{\rm B} = GM/c_s^2$, where $c_s = (\gamma\, k T_p/m_p)^{1/2}$ is the speed of sound evaluated assuming the equation of state of an ideal gas and $\gamma = 5/3$, for monoatomic gas. In geometrized units $R_{\rm B} \approx 6.5\,M\,T^{-1}_{p,12}$ and 
\begin{equation}
L_{\rm brem} \approx 6.7 \times 10^{44}\, {\rm erg\,s^{-1}} 
\epsilon^{1/2}_{-2}\,T^{-1/2}_{p,12} 
(1 + 4.4\, \epsilon_{-2}\,T_{p,12})_{5.4}\,\tau_T^2\,M_8^4 \,.
\label{eq_bremss}
\end{equation}
Here, $\epsilon_{-2} = \epsilon/10^{-2}$, $T_{p,12} = T_p/10^{12}$K, $M_8 = M/10^8\,M_\odot$, $\tau_{\rm T} = \kappa_{\rm T}\,\rho\,R_{\rm B}$ is the optical depth for Thomson scattering within the Bondi sphere, $\kappa_{\rm T}$ is the opacity to Thomson scattering, and $\rho$ is the gas density. The subscript ``5.4'' indicates that the expression in the brackets is normalized to this value. Note that $T_p \approx 10^{12}$K (corresponding to $kT_p \approx 100\,$MeV) is the maximum temperature that the ion plasma can reach at the innermost stable circular orbit of the MBH if all of its gravitational potential energy is converted to thermal energy, so that $GM m_p/r_{\rm ISCO} \approx kT_p$. Equation~\eqref{eq_bremss} implies a maximum bremsstrahlung luminosity that can be reached by an accretion flow as long as its optical depth $\tau_{\rm T} \lesssim 1$ (because the plasma of this temperature is fully ionized, we consider Thomson scattering to be the dominant source of opacity in this regime). Flows with a larger optical depth would be subject to radiation pressure, which could alter the kinematics of the gas or unbind it from the MBH altogether, potentially erasing any characteristic variability and suppressing the luminosity.

If the hot accretion flow is threaded by a strong magnetic field, a significant fraction of its luminosity could be emitted in the form of synchrotron radiation. Assuming a field strength $B \approx 10^4\,{\rm G}\,\beta_{10}^{-1/2}\,T_{p,12}\,\tau_T^{1/2}\,M_8^{-1/2}$, the synchrotron luminosity could reach
\begin{equation}
L_{\rm syn} \approx 4 \times 10^{46}\, {\rm erg\,s^{-1}} 
\beta_{10}^{-1}\, \tau_T^{2}\,M_8^4\,,
\label{eq_sync_hot}
\end{equation}
where $\beta = 8\pi\,p_{\rm th}/B^2 = 10\beta_{10}$ is the ratio of thermal to magnetic pressure in the gas, expected to reach values of $1-10$ in the central regions of RIAFs \citep{cao11}. The presence of the softer photons supplied in situ by synchrotron and bremsstrahlung emission would also give rise to inverse Compton radiation of similar magnitude
\begin{equation}
L_{\rm IC} \approx 2\, \epsilon_{-2}\,T_{p,12}\, \tau_T \,L_{\rm soft} \;,
\label{eq_ic_hot}
\end{equation}
where $L_{\rm soft}$ is the luminosity of soft radiation. This expression is evaluated for a thermal distribution of nonrelativistic electrons \citep[see equation 7.22 in][]{rybicki79_text} with temperature $T_e \approx 10^{10}\,$K and an emission region corresponding to the Bondi sphere associated with the MBH.

Where the high energy tail of protons reaches the threshold of $kT_p \approx 100\,$MeV an additional high energy process contributes to the radiative cooling: proton-proton collisions result in copious pion production, followed by pion decay to two $\gamma$-ray photons, $p + p \rightarrow p + p + \pi^0 \rightarrow p + p + 2\gamma$. Following \citet{colpi86}, who calculated the $\gamma$-ray emission from the $p-p$ collisions of a thermal distribution of protons in the vicinity of a single Kerr BH, we estimate
\begin{equation}
L_{pp} \approx 2 -13 \times 10^{40}\, {\rm erg\,s^{-1}} 
T_{p,12}\,\tau_T^2\,M_8 \,,
\end{equation}
where the two extreme values correspond to a static and maximally rotating BH, respectively. We expect the luminosity in the MBHB system to be closer to the higher value because the gas in the rotating and dynamic spacetime of the pair of orbiting black holes is very efficiently shock-heated to 100\,MeV. The emission of $\gamma$-rays due to pion decay is strongly suppressed in the limit $\tau_T \gtrsim 1$ due to the increased cross-section of $\gamma$-ray photons to electron-positron pair production, as well as the increased coupling between electron and proton plasma, which lowers $T_p$ below the energy threshold for pion production \citep{colpi86}. In calculating luminosities in this section, we assumed the gas to be optically thin within the Bondi radius, which sets an upper limit on the gas density of the hot accretion flow (such that $\rho < 2.6\times 10^{-12}{\rm g\,cm^{-3}} \tau_T\,M_8^{-1} $), and thus an upper limit on $L_{\rm brem}$, $L_{\rm syn}$, $L_{\rm IC}$, and $L_{pp}$.

The spectral energy distribution of these sources would be similar to a group of low-luminosity AGNs to which RIAF models have been applied \citep[see][for example]{nemmen06}. Spectral bands where these emission mechanisms are expected to peak in the reference frame of the binary are submillimeter (synchrotron), UV/X-ray (inverse Compton), $\sim100\,{\rm keV} - 1\,$MeV $\gamma$-ray (bremsstrahlung and inverse Compton), and $\sim 20$\,MeV (pion decay). Additional components could in principle arise and overtake the emission from the hot gas, such as wide-band non-thermal synchrotron emission (if active and persistent jets are present in the system), as well as the optical/UV emission associated with the accretion disk that may encompass the hot flow at larger radii. Now that we laid out the expectations for emission mechanisms and luminosity of RIAFs, we turn our attention to characteristics of such flows identified in simulations of MBHB mergers.

\begin{figure*}[t]
\centering{
\includegraphics[trim=0 0 0 0, clip, scale=0.43,angle=0]{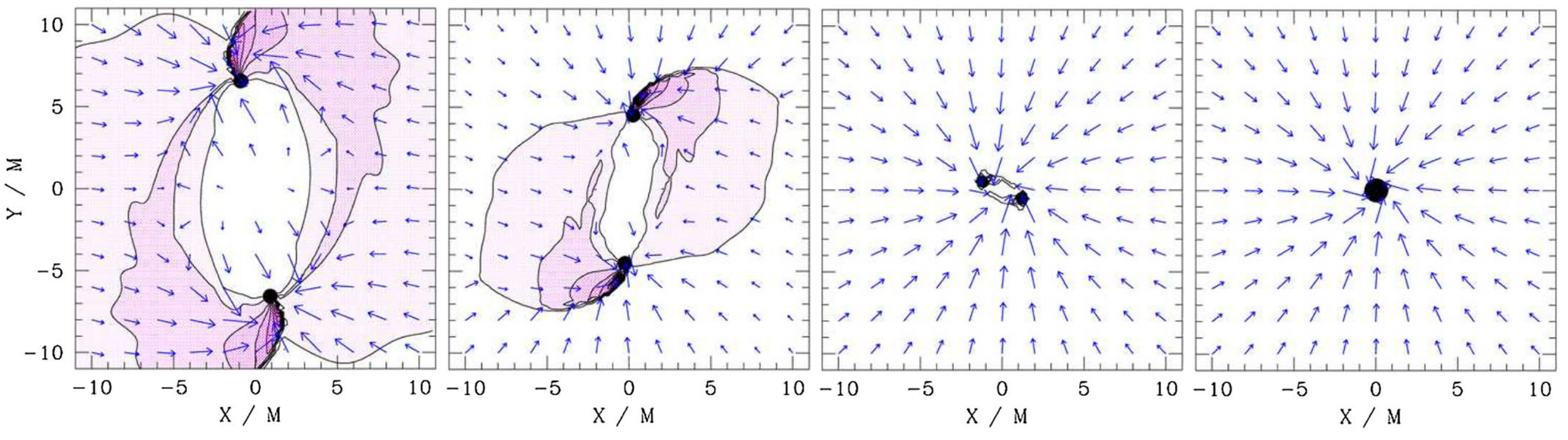} 
}
\caption{Contours tracing the amplitude of shock-heating of the gas in vicinity of the MBHB at $t = 14\,{\rm days}\, M_8$, $3\,{\rm days}\,M_8$, $2\,{\rm h} \,M_8$ before and $2\,{\rm days}\,M_8$ after the merger, respectively. Shocks are initially confined to the tidal wakes following the two MBHs on their orbit. The shocked gas is promptly accreted by the newly formed daughter MBH. Arrows denote velocity vectors. Figure adapted from \citet{farris11}.}
\label{fig20}      
\end{figure*}

MBHB mergers in radiatively inefficient accretion flows have been explored in simulations by multiple groups. The early work by \citet{vanmeter10} examined test particle orbits in the presence of a coalescing binary and concluded that there could be collisions at moderate Lorentz factor, which might lead to high-energy emission. A few subsequent works explored the fluid and emission properties of binary accretion flows resembling RIAFs in the context of general relativistic hydrodynamic simulations \citep{bode10, bode12, farris10, bogdanovic11}. They found that because of high thermal velocities, the radial inflow speeds of gas in the flow are comparable to the orbital speed at a given radius. This implies that in a hot gas flow, unlike the circumbinary disk scenario, binary torques are incapable of creating a central low density region, because the gas ejected by the binary is replenished on a dynamical timescale. As a consequence, the MBHB remains immersed in the flow until merger, which allows it to interact continually with the fluid and shock it to temperatures close to $\sim 10^{12}\,$K. Figure~\ref{fig20} shows that the shocks are initially confined to the tidal wakes following the two MBHs. Around the time of the merger, the shocked, high temperature gas is promptly accreted by the newly formed daughter MBH.

\begin{figure*}[t]
\centering{
\includegraphics[trim=0 0 0 0, clip, scale=0.5,angle=0]{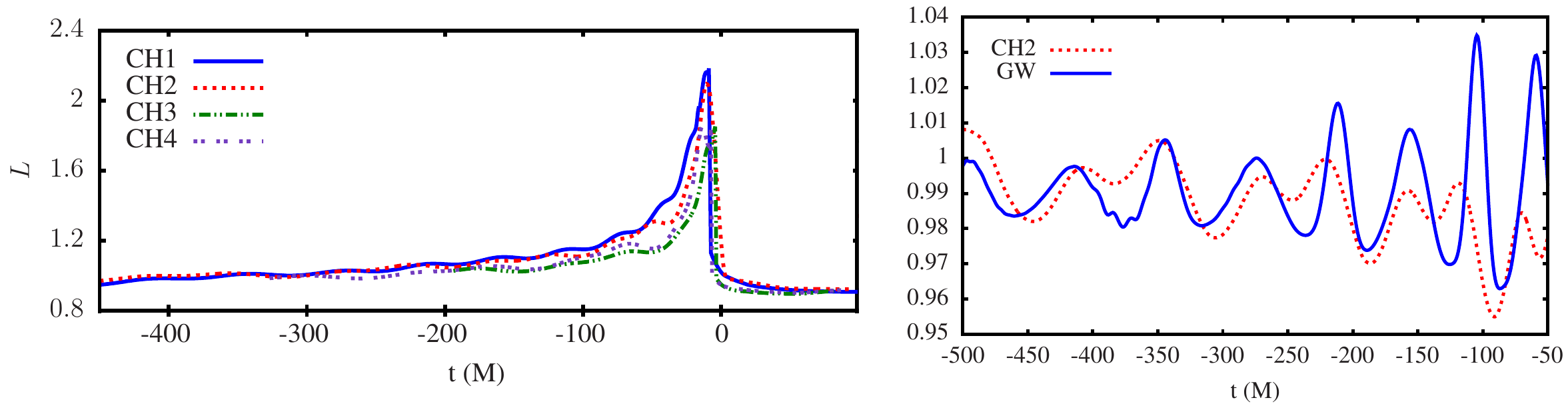} 
}
\caption{{\it Left:} Bremsstrahlung luminosity of hot accretion flows for different MBHB configurations, including mass ratios $q=1$ (CH1) and 1/2 (CH2, CH3, CH4) and different spin orientations (CH1, CH2 -- parallel to the orbital angular momentum; CH3, CH4 -- misaligned), normalized to the luminosity of a single MBH with corresponding mass. Coalescence occurs at $t=0$. {\it Right:} Closer inspection of the MBHB light curves on the left reveals a correlation between the EM signal (red, dotted line) and gravitational waveforms (blue, solid) during the inspiral, shown for a particular model (CH2) but present in all models to some degree. The quasi-periodic signal in the EM emission is obtained as a ratio of the beamed to unbeamed light curve and is calculated for a fiducial observer placed in the plane of the binary at infinity. Figure adapted from \citet{bode12}.}
\label{fig21}      
\end{figure*}

Both the appearance and disappearance of shocks give rise to a characteristic variability of the emitted light. This is illustrated in the left panel of Fig.~\ref{fig21}, which shows bremsstrahlung luminosity as a function of time, normalized by the light curve for a single MBH with equivalent mass (see Eq.~\eqref{eq_bremss}). The most characteristic feature is the broad peak in luminosity, whose growth coincides with the formation of the shocked region within the binary orbit. As the binary shrinks, the brightness of this region increases until the merger, at which point the final MBH swallows the shocked gas and the luminosity drops off precipitously. 

Interestingly, \citet{bode10, bode12} showed that in such systems, the changing beaming pattern of the orbiting binary surrounded by emitting gas can also give rise to modulations in the observed luminosity of the EM signal that is closely correlated with GWs (also referred to as the ``EM chirp"). This can be seen in the right panel of Fig.~\ref{fig21}, which shows the quasi-periodic signal in the EM emission (obtained as a ratio of the beamed to unbeamed light curve) superposed on the GW emission in arbitrary units.  It is worth noting that the amplitude of the EM variability depends sensitively on the thermodynamic properties of the gas disk and optical depth of the surrounding gas. In order for it to be identified as a unique signature of an inspiraling and merging MBHB in observations, the EM chirp would need to have a sufficiently large amplitude, probably comparable to or larger than the intrinsic variability of non-binary AGNs (which for example corresponds to $\sim 10\%$ in X-rays). 

The signatures discussed above are somewhat weaker for lower mass ratios, because orbital torques from an unequal-mass binary are less efficient at driving shocks in the gas, resulting in a less luminous, shorter lasting emission from this region. Similarly, in systems with generic spin orientations (CH3 and CH4), the luminosity peaks at lower values, relative to the $q = 1$ aligned binary. This is a consequence of the orbital precession present in binaries with misaligned spins, which further inhibits the formation of a stable shock region between the holes. The gradual rise and sudden drop-off in luminosity, however, seem to be a generic feature of all modeled light curves, regardless of the spin configuration and the mass ratios. They were observed for a relatively wide range of initial conditions \citep{bode10, farris10}, indicating that this is a robust signature of binary systems merging in hot accretion flows.

(b) {\it Mergers in radiatively efficient gas flows.} As noted in Sect.~\ref{sss_cbd}, simulations of inspiraling MBHBs indicate that the inner rim of the circumbinary disk is able to follow the binary to radii of $\sim 10\,M$ \citep{noble12, farris15b, bowen17, tang18}. If matter can keep up with the binary to even smaller separations, then the EM signatures would be a smooth extension of those discussed in the previous section (associated with the time-variable disks and streams). By extension, the spectra associated with gas that emits until the instant of merger would also be qualitatively similar to those shown in Fig.~\ref{fig17}, namely, a combination of thermal and coronal emission. 

\begin{figure*}[t]
\centering{
\includegraphics[trim=0 0 0 0, clip, scale=0.45,angle=0]{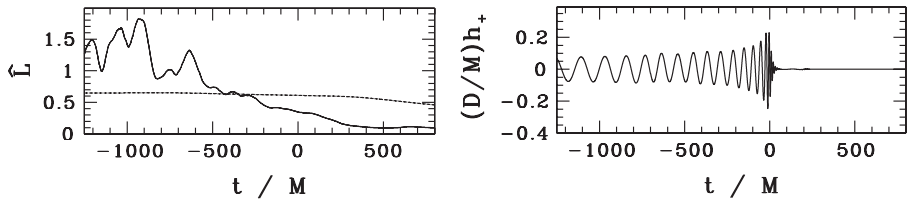} 
}
\caption{{\it Left:} Synchrotron (solid line) and bremsstrahlung (dashed, relatively flat line) luminosity associated with an equal-mass, non-spinning MBHB coalescing in a circumbinary disk. The synchrotron luminosity was calculated for $\beta =10$. The luminosity unit is $\hat{L} = L/[10^{46}\,M_8^3\,n_{12}^2\,{\rm erg\,s^{-1}}]$, where $n_{12} = n/10^{12} {\rm cm^{-3}}$ is the gas number density. {\it Right:} $h_+$ is the $+$ polarization of the GW signal measured by an observer looking down the polar axis at a distance $D$ from the MBHB. The initial separation of the binary is $a=10\,M$ and coalescence occurs at $t=0$. Figure adapted from \citet{farris11}.}
\label{fig22}      
\end{figure*}

Figure~\ref{fig22} shows the EM light curves and the GW signal calculated from one such system, where the initial separation of the binary is $a=10\,M$. The emission signatures shown in the figure were calculated from the first fully relativistic 3D hydrodynamic simulation of an equal-mass, non-spinning MBHB coalescing in a geometrically thin circumbinary disk with the initial scale height $h/r \sim 0.1$ \citep{farris11}. The most notable properties of the displayed EM light curves is that their luminosity decreases as a consequence of late-time decoupling of the MBHB from the circumbinary disk and that their variability does not trivially correspond to the GW signal.

A more optimistic conclusion was reached by \citet{tang18}, who used 2D viscous hydrodynamic simulations of an equal-mass MBHB surrounded by a radiatively efficient circumbinary disk with the assumed scale height $h/r = 0.1$. By evolving the inspiral of the MBHB from $a=60\,M$ to merger, they find that it continues to accrete efficiently, and thus remains luminous all the way to merger. Furthermore, \citet{tang18} find that these systems display strong periodicity at twice the binary orbital frequency throughout the entire inspiral. The quasi-periodic emission, modeled as modified blackbody radiation, is most pronounced in the X-ray band and associated with strong shocks at the inner rim of the circumbinary disk (at $\sim 2\,$keV) and the two mini-disks ($\sim 10\,$keV). Because a clear EM chirp, correlated with the GW emission, is present until the very end of inspiral and the EM emission can potentially reach Eddington-level luminosities\footnote{As noted before, if in the vicinity of the MBHB the gas density is high enough to make it optically thick, its characteristic luminosity is the Eddington luminosity, independent of the gas mass \citep{2010ApJ...709..774K}.}, these types of systems would represent the most promising  multimessenger MBHBs. 

In addition to the results themselves, one can also appreciate the range of predictions about the EM signatures found in simulations described above and more generally, in the literature, that stems from differences in the simulation setup: initial conditions, dimensionality and other technical details. They hint at a complex nature of these models, which are still work in progress, as well as the need to carefully examine the dependence of the EM signatures on additional physical phenomena, such as the magnetic fields and radiation.

\subsubsection{Effect of dynamical spacetime on magnetic fields and formation of jets}\label{SS_Bfields}

We now discuss the response of magnetic fields to the dynamical spacetime of a merging MBHB. In order to separate the effects of the spacetime and the gas on magnetic fields, it is useful to examine mergers in three different scenarios: (a) in (near) electrovacuum, (b) in radiatively inefficient, magnetized gas flows resembling RIAFs, and (c) in radiatively efficient, magnetized circumbinary disks.

(a) {\it Mergers in (near) electrovacuum}. The earliest group of papers on this topic explored the effect of the dynamical spacetime on pure magnetic fields in a vacuum \citep{palenzuela09,moesta10} and on magnetic fields within a tenuous plasma with zero inertia
\citep[in a so-called force-free approximation][]{palenzuela10_magnetospheres, palenzuela10_science, palenzuela10a, moesta12, alic12}. These setups are illustrative of conditions that might arise if the binary decouples from the circumbinary disk early in the inspiral but continues to interact with the fields anchored to it. They were the first to indicate that even if the gas is dilute and optically thin, the winding of the magnetic fields by the MBHB could lead to Poynting outflows and the formation of jets. 

\begin{figure*}[t]
\centering{
\includegraphics[trim=0 0 0 0, clip, scale=0.45,angle=0]{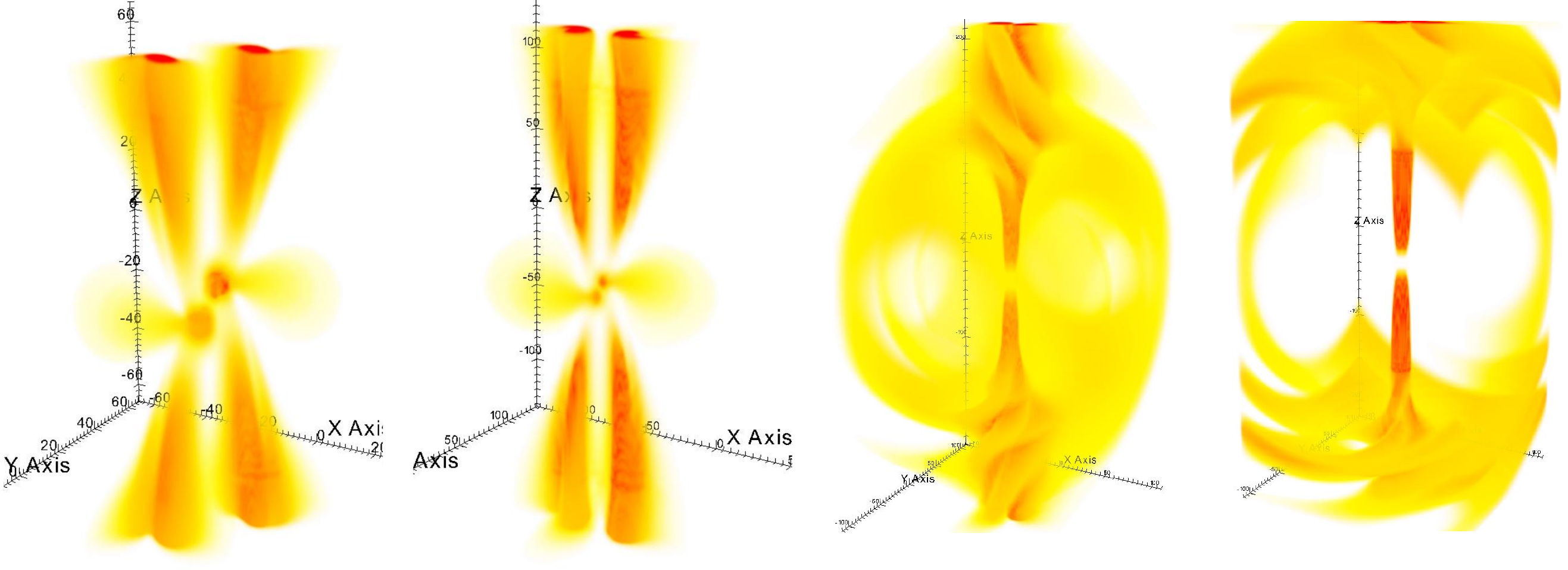} 
}
\caption{Poynting flux associated with the electromagnetic field just hours before and after the MBH coalescence. From left to right: $t=-11{\rm h}\, (M/10^8\,M_\odot)$, $t=-3.0{\rm h}\, (M/10^8\,M_\odot)$, $t=4.6{\rm h}\, (M/10^8\,M_\odot)$, and  $t=6.8{\rm h}\, (M/10^8\,M_\odot)$, all expressed in the frame of the MBHB.  Darker colors mark larger flux amplitudes. Figure adapted from \citet{palenzuela10_science}.}
\label{fig23}      
\end{figure*}

\begin{figure*}[t]
\centering{
\includegraphics[trim=0 0 0 0, clip, scale=0.45,angle=0]{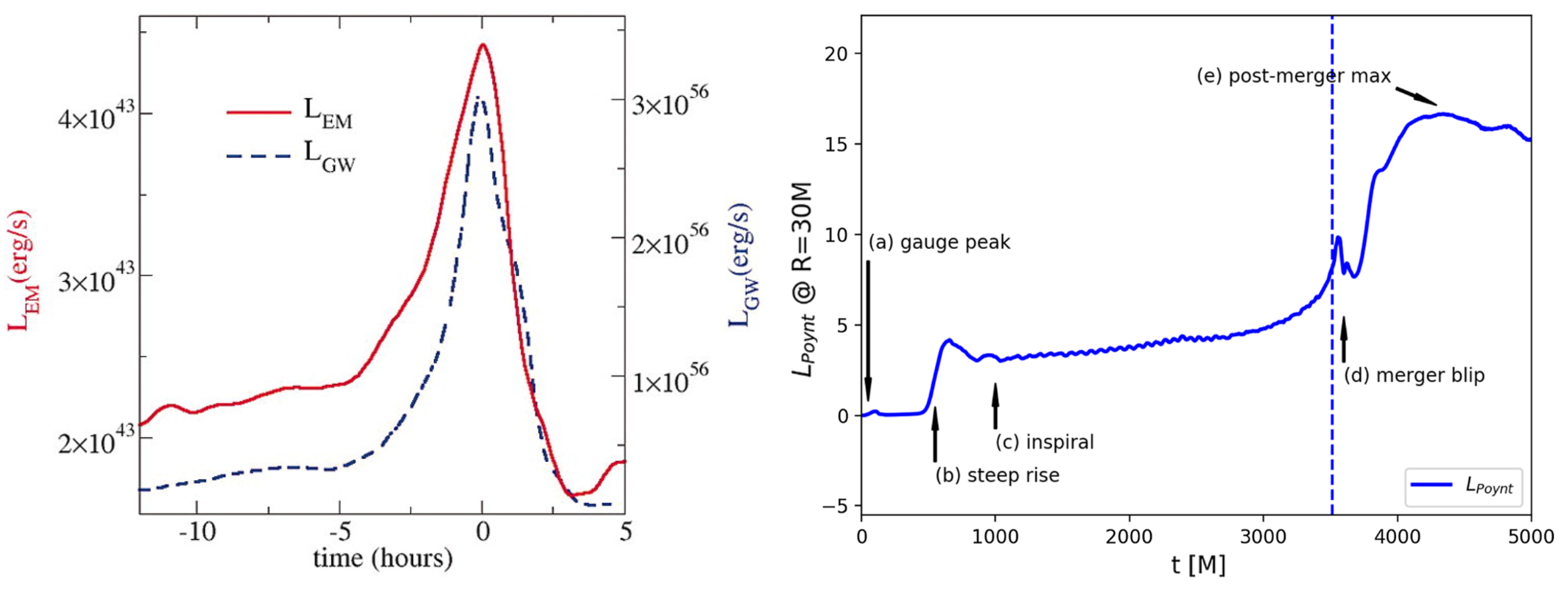} 
}
\caption{{\it Left}: Poynting luminosity as a function of time (red, solid line) associated with the MBH coalescence in near vacuum, calculated for $M=10^8\,M_\odot$ MBHB and $B_0=10^4\,$G. The blue dashed line and axis on the right mark the corresponding GW luminosity. Time $t=0$ corresponds to merger. Figure adapted from \citet{palenzuela10_science}. {\it Right}: Poynting luminosity associated with the MBH coalescence in a radiatively inefficient magnetized gas flow. The vertical dashed line corresponds to the instant of merger. For $M=10^8\,M_\odot$ and $\rho_0 = 10^{-13}\,{\rm g\,cm^{-3}}$, the units of time and luminosity correspond to $1 \, t = 0.14\,{\rm h}\,\, M_8$ and $1 \, L_{\rm Poynt} = 5.9\times 10^{44}\,{\rm erg\,s^{-1}}\,\rho_{-13}\,M_8^2$. Figure adapted from \citet{kelly17}.}
\label{fig24}      
\end{figure*}

This is illustrated in Fig.~\ref{fig23}, which shows the Poynting flux associated with the electric and magnetic fields just hours before and after the coalescence of two non-spinning MBHs \citep{palenzuela10_science}. This and other studies established that the winding of magnetic fields by the binary results in the enhancement of the magnetic energy density and the Poynting luminosity on the account of MBHB kinetic energy. The Poynting luminosity scales as 
\begin{eqnarray}
L_{\rm Poynt} &\propto& v^2\, B_0^2\, M^2 \,\,\, {\rm and} \\
L_{\rm Poynt, peak} &\approx & 3\times 10^{43}\,{\rm erg\,s^{-1}} \left(\frac{B_0}{10^4\,{\rm G}}\right)^2 \left(\frac{M}{10^8\,M_\odot}\right)^2   \,\,,
\label{eq_Lpoynt1}
\end{eqnarray}
where $v$ is the orbital speed of the MBHB, $B_0$ is the strength of the initial, uniform magnetic field and $L_{\rm Poynt, peak}$ is the peak value of the Poynting luminosity. Given that $v$ reaches maximum around merger, these works find that the Poynting luminosity also peaks at merger, as shown in the left panel of Fig.~\ref{fig24}. They also make other important points: (i) even if all of the electromagnetic energy is spent on charge acceleration and eventually reradiated as photon energy, the emitted luminosity is expected to be orders of magnitude lower than Eddington \citep[assuming $B_0\sim 10^4\,{\rm G}$;][]{moesta10} and (ii) the Poynting luminosity associated with the collimated dual-jet structure is $\sim 100$ times smaller than that emitted isotropically, making the detection of jets less likely \citep{moesta12, alic12}. This led to some early reservations about a detectability of the merger flare associated with the dual jet.

(b) {\it Mergers in radiatively inefficient, magnetized gas flows.}  More encouraging results were found by general relativistic magnetohydrodynamic (GRMHD) simulations of MBHBs immersed in uniform density plasma threaded by an initially uniform magnetic field \citep{giacomazzo12, kelly17, cattorini21}. Equivalently to the relativistic hydrodynamic simulations with uniform gas flows discussed in the previous section, these setups correspond to physical scenarios in which the magnetized gas flows inwards as a RIAF. The key finding of these simulations is that in the presence of gas, increase in the magnetic field strength and energy density is even more dramatic relative to the simulations of magnetic fields in (near) vacuum. This results in a commensurate increase in the Poynting luminosity, which according to \citet{kelly17} scales as
\begin{eqnarray}
L_{\rm Poynt} &\propto& \rho_0\, v^{2.7} M^2 \,\,\, {\rm and} \\
L_{\rm Poynt, peak} &\approx & 1.2\times 10^{46}\,{\rm erg\,s^{-1}} \left(\frac{\rho_0}{10^{-13}\,{\rm g\,cm^{-3}}}\right) \left(\frac{M}{10^8\,M_\odot}\right)^2   \,\,,
\label{eq_Lpoynt2}
\end{eqnarray}
where $\rho_0$ corresponds to the initial, uniform value of gas density, $v$ is the orbital speed of the MBHB, and $L_{\rm Poynt, peak}$ is again the peak value of the Poynting luminosity. A comparison with Eq.~\eqref{eq_Lpoynt1} reveals several important differences. One is that in the presence of gas, the resulting Poynting luminosity does not depend on the initial strength of the magnetic field \citep[this outcome was tested for field strengths in the range $B_0 = 3\times 10^3 - 3\times 10^4\,{\rm G}$;][]{kelly17}. The other is that the magnetic field strength exhibits super-quadratic growth on approach to merger, due to the accretion of gas, which further compresses the field lines near the horizon. As a result, the peak value of the Poynting luminosity is orders of magnitude higher than that inferred from simulations in near vacuum.

The right hand side of Fig.~\ref{fig24} shows the evolution of the Poynting luminosity found in this GRMHD simulation. After the initial magnetic field configuration settles into a quasi-steady state during the MBHB inspiral, $L_{\rm Poynt}$ gradually rises and exhibits a local peak around the time of the merger (marked as ``(d) merger blip" in the figure). This local peak is equivalent to the peak in luminosity evident in the left hand panel, found in simulations of MBHBs and fields in near vacuum. The most striking difference between the two scenarios is that in near vacuum the Poynting luminosity rapidly decays to the level consistent with that associated with a single spinning MBH, whereas in magnetized plasma it reaches a new post-merger maximum. The post-merger increase in $L_{\rm Poynt}$ to the level of the Eddington luminosity ($L_E = 1.3\times 10^{46}\,{\rm erg\,s^{-1}} M_8$) is a result of continued accretion onto the remnant, spinning MBH which leads to a late increase in magnetic field strength. 

All studies of MBHB mergers in radiatively inefficient, magnetized gas flows discussed here investigated equal mass binaries. This is a configuration which maximizes the winding and compression of magnetic field lines, and we thus expect it to result in the highest Poynting luminosities (see also the discussion of mergers in magnetized circumbinary disks next). The exact dependence of the Poynting luminosity on the MBHB mass ratio is yet to be determined in future studies. The dependance of the Poynting luminosity on the MBH spins was recently studied by \citet{cattorini21}, who examined configurations with equal spins and dimensionless magnitudes $s_1 = s_2 = 0.3$ and 0.6, parallel to the orbital angular momentum. They find that the peak luminosity, reached shortly after the merger (defined as ``(d) merger blip" in Fig.~\ref{fig24}), depends on the MBH spins and is enhanced by a factor of about 2 (about 2.5) for binaries with spin magnitudes 0.3 (0.6) relative to the nonspinning binaries.


\begin{figure*}[t]
\centering{
\includegraphics[trim=0 0 0 0, clip, scale=0.70,angle=0]{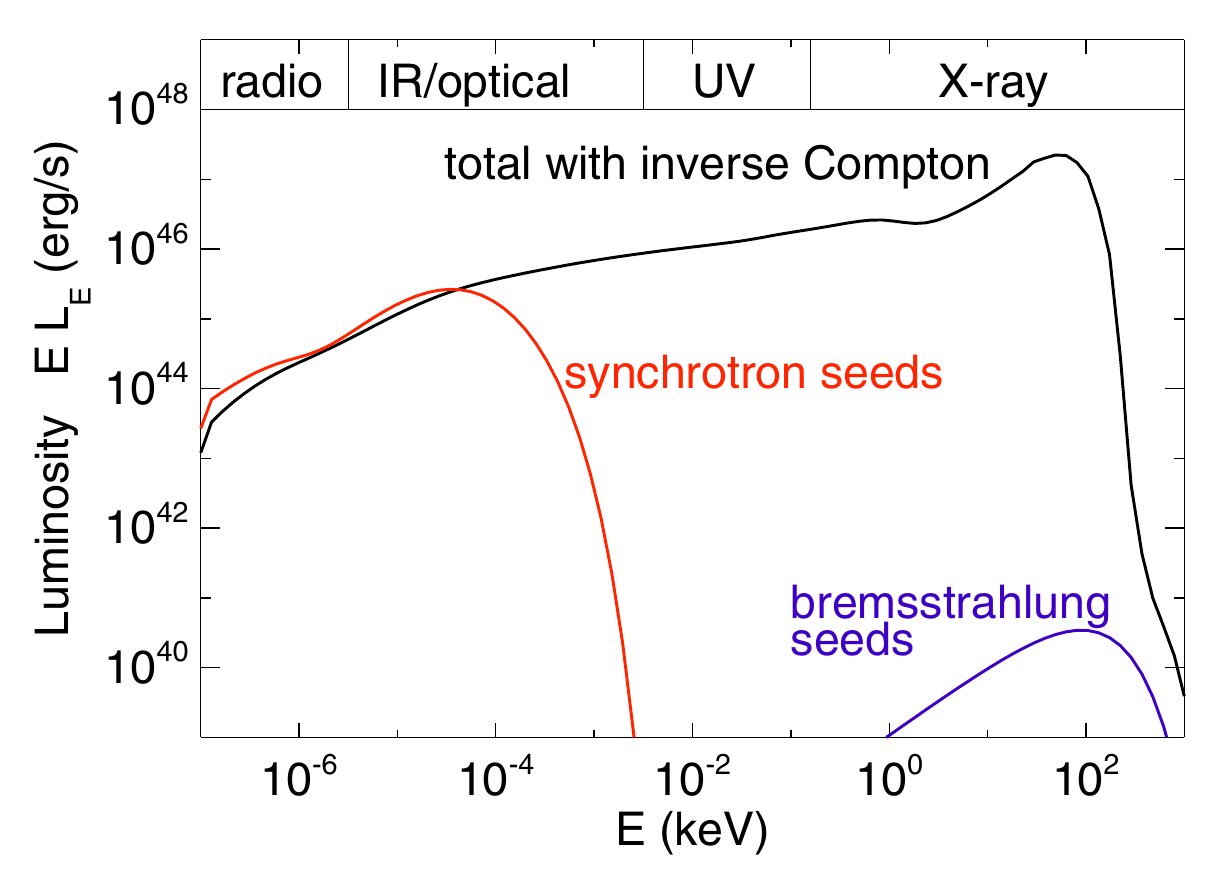} 
}
\caption{Broad-band spectrum inferred from the GRMHD simulation of a MBHB immersed in magnetized, uniform density plasma by \citet{giacomazzo12}. The spectrum is evaluated at the instant in time near the peak of the EM and GW emission and it corresponds to $10^7\,M_\odot$ MBHB, the gas with $kT_e = 100\,$keV and optical depth $\tau_T \approx 1$. Figure from \citet{schnittman13}. }
\label{fig25}      
\end{figure*}

Using a scaling with physical properties similar to Eq.~\ref{eq_sync_hot} and the assumption that the gas is optically thin to its own emission, \citet{kelly17} also estimated the synchrotron luminosity associated with a MBHB immersed in magnetized, initially uniform-density plasma. They found that the shape of the calculated light curve is qualitatively similar to that shown in the left panel Fig.~\ref{fig21}, i.e., the synchrotron luminosity gradually increases until the merger, at which point it drops off precipitously. Figure~\ref{fig25} illustrates the broad-band spectrum associated with this type of accretion flow, corresponding to the instant of super-Eddington luminosity at the peak of the EM and GW emission.
The spectrum was calculated by \citet{schnittman13}, who applied relativistic Monte-Carlo ray-tracing to the GRMHD simulation by \citet{giacomazzo12} as a post-processing step. In this approach, the ``seed" synchrotron and bremsstrahlung photons produced by the plasma are ray-traced through gravitational potential of the MBHB. They are allowed to scatter off the hot electrons in a diffuse corona (similar to the AGN coronae), giving rise to a power-law spectrum with cut-off around $kT_e = 100\,$keV, characteristic of inverse Compton scattering.


\begin{figure*}[t]
\centering{
\includegraphics[trim=0 0 0 0, clip, scale=0.70,angle=0]{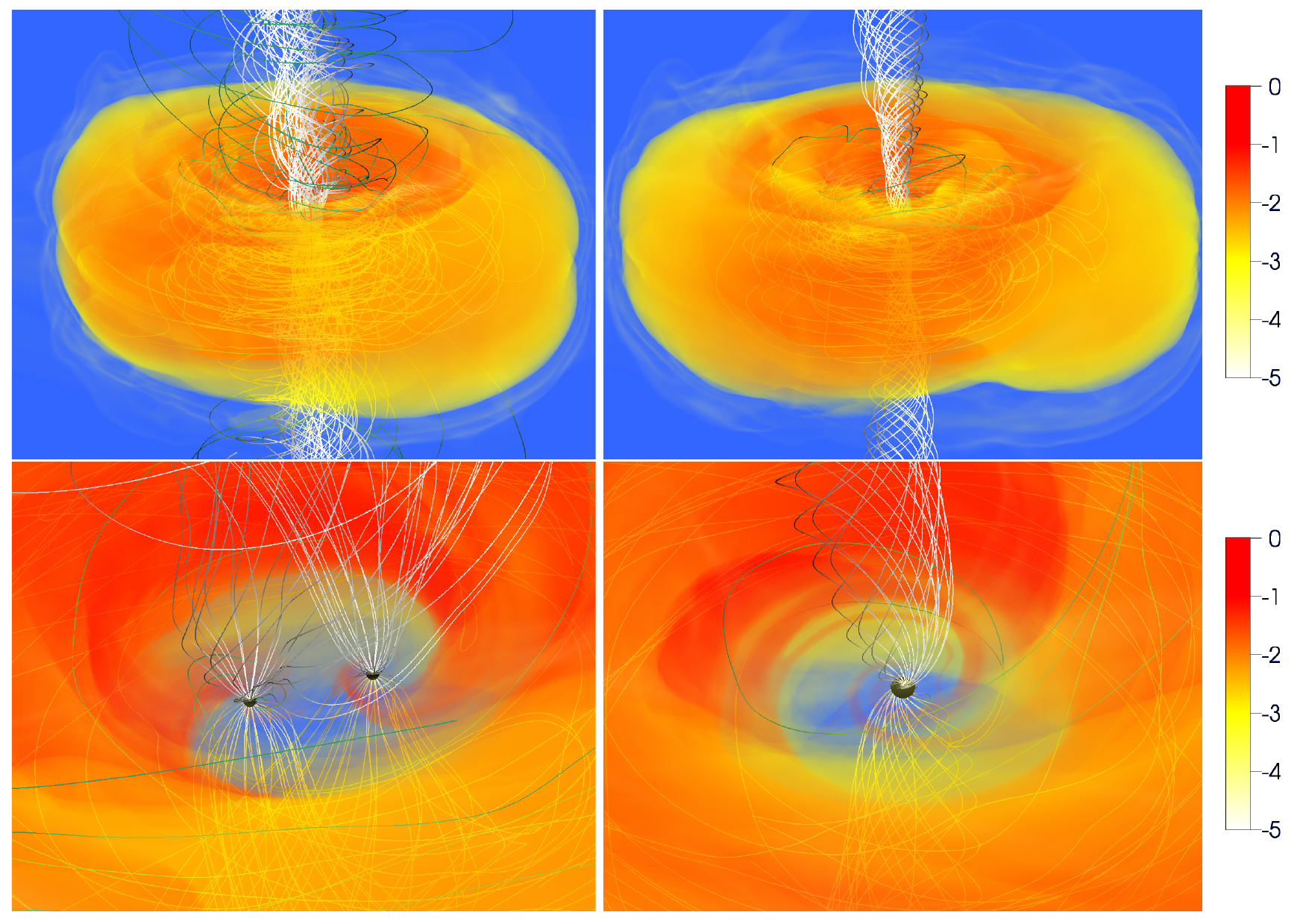} 
}
\caption{Instants before (left panels) and after (right panels) merger of a MBHB surrounded by a geometrically thick, magnetized circumbinary disk. The winding of magnetic field lines leads to the formation of a collimated jet. However, the diminished gas flow close to merger (visible in the bottom panels) provides weaker compression of magnetic field lines and leads to a more modest increase in the Poynting luminosity than in the case of RIAFs. The scale of the top panels is about $150\,M$. The bottom panels provide a closeup view within $20\,M$. Figure from \citet{gold14b}. }
\label{fig26}      
\end{figure*}

(c) {\it Mergers in radiatively efficient, magnetized circumbinary disks.} It is interesting to compare these results to the mergers of MBHBs immersed in magnetized disks. This setup was investigated by \citet{farris12}, \citet{gold14a, gold14b}, \citet{paschalidis21} and \citep{combi21}. Again in this case, the studies show that the winding of magnetic field lines leads to the formation of a collimated dual jet and an increase in the magnitude of the Poynting luminosity after merger (see Fig.~\ref{fig26}). The main difference from mergers in RIAFs is that the MBHB decouples from the circumbinary disk in the late stages of inspiral and just before merger (when $a < 10\,M$ in geometrically thick disks with $h/r \approx 0.3$). This leads to a decrease in the accretion rate and luminosity of the gas flow just before merger, which is most pronounced for equal-mass binaries \citep{gold14b}. This is qualitatively consistent with the behavior observed in early simulations of unmagnetized disks \citep{farris11,bode12} and discussed in Sect.~\ref{SS_matter}. The diminished gas flow provides weaker compression and collimation of magnetic field lines and leads to a less dramatic increase in the Poynting luminosity than in the case of RIAFs.
 
The dependence of the $L_{\rm Poynt, peak}$ on the MBHB mass ratio was explored in \citet{gold14b}, who found that it decreases with decreasing $q$, because the equal-mass setup maximizes the pre-merger winding of the magnetic field lines as well as the post merger gas inflow. Furthermore, the results of studies that consider mergers of spinning MBHs, with dimensionless spin magnitudes $s_1 = s_2 = 0.6$, parallel to the orbital angular momentum, show only modest enhancement (less than 50\%) in the $L_{\rm Poynt, peak}$ as a consequence of additional winding of magnetic field lines by the individual MBH spins before merger \citep{moesta12, alic12}. This is consistent with more recent findings by  \citet{paschalidis21} and \citet{combi21}, based on the GRMHD simulations of late inspiral in binaries with $a\lesssim 20\,M$. They report the enhancement of the Poynting flux in MBHB systems with $s_1 = s_2 = 0.75$ and $0.6$, respectively, aligned with the orbital angular momentum. In addition to these general trends, \citet{combi21} find that regardless of the spin magnitudes, late in the inspiral the Poynting flux is modulated at the beat frequency between the MBHB orbital frequency and the frequency of an overdense lump, driven by the quasi-periodicity of accretion onto the holes themselves. This is interesting because it indicates that quasi-periodicity may also be present in the emission from the jets associated with inspiraling MBHBs, if a sufficient amount of the Poynting flux is tapped to power acceleration of charged particles in the jets and their subsequent EM emission.

All studies discussed in this section (Sect.~\ref{SS_Bfields}) examine configurations in which the pre-merger MBH spins are zero, aligned or anti-aligned, and in which the fields lines are initially either poloidal or parallel to the orbital angular momentum of the binary, and thus to the spin axis of the remnant MBH. This is the setup that maximizes the pre-merger winding and enhancement of magnetic fields, and it is plausible that more tangled and intermittent initial fields may lead to lower $L_{\rm Poynt, peak}$ than the reported values. For example, \citet{palenzuela10_magnetospheres} and  \citet{kelly20} find that the steady-state (peak) Poynting luminosity depends strongly on the initial field angle with respect to the remnant MBH spin axis, with maximum luminosity achieved for perfect alignment and minimum luminosity achieved for the field perpendicular to the remnant's spin. Furthermore, they report that the proto-jet is formed along the remnant MBH spin-axis near the hole, while aligning with the asymptotic magnetic field at large distances. This raises interesting questions about what astrophysical processes set the geometry of magnetic fields on larger scales, beyond the influence of the MBH spin, and what is the resulting jet geometry in systems in which the remnant MBH undergoes a spin ``flip" relative to the spin (and possibly jet) axes of the pre-merger MBHs.

In summary, the GRMHD studies of MBHB mergers in magnetized environments carried out so far indicate that during and immediately after the merger $L_{\rm gas} < L_{\rm Poynt}$ in the radiatively inefficient flows, whereas $L_{\rm Poynt} < L_{\rm gas}$ in the radiatively efficient flows. The peak total luminosity ($L_{\rm gas} + L_{\rm Poynt}$) in either physical scenario can be comparable to, or larger than the Eddington luminosity. Thus, an opportunity for the EM luminous emission from jets exists, given that a sufficient fraction of the Poynting power is converted into charged particle acceleration and subsequently, photon luminosity. The picture emerging from these works is that radiatively inefficient accretion flows provide the best opportunity for collimation and enhancement of magnetic fields, and consequently for formation of the EM luminous jets coincident with the merger. In radiatively efficient accretion flows the luminous EM jets may still form but with some delay, corresponding to the viscous timescale for circumbinary gap refilling. In this case, the nonthermal EM emission from electrons accelerated by the forward shock in jets may emerge days to months after the MBH coalescence and appear as a slowly fading transient \citep{yuan21}. According to the same study, the multiwavelength emission from systems with post-merger accretion rates at the Eddington level (and thus, the EM luminous jets) may persist for months and be detectable out to $z\sim5-6$ in the radio band and to $z\sim1-2$ in the optical and X-ray bands.

Besides the EM emission, the relativistic jets launched close to the MBH coalescence can also produce neutrino counterpart emission originating from cosmic rays accelerated in the jet-induced shocks. For example, \citet{yuan20} find, under somewhat optimistic assumptions, that the post-merger high-energy neutrino emission ($E_\nu \gtrsim 1\,$PeV) from individual GW events may be detectable by the next generation IceCube-Gen2 detector, within 5-10 years of observation. If so, a truly {\it multiple} messenger detection of MBH coalescences may be possible in some cases.

\subsubsection{Retreat of the disk due to mass-energy loss}
\label{sssec:mass_loss}

When two black holes of comparable mass merge, several percent of the
mass-energy of the binary is radiated in GWs 
\citep[e.g.,][]{2008PhRvD..78b4017B,2008PhRvD..78h1501T,2008PhRvD..78h4030K,
2009PhRvD..80l4026R,2010CQGra..27k4006L,2012ApJ...758...63B}.
Most of the radiation is emitted in the last orbit or so, which
is much shorter than the orbital timescale of the inner edge
of the circumbinary disk.  Thus, from the standpoint of the disk, 
the mass-energy loss occurs impulsively, causing the disk orbits 
to adjust to the new central potential \citep{2007APS..APR.S1010B}.

\begin{figure*}[t]
\centering{
\includegraphics[trim=0 0 0 0, clip, scale=0.82,angle=0]{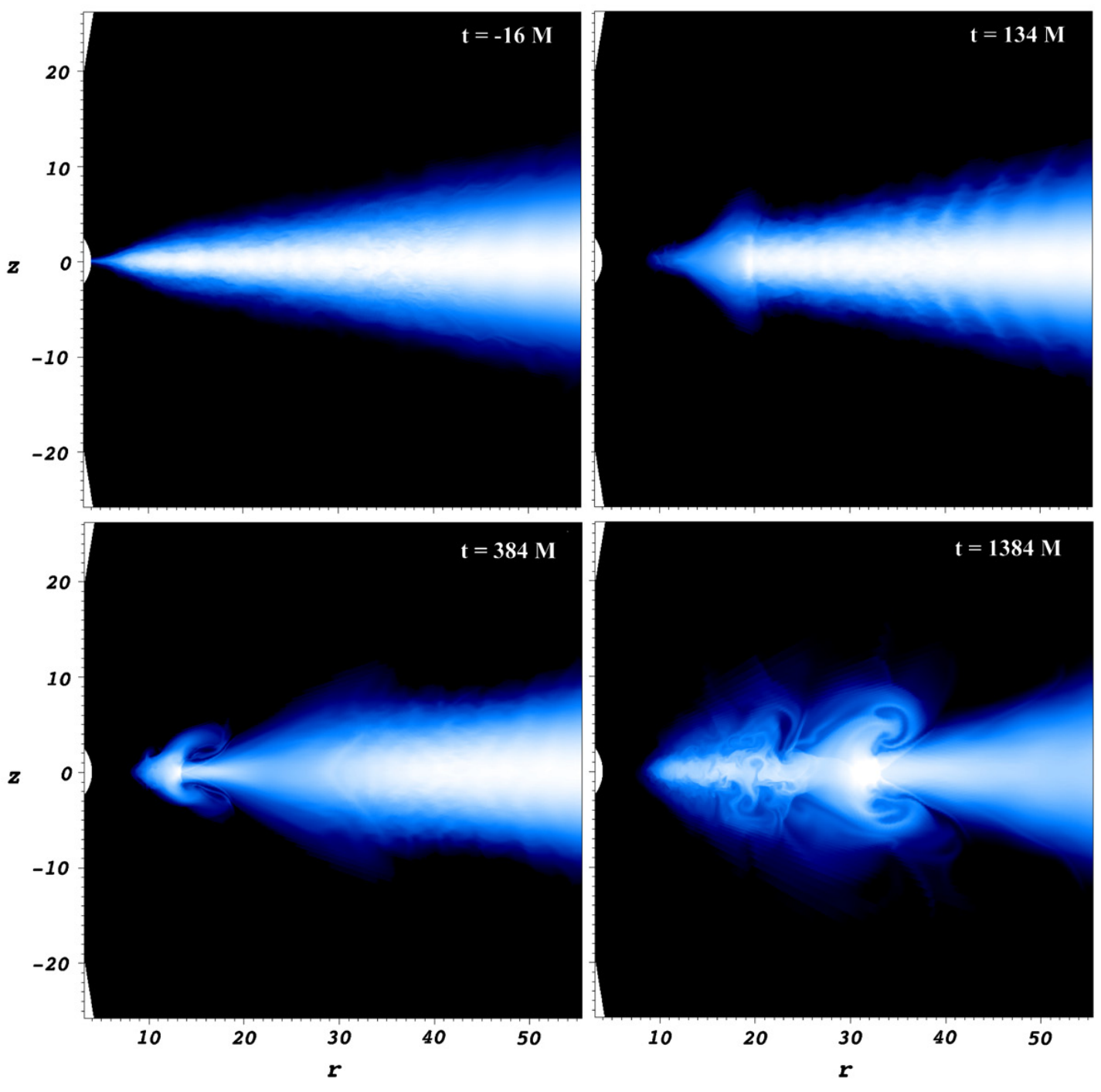} 
}
\caption{Snapshots of gas density showing oscillation of the accretion disk around the remnant MBH after coalescence, caused by the mass-energy loss of 10\%, assumed to be associated with the GW emission. The mass-loss event takes place at $t=0$. The initial wave is visible at $t=134\,M$, while the stronger shock is visible at $t=384\,M$ and $1384\,M$. Shocks and rarefactions propagate radially outward in the disk (from left to right in these images). The size of each panel corresponds to about 50\,M in height and radius. The colorbar is logarithmic, ranging from $10^{-4}$ (dark) to 1 (light) in units of the initial gas density. Figure adapted from \citet{2009ApJ...700..859O}. }
\label{fig27}      
\end{figure*}

\begin{figure*}[t]
\centering{
\includegraphics[trim=0 0 0 0, clip, scale=0.53,angle=0]{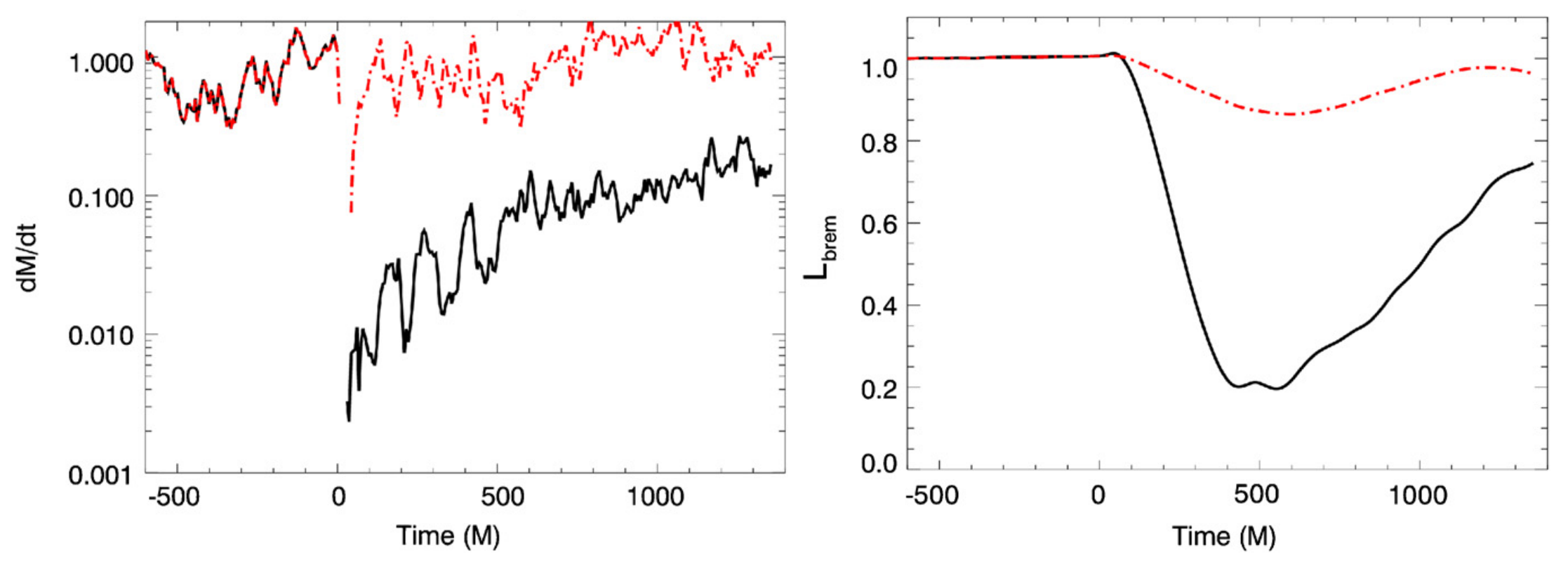} 
}
\caption{{\it Left:} Mass accretion rate across the innermost stable circular orbit of the remnant MBH  corresponding to the mass-loss values of 10\% (solid line) and 1\% (dot-dashed). The mass-loss event takes place at $t=0$. The 10\% mass-loss event shows an initial drop in the accretion rate, followed by a gradual resumption of accretion on the viscous timescale. {\it Right:} Bremsstrahlung luminosity measured between disk radii 5--35$\,M$ corresponding to the same mass-loss values as in the left panel. In both cases there is barely any increase in luminosity. There is, however, a drop in luminosity followed by a gradual recovery. The initial accretion rate and luminosity prior to binary merger are shown at $t < 0$ for reference. Both are shown in dimensionless units. Figure adapted from \citet{2009ApJ...700..859O}. }
\label{fig28}      
\end{figure*}

As discussed by \citet{2009ApJ...700..859O}, the most immediate
consequence of the mass-energy loss is that the inner edge of the
circumbinary disk moves outward, because the matter in the disk
maintains its angular momentum but suddenly finds itself around a lower-mass
object. Thus, its original radial location now becomes the pericenter
of an eccentric orbit. The gas therefore oscillates in radius,
with different radii having different oscillation periods.  
\citet{2007APS..APR.S1010B} suggested that the radial gradient in radial
epicyclic period would lead to shocks between annuli that would
enhance the emission in some energy bands. The oscillation of gas in radius is 
illustrated in Fig.~\ref{fig27}, from \citet{2009ApJ...700..859O}, who
explored this scenario with 3D hydrodynamic and MHD simulations. They found that the most observable effect was likely to be a {\it decrease} in the accretion rate and luminosity of the system,
followed by a gradual recovery to the original level of accretion and emission on a characteristic (viscous) timescale, as seen in Fig.~\ref{fig28}.

To explore this suggestion, we follow \citet{2009ApJ...700..859O} in
noting that the magnitude of the reaction of the circumbinary disk 
depends strongly on how relativistic the disk is.  Starting with a
Newtonian disk, we suppose that the fractional mass-energy loss 
during merger and ringdown is $\epsilon\ll 1$, so that if the original
mass-energy was $M_0$, the mass-energy after ringdown is $M=M_0(1-\epsilon)$.
A fluid element in the disk at radius $r$ has a specific angular momentum
of $\ell_0=\sqrt{GM_0r}$ both before and after the merger, such that if it
circularizes at constant angular momentum, its new radius and semimajor
axis will be
$a=r(1+\epsilon)$ to first order in $\epsilon$.  Because its pericenter
distance is $r=a(1-e)$, the initial eccentricity of the orbit is
$e=\epsilon$ to first order.  The energy release can be computed from
the general Newtonian formula for the angular momentum of an orbit
with eccentricity $e$: $\ell(e)=\sqrt{GMa(1-e^2)}$.  This implies that
during circularization the fractional energy release will be
$\Delta E/E_{\rm bind}=e^2=\epsilon^2$, where $E_{\rm bind}$ is the
binding energy at $r$.  Given the value of $\epsilon\sim 0.05$ or less for most
comparable-mass mergers, this suggests a local release of energy
that is $\sim$few$\times 10^{-3}$ times the local binding energy.

The time over which the annuli can release their energy is no shorter
than the time needed for neighboring annuli to go sufficently out of
phase with each other, so that their relative radial speed exceeds the sound
speed (if the relative speed is subsonic, then conversion of the motion
into heat is much less efficient). Given that the sound speed is
$h/r$ times the orbital speed in a Shakura-Sunyaev disk \citep{ss73}, this 
suggests that only fairly geometrically thin disks will have low enough
sound speeds for shocks to form.  More specifically, the total radial
distance moved during a radial epicycle is $2\epsilon r$ in half a period,
versus $\pi r$ moved orbitally in half a period.  The sound speed is
$(h/r)v_{\rm Kep}$ (where $v_{\rm Kep} $ is the circular orbital speed), so we compare $(h/r)v_{\rm Kep}$ to $(2/\pi)\epsilon v_{\rm Kep}$ to conclude
that $h/r<2\epsilon/\pi$ is a necessary criterion for formation of shocks.  This implies $h/r<1/(10\pi)$ if
$\epsilon=0.05$, suggesting a disk with an accretion rate well below the
Eddington rate \citep{ss73}.

The time for neighboring annuli to go out of phase by a radian is 
$T \sim 0.1P_{\rm orb}(r)/\epsilon$, so the maximum specific luminosity is 
\begin{equation}
L_{\rm max}=\frac{\Delta E}{T}\approx 10\,\epsilon^3 \frac{E_{\rm bind}}{P_{\rm orb}}\; .
\end{equation}
For example, $\epsilon=0.05$ implies $L_{\rm max}\approx 10^{-3}
E_{\rm bind}/P_{\rm orb}$.  The natural disk luminosity is 
\begin{equation}
L_{\rm disk}\sim \frac{E_{\rm bind}}{P_{\rm orb}(r)/[3\pi\alpha(h/r)^2]}
\sim 10^{-3}\, \frac{E_{\rm bind}}{P_{\rm orb}}\; ,
\end{equation}
where the expression in the denominator corresponds to the viscous timescale at radius $r$ and
we have used $\alpha=0.1$ for the Shakura-Sunyaev viscosity parameter and $h/r=1/(10\pi)$ from the
shock requirement above.  Thus for sufficiently thin disks the
modulation factor can be of order unity, but at the cost of a
generally low luminosity.  Note as well that if the disk is Newtonian,
the binding energy at the inner edge of the disk is low \citep[this would be
the case if the binary decouples from the disk when the binary separation
is of order $100\,M$, as suggested by][]{milosavljevic05}. 

If the circumbinary disk has followed the binary down to relativistic
separations, then the situation is different.  The primary reason is that
circular orbits near the innermost stable circular orbit have specific angular momenta that do not
vary much with radius, so a small change in the mass can lead to a
large change in radius.  For example, \cite{2009ApJ...700..859O} consider
 the effect of a fractional loss 
$\epsilon\ll 1$ of 
the mass-energy of the central object on an annulus initially at
radius $r=xM_0$ in Schwarzschild coordinates.  After the mass loss
(assumed impulsive), the new radius is determined by
\begin{equation}
{x^2\over{x-3}}={x^{\prime 2}(1-\epsilon)^2\over{x^\prime-3}}
\end{equation}
where $x^\prime\equiv r^\prime/M$, with $r^\prime$ being the radius
after circularization and $M=M_0(1-\epsilon)$.  The initial specific
energy of the orbit is
\begin{equation}
E_{\rm init}=\sqrt{(x-2)[x-2(1-\epsilon)]\over{x(x-3)}}
\end{equation}
and the final specific energy after circularization is
\begin{equation}
E_{\rm fin}={x^\prime-2\over{\sqrt{x^\prime(x^\prime-3)}}}\; .
\end{equation}
The fractional change in the binding energy is then
$\Delta E/E_{\rm tot}=(E_{\rm init}-E_{\rm fin})/(1-E_{\rm init})$ and 
the energy release of an annulus
initially near the ISCO is much greater than it would be in Newtonian
gravity.  For example, for $\epsilon=0.05$, the Newtonian fractional
energy release is $\Delta E/E_{\rm tot}=0.0025$, but for an annulus
initially at the ISCO of a Schwarzschild hole, $x=6$, $\Delta E/E_{\rm tot}=
0.0679$, nearly thirty times larger.  The spacetime will, of course,
be dynamic and thus not Schwarzschild, but we expect the general
principle of energy enhancement to apply.

Nonetheless, \cite{2009ApJ...700..859O} found that the prime result of mass
loss is not an enhancement, but a deficit of emission.  The reason
is that after mass loss the inner edge of the disk circularizes
at a radius larger (in gravitational units) that its original inner
edge.  Thus, until viscous effects can return the disk to its pre-merger radius, 
which can
take several hundred $M$ of time (amounting to days for a 
$10^8\,M_\odot$ central black hole), the efficiency of emission and
hence the luminosity will be less than it was pre-merger.  Subsequent
studies have confirmed this basic picture 
\citep{megevand09, corrales10, anderson10, 2012MNRAS.425.1958R,zanotti12}.
Whether this effect is
detectable depends on variables such as the primary band of emission
and the extinction of the center of the host galaxy.  It is also
conceivable that the eventual filling in of the circumbinary gap
would lead to the production of jets and thus radio emission where 
there was none previously, as discussed in Sect.~\ref{SS_Bfields}.  This would require that the source be
face-on to us, which would occur in only a small fraction of all
mergers, but it seems worth pursuing.

\begin{table}
\caption{Summary of the EM counterparts to GW emission before, during and immediately after the MBH merger discussed in Sect.~\ref{section:precursors} and \ref{section:coalescence}. The columns describe the wavelength band in which the EM emission is most likely to emerge (first), the type of the signature (second), and the origin of emission with a reference to the section where it was discussed (third).}
\label{tab:1}       
\begin{tabular}{lll}
\hline\noalign{\smallskip}
{\bf Wavelength band} & {\bf Signature} & {\bf Origin of emission} \\
\noalign{\smallskip}\hline\hline\noalign{\smallskip}
{\bf Radio, sub-mm} & Flare and possibly periodicity  & Radiatively inefficient binary accretion flows (Sect.~\ref{SS_matter})\\
		& & Jets in magnetized  accretion flows (Sect.~\ref{SS_Bfields})\\
\hline
{\bf IR, optical, UV} & Periodicity & Overdense ``lump" in the circumbinary disk (Sect.~\ref{sss_minidisks}) \\
			& Spectral inflection (``notch") & Low density cavity in the circumbinary disk (Sect.~\ref{sss_minidisks}) \\
\hline
{\bf X-ray} & Periodicity &  Modulating accretion through the mini-disks (Sect.~\ref{sss_minidisks})\\
		 &  &  Sloshing of gas between the mini-disks (Sect.~\ref{sss_minidisks})\\
		 &  &  Doppler-boosting of the mini-disk emission (Sect.~\ref{sss_other})\\
		 &  &  Self-lensing of the mini-disk emission (Sect.~\ref{sss_other})\\		 
		 & Relativistic Fe\,K$\alpha$ emission lines &  Mini-disk reflection spectrum (Sect.~\ref{sss_other})\\
		 & Dimming &  Retreat of the circumbinary disk at merger (Sect.~\ref{sssec:mass_loss})\\		 
 \hline
{\bf Hard X-ray, $\gamma$-ray} & Periodicity & Hotspots in the mini-disks (Sect.~\ref{sss_minidisks})  \\ 
			& Flare and possibly periodicity & Radiatively inefficient binary accretion flows (Sect.~\ref{SS_matter}) \\
\noalign{\smallskip}\hline
\end{tabular}
\end{table}

In Sect.~\ref{section:precursors} and \ref{section:coalescence} we discussed a number of signatures associated with the binary accretion flows that can provide the EM counterparts to their GW emission. In Table~\ref{tab:1} we provide a summary of these signatures and their origins according to the wavelength band in which they are most likely to emerge. Most proposed EM counterparts involve variability on relatively short time scales and are expected to emerge in high-energy bands, indicating perhaps the need for a future X-ray timing mission that could carry out the EM followup.

In the next section we turn to signatures that might be visible months to millions of years after the merger, including kicks and afterglows.

\subsection{Afterglows and other post-merger signatures}
\label{section:postmerger}

In addition to the coincident EM counterparts to MBH mergers that may be observable as near-simultaneous multimessenger events, EM afterglows could appear on longer ($\sim$ months to decades) timescales after merger. The enormous release of energy and momentum during a MBH merger provides numerous possible mechanisms for perturbing the surrounding gas disk, if present. These sudden perturbations may induce oscillations, density waves, or shocks in the disk, all of which could produce some type of EM flare that is distinct from standard accretion emission. Here we discuss various physical mechanisms that may produce EM afterglows in the aftermath of MBH mergers, which could be crucial for localization of GW sources detected with LISA. We also discuss signatures of past MBH mergers that may form and persist on much longer timescales (thousands to millions of years), including signatures of GW-recoiling MBHs. Even though these long-timescale signatures cannot be associated with a specific GW observation, they could still place valuable constraints on the overall population and characteristics of merging MBHs.

\subsubsection{Afterglows from circumbinary gap refilling}
\label{subsection:cbd_afterglows}

In Sect.~\ref{sssec:mass_loss} we discussed the immediate response of a gas disk to the MBH mass loss radiated away in GWs, which should produce a short-term decrease in the accretion luminosity as the disk retreats from the lower-mass MBH. After this initial phase, however, the disk will continue to refill the pre-merger circumbinary gap on a viscous timescale; completely refilling the gap will generally take months to years in the source rest frame, depending on the MBHB mass and decoupling radius. Throughout this gap refilling phase, a steady increase in the luminosity and photon energy of the emission is expected, as this inner portion of the accretion disk will radiate at higher energies with peak emission in the hard UV and X-ray bands \citep[e.g.,][]{milosavljevic05, dotti06, tanmen10, tanaka10, shapiro10}. This scenario is therefore a candidate for producing an observable EM afterglow that could be seen in the months to years following a GW detection of a MBH merger. 

Relatedly, \citet{schnittman08} argue that the large optical depth of the disk will thermalize energy from merger-induced perturbations, such that most of the afterglow emission should occur in the IR. For LISA sources with masses $\sim 10^6\,M_{\odot}$, the timescale for this IR afterglow could be years to decades. This possibility is of particular interest because, unlike UV and soft X-ray emission, IR flares would be much less susceptible to attenuation by gas and dust. 

If the accretion rate from the pre-merger circumbinary disk is near the Eddington luminosity prior to the decoupling of the MBHB from the disk, some estimates indicate that the gap-refilling afterglow could be quite luminous, perhaps even greater than the Eddington luminosity in extreme cases \citep[][]{tanmen10, shapiro10}. However, the characteristics and detectability of this accretion flow are uncertain. For one, \citet{tanaka10} find that the afterglow luminosity is sensitive to the circumbinary disk properties, especially the surface density and the ratio of the viscous stress to gas pressure. 
More importantly, recent simulations of circumbinary disk evolution demonstrate that gas streams can readily flow through the circumbinary gap to fuel the mini-disks around individual MBHs at rates comparable to the accretion rate in a disk with no gap at all \citep[e.g.,][]{dorazio13, farris14, shi15, tang18, bowen17}. The pileup of gas at the edge of the circumbinary gap predicted by earlier work may therefore be fairly minimal. While this greatly increases the possibilities for MBHBs to produce distinctive EM counterparts \textit{prior} to merger, the continuous feeding of the mini-disks around each MBH would reduce the spectral and luminosity contrast between a post-merger EM flare and the emission from pre-merger accretion \citep[e.g.,][]{shi12, noble12, fontecilla17, bowen18, bowen19}. 

\begin{figure}
\centering
\includegraphics[width=0.49\textwidth, trim={0 0 2.57in 0},clip]{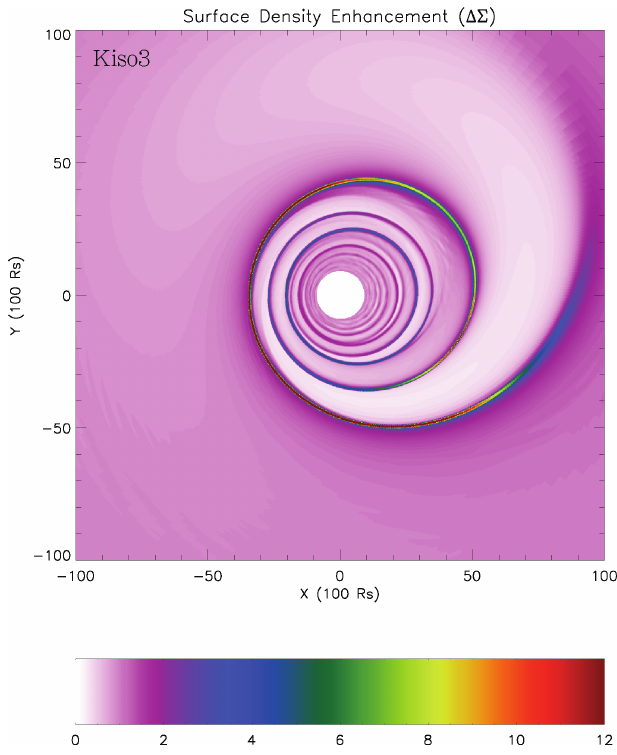}
\includegraphics[width=0.485\textwidth]{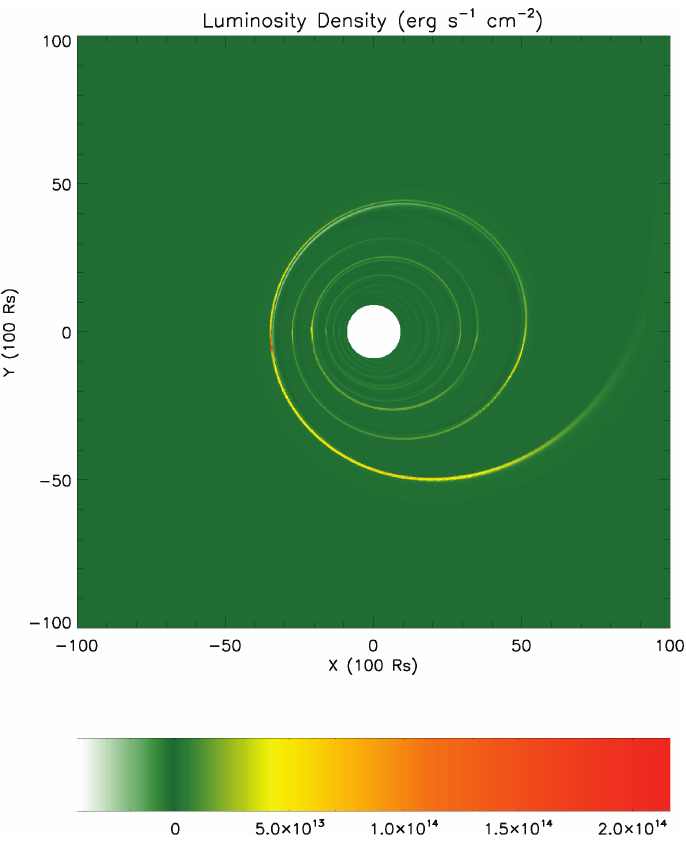}
\caption{Hydrodynamic simulation of the accretion disk response to a GW recoil kick oriented in the plane of the disk. The recoil kick velocity is 530 km s$^{-1}$, and the snapshot is taken 210 d after the merger and recoil. \textit{Left:} The contour image shows the surface density distribution, in which overdensities form a clear spiral pattern in response to the recoil perturbation that produces corresponding shocks. \textit{Right:} For the same simulation snapshot, the contours show the luminosity density of the gas, defined as the energy radiated per unit area per unit time. A spiral pattern corresponding to the shocked gas is again clearly apparent. Figure adapted from \citet{corrales10}. \label{fig:recoil_shocks}}
\end{figure}

\subsubsection{Afterglows from GW recoil kicks}
\label{subsection:recoil_afterglows}

As described in Sect.~\ref{ssection:GWphase}, asymmetric GW emission from an unequal-mass, misaligned, spinning MBHB merger can impart a recoil kick to the merged MBH of hundreds to thousands of km s$^{-1}$. In the frame of the kicked MBH, this causes a velocity $\mathbf{v_{\rm kick}}$ to be added to the instantaneous velocity of the orbiting gas. The perturbation to the disk will induce strong density enhancements, which can shock the gas and produce luminous afterglows on timescales of weeks, months, or years after the MBH merger, depending on the disk and recoiling MBH properties \citep[Fig.~\ref{fig:recoil_shocks};][]{lippai08, schnittman08, megevand09, rossi10, corrales10, zanotti10, meliani17}. 

Using a collisionless particle simulation, \citet{lippai08} demonstrated the formation of tightly-wound spiral density caustics resulting from a GW recoil kick oriented in the disk plane. In contrast, for a kick oriented perpendicularly to the disk, the resulting concentric density enhancements are weaker and delayed to potentially $\gtrsim$ 1 year after the merger. This general picture of recoil-induced disk perturbations, and the dependence on kick direction, has been substantiated by later studies using full hydrodynamics simulations \citep[Fig.~\ref{fig:recoil_shocks};][]{megevand09, rossi10, corrales10, zanotti10, zanotti12, ponce12}. 

\citet{rossi10}, \citet{corrales10}, and \citet{zanotti10} all reach similar conclusions regarding the luminosity and timescale of a recoil-induced afterglow. They find that for a $10^6 M_{\odot}$ MBH, the luminosity should rise steadily over a timescale of $\sim$ months, reaching a peak of $\sim$ a few $\times 10^{43}$ erg s$^{-1}$, or $\sim 10\%$ of the Eddington luminosity at that MBH mass. This emission is likely to peak in the UV or soft X-ray bands. Variability is also a potential signature of these afterglows \citep{anderson10, rossi10, ponce12}. 

A number of complicating factors may make these recoil afterglows difficult to detect in practice. \citet{rossi10} explored a range of recoil kick angles relative to the disk plane and concluded that the energy available for dissipation depends strongly on this angle, by up to three orders of magnitude. They also found that the observability of the afterglow depends on the mass of the disk at the radii where most energy is deposited -- if the disk is truncated beyond the self-gravitating regime, then the afterglow luminosity is expected to be small, possibly undetectable. 

The recoil velocity is an important factor as well: clearly, if all else is equal, higher-velocity recoils will produce stronger disk perturbations. The brightest afterglows would presumably result from high-velocity recoils oriented in the plane of the disk. However, the largest component of a high recoil velocity (arising from misaligned MBH spins) will generally be perpendicular to the binary orbital plane prior to merger. Contributions to the recoil velocity directed into the orbital plane are more likely to dominate in the case of aligned MBH spins, which produce much smaller kick speeds \citep[e.g.,][]{baker08, vanmeter10, lousto12}. The MBH binary orbital plane need not be aligned with the plane of the circumbinary disk, of course, and it is reasonable to envision that such a scenario could occur in systems where the MBH spins remain misaligned with each other and with the disk all the way to decoupling.  

Another possible source of recoil-induced EM flares is the fallback of marginally-bound gas after the recoiling MBH and its bound accretion disk leave the galaxy nucleus \citep{shibon08}. This fallback could shock the surrounding gas, leading to a flare that should peak in the soft X-ray bands. The timescale for this type of flare ($\sim 10^3$--$10^4$ yr) would not be accessible as an EM counterpart to a GW source on human timescales, but some of these could be detected in AGN surveys if they are sufficiently luminous. \citet{shibon08} argue that the reprocessing of the soft X-ray flare by infalling material could produce a distinct UV/optical emission line spectrum.

Finally, attenuation by gas and dust is again likely to be a serious concern for observing many of these UV and soft X-ray afterglows, particularly because galactic nuclei often have high extinction. The possibility that some of this emission could be reprocessed and detected as an IR afterglow is therefore an appealing one to explore \citep{schnittman08}. We emphasize, however, that the detailed light curves and spectral evolution of any post-merger EM afterglows are still quite uncertain, so the possibility of observable flares at other wavelengths cannot be discounted. 

\subsubsection{Post-merger signatures of AGN jets}
\label{subsection:postmerger_jets}

The morphology of AGN radio jets may contain signatures of a recent MBH merger. If a MBH has an associated radio jet prior to merging with another MBH, the rapid reorientation of its spin induced by the merger  (``spin flip") could create a second, younger radio jet that is misaligned with the older, pre-merger radio lobes \citep[e.g.,][]{merritt02, dennett02}. \citet{liu04} suggest that similar features could be produced by the rapid reorientation of a BH by a misaligned accretion disk, via the Bardeen-Petterson effect. Such ``X-shaped" radio sources are indeed observed in hundreds of AGNs \citep[e.g.,][]{murgia01, merritt02, liu04, lal07, roberts15, bera20}. Numerous studies have concluded that the morphology and spectral features of some of these sources favor a scenario in which BH spin reorientation occurred on a short timescale \citep[$\lesssim 10^5$ yr; e.g., ][]{cheung07, mezcua11, gopal12, hernandez17, saripalli18}. However, other studies find that these X-shaped radio sources are more likely to arise from physics that has nothing to do with MBH spins, such as the backflow of jet material after it collides with a dense intergalactic medium, or the expansion of a jet-inflated cocoon along the minor axis of an elliptical galaxy \citep[e.g.,][]{leahy84, capetti02, kraft05, hodges10, hodges11, rossi17, joshi19, cotton20}. 

MBH mergers could also lead to the interruption of AGN jets, owing to either a drop in accretion following decoupling of a MBH binary from a circumbinary disk or following the MBH mass loss at merger. \citet{liu03} suggest that examples of these interrupted jets could be seen in double-double radio galaxies, where a younger set of radio jets is aligned with older radio lobes along the same axis. The inferred interruption timescales for many of these objects are quite long, however ($\sim$ Myr). Alternately, if a jet (re)launches after the circumbinary disk responds to the merger, this event could be observed as an EM counterpart to a GW detection on timescales of weeks to years \citep[e.g.,][]{ravi18, blecha18b, yuan21}.

\subsubsection{Long-lived accretion signatures of GW recoil}
\label{subsection:long_term_recoil}

Another interesting possibility is the detection of GW recoiling MBHs long after the merger has occurred---up to millions or tens of millions years later.  This scenario depends strongly on the pre-merger MBHB spins; perfectly-aligned spins will yield a maximum kick velocity $< 200$ km s$^{-1}$, while highly misaligned, rapidly spinning MBHs are necessary to produce kicks of up to $4000-5000$ km s$^{-1}$ \citep[e.g.,][]{campanelli07a, lousto12}. If MBHB spins are highly aligned prior to merger, the post-merger recoiling MBH will not stray far from the galactic center in most galaxies, and long-lived signatures of GW recoil will be difficult if not impossible to observe. As discussed in Sect.~\ref{sssection:implications_mbhb}, MBHB spin evolution is poorly constrained, and the degree of spin misalignment is sensitive to the nature of the accretion flow and the degree to which the MBH binary environment is gas-dominated \citep[e.g.,][]{king06, bogdanovic07, lodato13, miller13, gerosa15, sayeb21}. 

If an accretion disk is present at the time of the MBHB merger, it will generally remain bound to the merged MBH following a GW recoil event. A rough approximation of the radius within which gas remains bound to the ejected MBH is given by $r_{\rm ej} \sim G M_{\rm BH} / v_{\rm kick}^2$. This suggests that material within $\sim 10^4 - 10^5 R_g$ will remain bound to the MBH for all but the most extreme kick speeds of several thousand km s$^{-1}$. If one assumes a thin, viscous, radiatively-efficient accretion disk, this implies that the typical mass of gas carried along will be $M_{\rm ej}/M_{\rm BH} \sim$ a few percent \citep[e.g.,][]{loeb07, bleloe08}.

For these disk masses, simple estimates of the lifetime of the ejected accretion disk obtained from the viscous timescale $t_{\rm visc}$ or a $M_{\rm ej}/\dot M$ scaling yield recoiling AGN lifetimes of $\sim 1-10$ Myr \citep[][]{loeb07, bleloe08, volmad08}. In reality, the accretion rate should decline monotonically as the isolated accretion disk begins to diffuse outward  \citep[e.g.,][]{lynpri74, pringle81}. It is reasonable to assume that the accretion disk will cease to be fed at any significant rate, as the $v^{-3}$ scaling of Bondi-Hoyle accretion should prevent much additional gas from being swept up by the recoiling MBH. For an $\alpha$-disk, a self-similar solution for the evolution of the isolated disk surface density can be obtained, yielding a time-dependent accretion rate
\begin{eqnarray}
\dot M(t) = \dot M_{\rm mrg} \left ( {t \over t_0} \right )^{-19/16},\\
t_0 = {3 \over 16} {M_{\rm ej} \over \dot M_{\rm mrg}},
\end{eqnarray}
where $\dot M_{\rm mrg}$ is the accretion rate at the time of BH merger and $M_{\rm ej}$ is the accretion disk mass ejected along with the MBH \citep[][]{cannizzo90, pringle91, blecha11}. This can yield recoiling AGN lifetimes of tens of Myr up to $\sim 100$ Myr \citep{blecha11, blecha16}. Such objects could be seen as ``offset AGNs", which would be distinguishable from ``normal" AGNs if they are observed to be spatially and/or spectroscopically offset from their host galaxy \citep[e.g.,][]{madqua04, loeb07, volmad08, komossa08, blecha11, guedes11}.

\begin{figure}
\centering
\includegraphics[width=\textwidth]{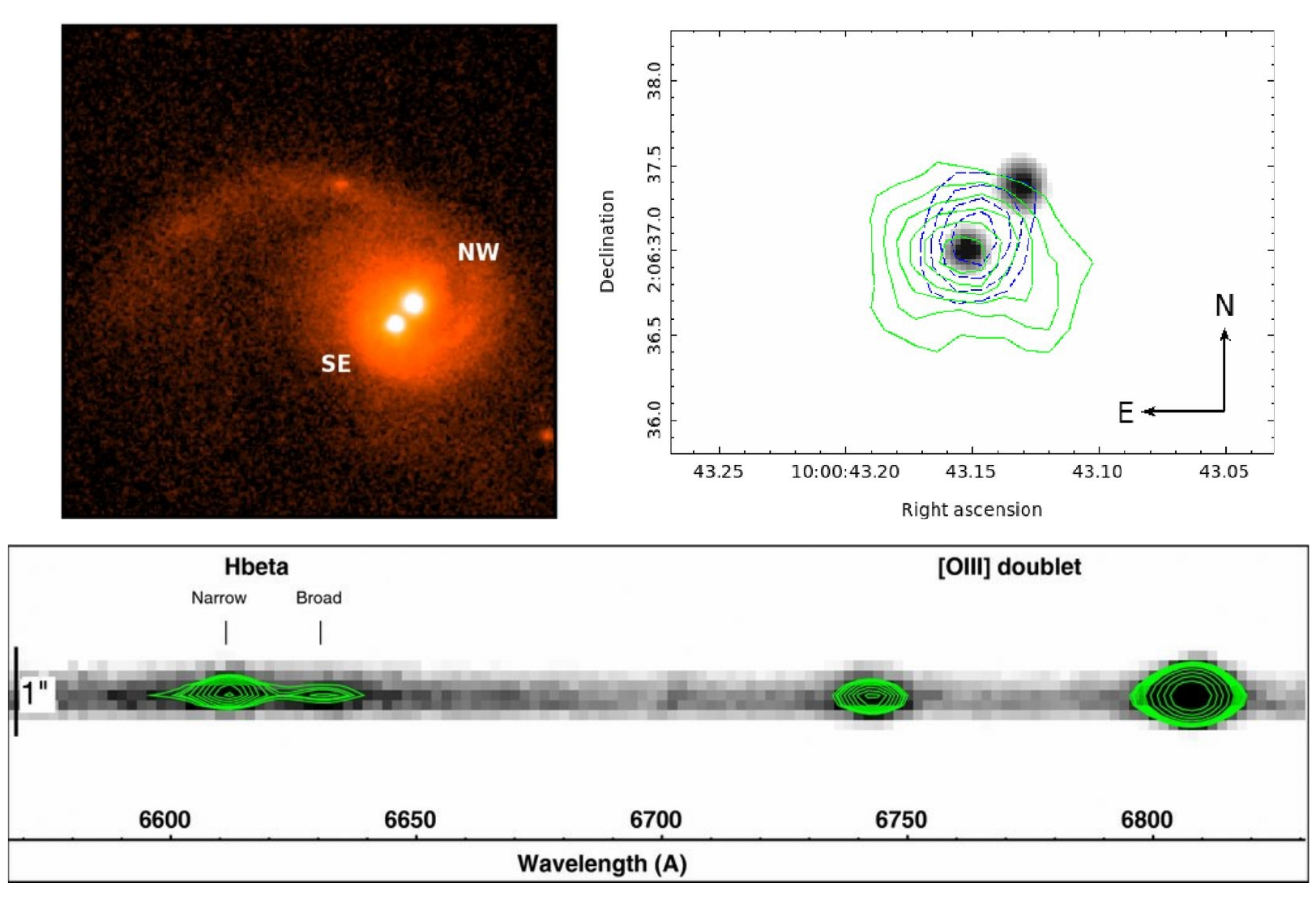}
\caption{Example of a candidate recoiling AGN that has both spatial and spectroscopic offsets. \textit{Top left:} \textit{HST}/ACS F814W image of the recoiling AGN candidate CID-42 and its host galaxy. The SE point source is the recoil candidate, while the NW source is the putative ``empty" nucleus. \textit{Top right:} the grayscale shows the \textit{HST}/ACS image scaled to highlight the two bright optical nuclei, the green contours show the \textit{Chandra} X-ray image \citep[from][]{civano12b}, and the blue contours show the VLA 3 GHz image. The X-ray and radio emission is clearly associated with the candidate recoiling AGN, with no evidence for an AGN in the NW nucleus. \textit{Bottom:} 2D IMACS spectrum of CID-42, focused on the H$\beta$ and [OIII] doublet region. A clear offset of 1360 km s$^{-1}$ is apparent between the broad and narrow components of the H$\beta$ line. Figure adapted from \citet{civano10} and \citet{novak15}.\label{fig:cid42}}
\end{figure}

The region that remains bound to the recoiling MBH will generally encompass the broad emission line region in addition to the accretion disk. Offset broad lines (BLs) in AGN spectra are therefore one potential signature of a recoiling AGN. Specifically, an AGN that is observed to have broad lines (BLs) offset from narrow emission lines (produced at much larger spatial scales) or from host stellar absorption lines could indicate a bulk velocity of the AGN relative to the galaxy (Fig.~ \ref{fig:cid42}, bottom panel). Because these emission lines are highly broadened and often asymmetric, however, accurate measurement of small centroid shifts is not trivial. The large majority of recoil kicks should be less than a few hundred km s$^{-1}$, even if the progenitor binary MBHs are misaligned and rapidly spinning. In this case, the BL offset will be very difficult if not impossible to detect robustly, given the typical BL FWHM of $\gtrsim 1000$ km s$^{-1}$. Moreover, if $v_{\rm kick} \ll v_{\rm esc}$ for the host galaxy, the recoiling AGN will soon begin to decelerate, further reducing the chances of observing a spectroscopic BL shift in its spectrum.

Nonetheless, the distribution of recoil velocities should have a tail of high-velocity kicks ($\gtrsim 1000$ km s$^{-1}$) as long as spins are not always highly aligned prior to merger \citep[][]{bogdanovic07, schnittman07, lousto12}. One of the first GW recoil candidates was discovered via its unusual quasar spectrum, which has BLs offset by $> 2600$ km s$^{-1}$ from its narrow lines \citep[NLs;][]{komossa08a}. Alternate scenarios for this object have been proposed, including a sub-pc MBH binary or a superposition of a quasar and a galaxy \citep[e.g.,][]{dotti09a, bogdanovic09, vivek09, shields09b, heckman09, decarli09, decarli10b}. The detection of galactic-scale molecular gas at the redshift of the BLs strongly disfavors the recoiling and binary MBH scenarios, but the origin of this system's unusual spectrum remains unknown \citep{decarli14}. Several other recoil candidates have been identified via BL offsets as well \citep[e.g.,][]{civano10, robinson10, hogg21}, and dedicated searches for BL offsets in SDSS quasar spectra revealed 88 systems with $> 1000$ km s$^{-1}$ offsets \citep[][]{eracleous12}. However, some of these are also candidate mpc MBH binary systems with one active MBH, and many are likely explained by less exotic scenarios such as high-velocity AGN outflows. Continued spectroscopic monitoring of these systems has already ruled out the binary or recoiling MBH scenarios for some of these objects \citep{runnoe15, runnoe17}. Relatedly, follow-up of offset BL AGNs identified in another systematic search of SDSS quasar spectra \citep{tsal11} reveals that the mid-infrared spectra of these objects resemble typical AGNs, suggesting that they are surrounded by a hot dusty ``torus'' component \citep{lusso14}. This may indicate that they are unlikely to be recoiling AGNs that have left their host nucleus \citep{hao10, guedes11}.

Once a recoiling AGN travels far enough from its host nucleus, it could be resolved as a distinct point source (i.e., a spatially offset AGN; Fig.~\ref{fig:cid42}, top panels). Of course, how far is ``far enough'' depends sensitively on the source distance, the spatial resolution of the observation, and the accuracy with which the host centroid can be pinpointed. \citet{blecha16} predict that large surveys with the Vera Rubin Observatory (formerly LSST), the Nancy Grace Roman Space Telescope (formerly WFIRST), and Euclid could potentially detect hundreds of spatially offset AGNs, if pre-merger MBH spins are not always highly aligned. Numerous candidates have been identified with $\sim$ kpc-scale spatial offsets \citep[][]{jonker10, koss14, markakis15, kalfountzou17, kim17}. In many cases, however, the recoiling AGN is still superimposed on the host galaxy at these separations, making these candidates difficult to distinguish from an \emph{infalling} AGN in an ongoing galaxy merger (Fig.~\ref{fig:cid42}). Depending on the specific configuration, host morphology may provide some circumstantial evidence to support one of these scenarios. A recoil candidate was recently identified in the nucleus of a brightest cluster galaxy (BCG) by \citet{condon17} in a VLBA search for offset AGNs, but the presence of a radio galaxy around the offset source indicates that it is instead a heavily tidally stripped galaxy moving away from the BCG center following a pericentric passage. In more typical cases, disturbed morphology in a host galaxy may provide supporting evidence for a kicked MBH. However, the presence or absence of disturbed morphology cannot on its own distinguish between a recoiling MBH and an inspiraling dual MBH, as tidal features are typically present for hundreds of Myr before and after the MBH merger. In contrast, the variable object SDSS 1133 is well separated from a nearby dwarf galaxy and has exhibited stochastic variability over a $>60$ year time baseline \citep{koss14}. Its spectral features are quite unusual for an AGN, though, so this object may instead be a rare luminous blue variable star exhibiting repeated, extreme, non-terminal outbursts \citep[][]{koss14, burke20, kokubo21, ward21}.

Arguably most promising are those recoil candidates that have both spatial and spectroscopic offset signatures \citep[Fig.~\ref{fig:cid42};][]{civano10, civano12b, blecha13a, novak15, chiaberge17, chiaberge18, hogg21, jadhav21}. Even in these cases, however, current data cannot exclude an inspiraling, kpc-scale MBH pair in which one MBH is quiescent and the other has BL offsets driven by high-velocity outflows. High-quality stellar or gas kinematics data from ALMA, JWST, or optical integral field units can provide an additional means for distinguishing between recoiling AGN and inspiraling dual AGN scenarios for ambiguous candidates. 

The possibility of offset AGN detections in time-domain surveys such as ZTF and the Vera Rubin Observatory is another promising avenue, because this allows one to distinguish spatially offset point sources that are AGN-like (i.e., stochastically variable) from transient sources such as supernovae \citep[][]{kumar15, ward21}. A recent search for offset AGNs in ZTF data revealed nine such candidates, in addition to 52 AGNs in ongoing galaxy mergers \citep{ward21}. Furthermore, a technique called ``varstrometry" has recently been developed to utilize variability information in high-astrometric-precision data from Gaia \citep{hwang20}. The stochastic variability observed from a spatially offset AGN will produce astrometric jitter in the photocenter of an unresolved AGN-host system. \citet{shen19} apply this principle to analysis of broad-line AGNs at $0.3 < z < 0.8$ in the Gaia DR2 data and find that 99\% are within 1 kpc of their host, 90\% are within 500 pc, and 40\% are centered to within 100 pc. They conclude that large spatial offsets are rare at low redshift.  

Searches for spatial offsets on much smaller ($\sim$ parsec) scales have also been carried out. The majority of recoiling MBHs will not be ejected entirely from their host galaxies, so they will eventually return to the galactic nucleus. Upon their return, they may undergo long-lived, small-amplitude oscillations about the galactic center, particularly if the central potential is shallow \citep[e.g.,][]{guamer08, bleloe08, volmad08}. Studies of elliptical galaxies with shallow central core profiles have indeed identified numerous AGNs with parsec-scale offsets from the optical photometric center of the host galaxy \citep[][]{batcheldor10, lena14, menezes14, makarov17}. However, some of these apparent offsets may be attributable to photometric and astrometric uncertainties \citep{jones19}. In particular, many of these objects contain prominent AGN jets, which could create the false appearance of an AGN offset \citep[as was recently argued for the apparent offset in M87;][]{loppri18}. 

As with binary MBH candidates \citep[e.g.,][]{burke18_ngVLA}, future high-sensitivity, high-resolution radio observations provide another potential avenue for confirming or ruling out existing GW recoil candidates, and for discovering new candidates \citep[e.g.,][]{blecha18b}. The ngVLA, particularly with a long-baseline configuration,  will provide strong constraints on AGN offsets and the possible presence of a secondary, faint AGN in the putative ``empty" host nucleus. In fact, if pre-merger MBH spins are nearly always highly aligned prior to merger, the ngVLA may provide the only means of detecting and confirming candidate recoiling AGNs. The broad frequency coverage of the ngVLA will also help to distinguish between genuine offset AGNs and transient jet features that may mimic these objects. 

A long baseline option ($>$ 1000 km) for the ngVLA would provide another intriguing possibility as well: detecting proper motions of rapidly-recoiling AGNs in the nearby Universe. Relative astrometric precision of $<$ 1\% of the beam FWHM would enable proper motions of 1 $\mu$as yr$^{-1}$ to be detected out to $\sim$ 200 Mpc over a 5-10 year time baseline for transverse recoil velocities $\gtrsim$ 1000 km s$^{-1}$ \citep[][]{blecha18b}. Achieving this astrometric precision for single objects (as opposed to binary MBHs) would be non-trivial, however. 

We note one additional potential gaseous EM signature of GW recoil. A recoiling MBH moving through a hot gaseous background can create density and temperature perturbations, leading to an observable enhancement in bremsstrahlung emission in the form of wakes trailing the BH \citep{devecchi09}. Recoiling AGNs that are still embedded in dense, gaseous galactic nuclei can also create hot, low-density wakes trailing their orbit, by imparting feedback energy to their surroundings \citep[][see also Fig.~\ref{fig10}]{sijacki11}. Observing the latter type of density wakes would likely be quite difficult, but as mentioned in Secti.~\ref{sssection:implications_mbhb}, these low-density regions will impact the recoiling MBH dynamics by significantly reducing gas dynamical friction.

\subsubsection{Stellar signatures of GW recoil}
\label{sssec:stellar_recoil}

\citet{stone11} showed that stellar tidal disruption events following a GW recoil kick could produce EM counterparts to MBH mergers on timescales of years to decades after coalescence, for MBH masses $\lesssim 10^7 M_{\odot}$. The instantaneous velocity imparted to a merged MBH by a GW recoil kick will immediately alter the phase space of stellar orbits in the frame of the recoiling MBH, which can have the effect of promptly refilling its stellar loss cone. As a result, \citet{stone11} find that in the years to decades following the MBH merger, the peak rate of tidal disruption events can be $\sim 10^4$ times higher than the typical rate in galaxies with single, non-recoiling MBHs. 

Stars could also produce long-lived EM signatures of GW recoil on Myr to Gyr timescales; these are likely the only possible EM counterpart for MBHs that are ejected without an accretion disk or that have exhausted their supply of bound gas. Most MBHs should have a tightly-bound stellar cusp that will be carried along with the MBH in the event of a recoil kick \citep[e.g.,][]{guamer08, kommer08a}. These ``hypercompact stellar systems" (HCSSs) around recoiling MBHs will generally have up to  $\sim$ 1\% of the MBH mass \citep[e.g.,][]{merritt09}. \citet{kommer08a} predict that the tidal disruption rates for these systems are only somewhat smaller than the rates for non-recoiling MBHs. Such events would exhibit the distinctive signatures of a tidal disruption flare that is offset from the nucleus of the host galaxy, or even a flare that occurs in intergalactic space. \citet{li12} find that tidal disruptions should also occur for bound recoiling MBHs when they pass through the galactic center, but at significantly lower rates than for non-recoiling MBHs. 

The masses and luminosities of HCSSs are expected to be similar to globular clusters, or perhaps ultracompact dwarf galaxies in the most optimistic scenarios. If observed, they would be distinguishable from other stellar systems by their extreme velocity dispersions (comparable to $v_{\rm kick}$). In particular, the extreme compactness and velocity dispersions of these systems should distinguish them from wandering MBHs that are expected to be produced by inefficient dynamical friction in low-mass galaxies \citep[e.g.,][]{tremmel18, reines20, bellovary21}. \citet{merritt09} predict that roughly 100 HCSSs could be detectable around the Virgo cluster, a few of which could be bright enough to be detectable. \citet{oleary09} estimate that a similar number of free-floating MBHs with HCSSs may be present in the MW halo---remnants from past recoil events throughout the assembly history of the MW. A search of the SDSS revealed about 100 recoiling cluster candidates, but detailed spectroscopic follow-up of each would be required to determine whether they are true HCSSs \citep[][]{oleary12}. The existence of dozens to hundreds of free-floating MBHs in galaxy halos has also been predicted using cosmological simulations that include GW recoil \citep[e.g.,][]{bellovary10}. Detecting and confirming the stellar counterparts to these wandering MBHs would be a significant challenge, but HCSSs around $\sim 10^5 M_{\odot}$ MBHs in the MW halo could be resolved with \textit{Euclid} \citep{lena20}. Notably, their long lifetimes and predicted ubiquity likely mean that HCSSs are actually the most common EM counterparts to recoiling MBHs. 

Finally, recoiling MBHs could leave imprints on the stellar population in their host galaxies. Repeated passages of a recoiling MBH through its host nucleus could increase the size of shallow-density stellar cores by exacerbating the ``core-scouring'' effect of stellar scattering during binary inspiral \citep{boylan04, guamer08, guedes09}. If a MBH merger takes place in a gas-rich environment, however, the opposite effect may occur: the displacement of the MBH---and its AGN feedback---from the galactic nucleus can allow higher rates of star formation, producing a denser stellar cusp than would form in the presence of a stationary MBH \citep[]{blecha11}. GW recoil events can also contribute to scatter in the BH-bulge relations, and they can produce a population of ``empty'' galactic nuclei with no BH at all \citep[e.g.,][]{volonteri06, volonteri07, volonteri10, blecha11, choksi17}. If superkicks of thousands of km s$^{-1}$ occur in nature, even brightest cluster galaxies (BCGs) should have an occupation fraction less than unity, which could be tested with a survey of BCGs using thirty-meter-class telescopes \citep{gerosa15c}.

\subsection{Questions uniquely addressed with detection of EM and GW signals from MBHBs}\label{section:multimessenger}

\begin{figure}[t]
\includegraphics[width=\textwidth]{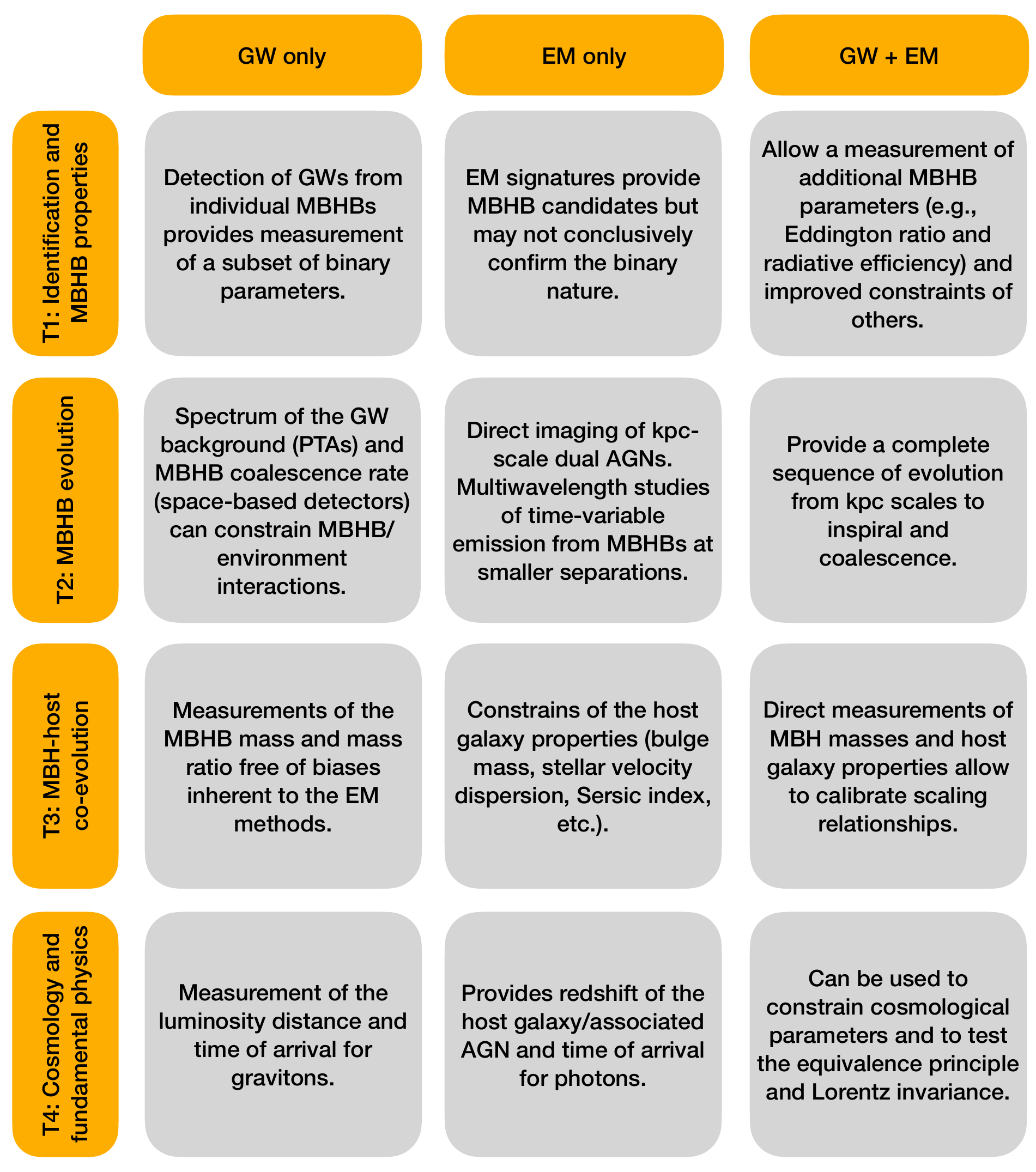}
\caption{Summary of major advances achievable through GW and EM observations of MBHBs.}
\label{fig_multi_chart}
\end{figure}

Having reviewed potential simultaneous and non-simultaneous EM counterparts to GW emission from MBHBs, we now consider what astrophysical questions can be uniquely addressed with the detection of both messengers. We model our discussion after \citet{kelley19_astro2020} and identify four topics that would benefit the most from such detections. A chart in Fig.~\ref{fig_multi_chart} summarizes major advances achievable through multimessenger observations of MBHB.
%

\begin{description}
\item[ {\bf T1.}] {\bf Identification and properties of individual MBHBs.} As discussed in Sect.~\ref{section:precursors}, \ref{section:coalescence} and \ref{section:postmerger} there are multiple promising EM signatures that can arise as counterparts to inspiraling and merging MBHBs. The lessons learned from the EM searches so far is that ``regular" AGNs may mimic some of these signatures, making the EM-only detections ambiguous (see Sect.~\ref{section:observations}). Multimessenger detections of MBHBs will be crucial in order to understand the unique aspects of EM signatures associated with MBHBs heading for coalescence, or with their merger remnants. Combined MBHB mass and mass ratio measurements (from GWs) and AGN luminosity (EM) can be used to estimate the Eddington luminosity ratio and radiative efficiency of the holes, thus enabling independent constraints on the properties of the spin. In the case of the PTA binaries, this combination of the EM and GW signatures will provide a unique way to learn about the properties of individual MBHBs, since GW alone will not place strong constraints on the binary parameters \citep{arzu14, shannon15, lentati15, liu21}. In the case of MBHBs detectable by the space-based GW observatories, the constraints on the MBHB orbit, mass ratio and the spins obtained from the two messengers will provide independent measurements that can be combined to increase the precision of the result.
\item[{\bf T2.}] {\bf Interactions of MBHBs with their environments.} Theoretical models predict that MBHBs evolve toward the GW regime through complex interactions with their galactic environment. On larger, galactic scales, these involve the stellar and gas dynamical friction (Sect.~\ref{sssection:df}). In galactic cores, they can include three-body scattering of stars,  interactions with additional MBHBs from subsequent mergers (Sect.~\ref{sssection:3body}), and interactions with the nuclear gas disk (Sect.~\ref{sssection:circumbinary}). For more massive MBHs, the detection of the GW background by the PTA observatories will strongly constrain interactions that take place in the galactic cores, since they affect the shape of the GW background spectrum \citep{rajagopal95,jaffe03,wyithe03,enoki04,sesana04}. For lower mass MBHs detected by the space-based GW observatories, like LISA, dynamical friction determines the total evolution time, and hence the cosmological coalescence rate of MBHBs, whereas the impact of the physical mechanisms that operate at smaller orbital separations is small on average \citep{volonteri20, li22a}. This means that the MBHB coalescence rate, obtained from the space-based GW measurements, will provide statistical constraints on the efficiency of dynamical friction in merger galaxies. Combined with the multiwavelength EM studies, which can supply spatially resolved, kpc-scale dual AGNs and time-variable emission from MBHBs at smaller separations, the two messengers can provide a complete sequence of evolution from the kpc scales all the way to inspiral and coalescence.
\item[{\bf T3.}] {\bf Co-evolution of MBHs with their host galaxies.} MBH-galaxy scaling laws, like the $M-\sigma$ relation, indicate that the growth of MBHs is intertwined with their hosts (see Sect.~\ref{ssection:rates_z} and references therein). PTA upper limits on the GW background already put constraints on these scalings for heavier MBHs \citep{simon16}, which will become more stringent once a GW background is detected. On the lower mass end, mass measurements obtained from space-based GW detections of individual MBHBs will provide a test of the mass measurements obtained from the EM observations. Together with information about the host galaxy properties available from the EM observations, (i.e., bulge mass, stellar velocity dispersion, Sersic index, etc.)  these will help to identify biases in scaling relationships, particularly for relatively weakly constrained populations of MBHs with the lowest and highest masses \citep{bennert11,mcconnell13,graham13,reines15,shankar16}.
\item[{\bf T4.}] {\bf MBHBs as probes of cosmology and fundamental physics.}  As discussed in Sect.~\ref{section:introduction}, the identification of the MBHB host galaxy provides an opportunity to turn merging MBHBs into standard sirens \citep{schutz86,holz05}, as the LIGO-Virgo Collaboration (LVC) did with the detection of a double neutron star merger \citep{abbott17_hubble}. The measurements by the PTAs and the space-based observatories will complement those by LVC, extending them to higher redshift and higher compact object masses. Coincident EM and GW detections can also be used to measure differences in the arrival times of photons and gravitons from the same source, and in such way constrain the mass of the graviton and point to possible violations of the equivalence principle and Lorentz invariance in the gravitational sector \citep{kocsis08,hazboun13,yagi16,abbott16_tests}.
\end{description}

\section{Conclusions}
\label{section:conclusions}

Future detections of EM and GW signatures from MBHBs would launch a new era of multimessenger astrophysics by expanding this growing field to the low-frequency GW regime. Such observations would provide unprecedented understanding of the evolution of MBHs and galaxies, they would constitute fundamentally new probes of cosmology, and they would enable unique tests of gravity. The next two decades are expected to open the door to the first coincident detections of EM and GW signatures associated with MBHBs heading for coalescence. 

Even prior to this, however, we may expect significant progress in observational studies of MBHBs. EM searches for gravitationally bound MBHBs may yield convincing cases for close binaries, and PTAs should detect nHz GWs from MBHBs or place stringent constraints on their numbers. PTA upper limits on the stochastic GW background amplitude from MBHBs have already placed meaningful constraints on some models of MBHB evolution and on statistical populations of candidate MBHBs identified in EM surveys. The upcoming Rubin Observatory will revolutionize time-domain astronomy, revealing a wealth of new EM MBHB candidates with periodically variable lightcurves and perhaps self-lensing spikes. The large number of TDEs that the Rubin Observatory will detect could also place some indirect constraints on the population of inspiraling MBHBs. Moreover, Rubin will identify many candidate recoiling AGNs offset from their host nuclei, and varstrometry measurements with Gaia will continue to provide complementary constraints on AGN offsets. 

New spectroscopic evidence for MBHBs is expected in the next decade as well. The Black Hole Mapper multi-epoch spectroscopic survey planned as part of SDSS-V will identify candidate MBHBs with periodically variable broad lines, and it will also expand our understanding of other sources of variability in broad-line AGNs. XRISM will resolve X-ray AGN spectra in unprecedented detail, including possible variability signatures of MBHBs in Fe K$\alpha$ line profiles.

A relevant question prior to the launch of the LISA mission is whether, in the absence of a GW precursor, the flares and variability from a MBHB coalescence can be detected in stand-alone EM observations. Given that more massive binary systems are also expected to be more luminous, a future serendipitous discovery of a $\gtrsim 10^7\,M_\odot$ MBHB coalescence by a burst alert mission cannot be excluded. A more systematic search will require deep monitoring of the transient sky with multiwavelength synoptic sky surveys. While this biases EM searches toward high binary masses, these avenues are complementary to future GW observations that will likely be optimized to search for the the lower- and higher-mass end supermassive MBHBs. Both will be required in order to eventually understand the properties of the MBHB population and their role in the evolution of galaxies and structure in the Universe.

At present, however, the uncertainty regarding the nature and observability of EM counterparts to MBHB coalescences is the largest barrier to identifying and interpreting their future multimessenger signals.  Unlike GW signatures, EM signatures have never been calculated from first principles for MBHB (and many other astrophysical) systems, due to the numerical complexity associated with propagation of radiation in matter. The first challenge for theoretical models is therefore to better predict the typical luminosities and distinctive spectral or variability signatures of possible EM counterparts. Furthermore, EM signals from coalescing MBHBs must be detectable against emission from and attenuation by the their host galaxies, and their variability must be distinguishable from that of normal AGNs. These are challenges for both theory and observations; both improved predictions from simulations and an advanced understanding of AGN variability from upcoming surveys will be key goals over the next decade. Finally, large uncertainties also remain regarding the formation timescales for close MBH pairs. While theoretical models generally agree that a ``worst-case'' scenario in which all MBHBs stall outside the GW regime is quite unlikely, the uncertainty in MBH merger rates is still a significant unknown for this coming era of multimessenger astrophysics. A combination of GW and EM observations along with improved theoretical models and simulations will be crucial for reducing these uncertainties in the coming years.

Despite these challenges, new facilities planned for the 2030s will open numerous promising avenues for EM studies of MBHBs. The ngVLA could reveal spatially resolved MBHBs on sub-pc scales, and proper motions may even be observable for some orbiting MBHBs. Athena will enable new high-resolution searches for MBHB signatures in Fe\,K$\alpha$ profiles. Moreover, simply monitoring variability in MBHB candidates over an additional decade will allow any periodic signals on $\sim$ years timescales to be observed over multiple complete cycles. PTA sensitivity similarly increases with longer time baselines. By the 2030s, PTAs are expected to have either detected a stochastic GW background from MBHBs or obtained strict upper limits that may strain existing theoretical models. Continuous-wave detections of $\sim$nHz GWs from \emph{single} MBHB sources are also possible on this timescale. Such sources would offer particular advantages for multimessenger studies, as their relatively nearby location and large MBH masses would enhance the prospects for identification of a corresponding EM signal. In addition to the enormous implications of such a multimessenger detection for astrophysics and cosmology, a nearby multimessenger MBHB source could provide invaluable insight into the possible EM signatures of MBHB mergers detected with LISA and other spaced-based GW observatories at much larger cosmological distances. Finally, space-based GW observatories will offer the possibility of observing coincident EM and GW signals from the actual coalescence of MBHBs.

%


\begin{acknowledgements}

We thank Manuela Campanelli, Xian Chen, Yi-Ming Hu, Julian Krolik and Carlos Lousto for their input and the anonymous referees for careful reading of the manuscript and constructive comments. T.B. acknowledges the support by the National Aeronautics and Space Administration (NASA) under award No. 80NSSC19K0319 and by the National Science Foundation (NSF) under award No. 1908042. T.B. and MCM acknowledge partial support by the National Science Foundation under grant No. NSF PHY-1748958 during their visit to the Kavli Institute for Theoretical Physics in Santa Barbara, CA. L.B. acknowledges support by NASA under award No. 80NSSC20K0502, by NSF under award No. AST-1909933, and by the Research Corporation for Science Advancement under Cottrell Scholar Award No. 27553.
\end{acknowledgements}

\phantomsection
\addcontentsline{toc}{section}{References}

\bibliographystyle{spbasic-FS} 
\bibliography{LRR_mcm_no_duplicates,LRR_tb,LRR_lb}

\end{document}